\documentclass[journal]{IEEEtran}
\usepackage{cite}
\usepackage[dvips]{graphicx}
\usepackage[cmex10]{amsmath}
\usepackage{color}
\usepackage{bm}
\usepackage{theorem}
\usepackage{amscd}
\usepackage{amssymb}

{\theorembodyfont{\normalfont}
\theoremheaderfont{\normalfont\it}
\newtheorem{thm}{ Theorem}
\newtheorem{dfn}[thm]{ Definition}
\newtheorem{lmm}[thm]{ Lemma}

}

{\theorembodyfont{\normalfont}
\theoremheaderfont{\normalfont\it}
\newtheorem{prf}{ Proof:}}

{\theorembodyfont{\normalfont}
\theoremheaderfont{\normalfont\it}
\newtheorem{rmk}{Remark:}}

{\theorembodyfont{\normalfont}
\theoremheaderfont{\normalfont\it}
}

{\theorembodyfont{\normalfont}
\theoremheaderfont{\normalfont\it}
}

{\theorembodyfont{\normalfont}
\theoremheaderfont{\normalfont\it}
}

\newcommand{\bra}[1]{\mbox{$\left\langle#1\right|$}}
\newcommand{\ket}[1]{\mbox{$\left|#1\right\rangle$}}
\newcommand{\inpro}[2]{\mbox{$\left\langle#1|#2\right\rangle$}}

\newcommand{\proj}[1]{\mbox{$\ket{#1}\!\bra{#1}$}}

\begin{document}

\title{Markovianizing Cost of Tripartite Quantum States}

\author{Eyuri Wakakuwa, Akihito Soeda and Mio Murao

\thanks{This work is supported by the Project for Developing Innovation Systems of MEXT, Japan and JSPS KAKENHI (Grant No.~23540463, No.~23240001, No.~26330006, and No.~15H01677).  We also gratefully acknowledge to the ELC project (Grant-in-Aid for Scientific Research on Innovative Areas MEXT KAKENHI (Grant No.~24106009)) for encouraging the research presented in this paper. This work was presented in part at ISIT 2015.}
\thanks{E. Wakakuwa is with the Department of Communication Engineering and Informatics, Graduate School of Informatics and Engineering, The University of Electro-Communications, Japan (email: wakakuwa@quest.is.uec.ac.jp).}
\thanks{A. Soeda is with the Department of Physics, Graduate School of Science, The University of Tokyo, Japan.}
\thanks{M. Murao is with the Department of Physics, Graduate School of Science, The University of Tokyo, and is with Institute for Nano Quantum Information Electronics, The University of Tokyo.}
}

\maketitle

\begin{abstract}
We introduce and analyze a task that we call {\it Markovianization}, in which a tripartite quantum state is transformed to a quantum Markov chain by a randomizing operation on one of the three subsystems. We consider cases where the initial state is the tensor product of $n$ copies of a tripartite state $\rho^{ABC}$, and is transformed to a quantum Markov chain conditioned by $B^n$ with a small error, using a random unitary operation on $A^n$. In an asymptotic limit of infinite copies and vanishingly small error, we analyze the {\it Markovianizing cost}, that is, the minimum cost of randomness per copy required for Markovianization. For tripartite pure states, we derive a single-letter formula for the Markovianizing costs. Counterintuitively, the Markovianizing cost is not a continuous function of states, and can be arbitrarily large even if the state is close to a quantum Markov chain. Our results have an application in analyzing the cost of resources for simulating a bipartite unitary gate by local operations and classical communication.
\end{abstract}


\section{Introduction}

Tripartite quantum states for which the quantum conditional mutual information (QCMI) is zero are called {\it short quantum Markov chains}, or {\it Markov states} for short\cite{hayden04}.  They play important roles, e.g., in analyzing the cost of quantum state redistribution \cite{deve08_2,jon09}, investigating effects of the initial system-environment correlation on the dynamics of quantum states \cite{francesco14}, and computing the free energy of quantum many-body systems \cite{poulin11}. 

In analogy to the quantum mutual information (QMI) of a bipartite state quantifying a distance to the closest product states, it would be natural to expect a similar relation between QCMI of a tripartite state and Markov states. However, this conjecture has been falsified\cite{ibinson08}(see also \cite{brandao11,li14}). The recent results show that the relation between QCMI and Markov states is not so straightforward\cite{ibinson08,brandao11,li14,liwinter14,fawzi15}, particularly when compared to the relation between QMI and product states.

From an operational point of view, QMI quantifies the minimum cost of randomness required for destroying the correlation between two quantum systems in an asymptotic limit of infinite copies \cite{berry08}. This fact and its variants including single-shot cases are called {\it decoupling theorems}, and have played a significant role in the development of quantum information theory for a decade \cite{horo05,horo07,abey09,fred10,berta11,patr08}. In a simple analogy, one may ask the following question: Is QCMI equal to the minimum cost of randomness required for transforming a tripartite state to a Markov state?

In this paper, we address this question, and answer in the negative. We derive a single-letter formula for the ``Markovianizing cost'' of pure states, that is, the minimum cost of randomness per copy required for Markovianizing tripartite pure states in the asymptotic limit of infinite copies. The obtained formula is not equal to QCMI, or not even a continuous function of states. Moreover, the Markovianizing cost of a state can be arbitrarily large, regardless of how close the state is to a Markov state. In the proof, we improve a random coding method using the Haar distributed random unitary ensemble, which is widely used in the proof of the decoupling theorems, by incorporating the mathematical structure of Markov states. 

There are two ways for defining the property of tripartite quantum states being ``approximately Markov'': one by the condition that the state is close to a Markov state, on which our definition of Markovianization in this paper is based; and the other by the condition that the state is {\it approximately recoverable}\cite{fawzi15}, i.e., there exists a quantum operation ${\mathcal E}:B\rightarrow BC$ such that $\rho^{ABC}\approx{\mathcal E}(\rho^{AB})$. Ref.\cite{fawzi15} proved that the latter condition has a direct connection with QCMI, namely, small QCMI implies recoverability with a small error. 

In \cite{waka15_rec}, we introduce another formulation of the Markovianizing cost by employing the concept of recoverability, and prove that the cost function is equal to the one obtained in this paper for pure states. We then apply the results in analyzing the cost of entanglement and classical communication for simulating a bipartite unitary gate by local operations and classical communication\cite{waka15_2}.  As a consequence, we prove in \cite{waka16_tradeoff} that there is a trade-off relation between the entanglement cost and the number of rounds of communication for a two-party distributed quantum information processing.

The structure of this paper is as follows. In Section \ref{sec:preliminaries}, we review mathematical theorems regarding the structure of quantum Markov chains, which are extensively used in this paper. In Section \ref{sec:results}, we introduce the formal definition of Markovianization, and describe the main results. Outlines of proofs of the main results are presented in Section \ref{sec:outline}. In Section\;\ref{sec:prop}, we describe properties of the Markovianizing cost. In Section \ref{sec:examples}, we calculate the Markovianizing cost of particular classes of tripartite pure states to illustrate its properties. Conclusions are given in Section \ref{sec:conclusion}. See Appendices for detailed proofs.\\

{\it Notations.}  A Hilbert space associated with a quantum system $A$ is denoted by ${\mathcal H}^A$, and its dimension is denoted by $d_A$. For $\rho\in{\mathcal S}({\mathcal H}^A)$, we denote ${\rm supp}[\rho]\subseteq{\mathcal H}^A$ by ${\mathcal H}^A_\rho$. A system composed of two subsystems $A$ and $B$ is denoted by $AB$. When $M$  and $N$ are linear operators on ${\mathcal H}^A$ and ${\mathcal H}^B$, respectively, we denote $M\otimes N$ as $M^A\otimes N^B$ for clarity. We abbreviate $|\psi\rangle^A\otimes|\phi\rangle^B$ as $|\psi\rangle^A|\phi\rangle^B$. The identity operator on a Hilbert space is denoted by $I$. We denote $(M^A\otimes I^B)\ket{\psi}^{AB}$ as $M^A\ket{\psi}^{AB}$, and $(M^A\otimes I^B)\rho^{AB}(M^A\otimes I^B)^{\dagger}$ as $M^A\rho^{AB}M^{A\dagger}$. We abbreviate $M^{AB}(\rho^A\otimes I^B)M^{\dagger AB}$ as $M^{AB}\rho^AM^{\dagger AB}$. When ${\mathcal E}$ is a quantum operation on $A$, we denote $({\mathcal E}\otimes{\rm id}^B)(\rho^{AB})$ as $({\mathcal E}^A\otimes{\rm id}^B)(\rho^{AB})$ or ${\mathcal E}^A(\rho^{AB})$.  For $\rho^{AB}$, $\rho^{A}$ represents ${\rm Tr}_B[\rho^{AB}]$. We denote $|\psi\rangle\!\langle\psi|$ simply as $\psi$. A system composed of $n$ identical systems of $A$ is denoted by $A^n$ or $\bar{A}$, and the corresponding Hilbert space is denoted by $({\mathcal H}^A)^{\otimes n}$ or ${\mathcal H}^{\bar A}$. The Shannon entropy of a probability distribution is denoted as $H(\{p_j\}_j)$, and the von Neumann entropy of a state $\rho^A$ is interchangeably denoted by $S(\rho^A)$ and $S(A)_\rho$.  $\log{x}$ represents the base $2$ logarithm of $x$.

\hfill

\section{Preliminaries}

\label{sec:preliminaries}
In this section, we present a decomposition of a Hilbert space called the {\it Koashi-Imoto (KI) decomposition}, which is introduced in \cite{koashi02} and is extensively used in the following part of this paper. We then summarize a result in \cite{hayden04}, which states that the structure of Markov states is characterized by the KI decomposition.

\subsection{Koashi-Imoto Decomposition}
For any set of states on a quantum system, operations on that system are classified into two categories: one that do not change any state in the set, and the other that changes at least one state in the set. It is proved in \cite{koashi02} that there exists an effectively unique way of decomposing a Hilbert space into a direct-sum form, in such a way that all quantum operations that do not change a given set of states have a simple form with respect to the decomposition. We call this decomposition of the Hilbert space as the {\it Koashi-Imoto decomposition}, or the {\it KI decomposition} for short. As we verify in Remark in this section, the KI decomposition is equivalently be represented in the form of a tensor product of three Hilbert spaces. Theorem 3 in \cite{koashi02}, which proves the existence of the KI decomposition, is described in this tensor-product form as follows. 

\begin{thm}\label{thm:kidec}(\!\! \cite{koashi02}, see also Theorem 9 in \cite{hayden04})
Consider a quantum system $A$ described by a finite dimensional Hilbert space ${\mathcal H}^A$. Associated to any set of states ${\mathfrak S}:=\{\rho_k\}_k$ on $A$, there exist three Hilbert spaces ${\mathcal H}^{a_{\scalebox{0.45}{$0$}}}$, ${\mathcal H}^{a_{\scalebox{0.45}{$L$}}}$, ${\mathcal H}^{a_{\scalebox{0.45}{$R$}}}$, an orthonormal basis $\{\ket{j}\}_{j\in J}$ of ${\mathcal H}^{a_{\scalebox{0.45}{$0$}}}$ and a linear isometry $\Gamma$ from ${\mathcal H}^A_{\scalebox{0.6}{${\mathfrak S}$}}:={\rm supp}(\sum_k\rho_k)\subseteq{\mathcal H}^A$ to ${\mathcal H}^{a_{\scalebox{0.45}{$0$}}}\otimes{\mathcal H}^{a_{\scalebox{0.45}{$L$}}}\otimes{\mathcal H}^{a_{\scalebox{0.45}{$R$}}}$, such that the following three properties hold. (For later convenience, we exchange labels $L$ and $R$ in the original formulation.)

\begin{enumerate}

\item The states in ${\mathfrak S}$ are decomposed by $\Gamma$ as
\begin{eqnarray}
\Gamma\rho_k\Gamma^\dagger=\sum_{j\in J}p_{j|k}\proj{j}^{a_{\scalebox{0.45}{$0$}}}\otimes\omega_{j}^{a_{\scalebox{0.45}{$L$}}}\otimes\rho_{j|k}^{a_{\scalebox{0.45}{$R$}}}
\label{eq:kidecset}
\end{eqnarray}
with some probability distribution $\{p_{j|k}\}_{j\in J}$ on $J:=\{1,\cdots,{\rm dim}{\mathcal H}^{a_{\scalebox{0.45}{$0$}}}\}$, states $\omega_{j}\in{\mathcal S}({\mathcal H}^{a_{\scalebox{0.45}{$L$}}})$ and $\rho_{j|k}\in{\mathcal S}({\mathcal H}^{a_{\scalebox{0.45}{$R$}}})$.

\item A quantum operation $\mathcal E$ on ${\mathcal S}({\mathcal H}^A_{\scalebox{0.6}{${\mathfrak S}$}})$ leaves all $\rho_k$ invariant if and only if there exists an isometry $U:{\mathcal H}_{\scalebox{0.6}{${\mathfrak S}$}}^A\rightarrow{\mathcal H}_{\scalebox{0.6}{${\mathfrak S}$}}^A\otimes{\mathcal H}^E$ such that a Stinespring dilation of $\mathcal E$ is given by ${\mathcal E}(\tau)={\rm Tr}_E[U\tau U^{\dagger}]$, and that $U$ is decomposed by $\Gamma$ as
\begin{eqnarray}
(\Gamma\otimes I^E)U\Gamma^\dagger=\sum_{j\in J}\proj{j}^{a_{\scalebox{0.45}{$0$}}}\otimes U_j^{a_{\scalebox{0.45}{$L$}}}\otimes I_j^{a_{\scalebox{0.45}{$R$}}}.
\label{eq:kidecofsd}
\end{eqnarray}
Here, $I_j$ are the identity operators on ${\mathcal H}_j^{a_{\scalebox{0.45}{$R$}}}:={\rm supp}\sum_k\rho_{j|k}$, and $U_j:{\mathcal H}_j^{a_{\scalebox{0.45}{$L$}}}\rightarrow{\mathcal H}_j^{a_{\scalebox{0.45}{$L$}}}\otimes{\mathcal H}^{E}$ are isometries that satisfy ${\rm Tr}_E[U_j\omega_jU_j^{\dagger}]=\omega_j$ for all $j$, where ${\mathcal H}_j^{a_{\scalebox{0.45}{$L$}}}:={\rm supp}\:\omega_{j}$.
\item $\Gamma$ satisfies
\begin{eqnarray}
{\rm img}\Gamma=\bigoplus_{j\in J}{\mathcal H}_j^{a_{\scalebox{0.45}{$0$}}}\otimes{\mathcal H}_j^{a_{\scalebox{0.45}{$L$}}}\otimes{\mathcal H}_j^{a_{\scalebox{0.45}{$R$}}},\label{eq:imgGamma}
\end{eqnarray}
where ${\mathcal H}_j^{a_{\scalebox{0.45}{$0$}}}$ are one-dimensional subspaces of ${\mathcal H}^{a_{\scalebox{0.45}{$0$}}}$ spanned by $|j\rangle$.

\item ${\mathcal H}^{a_{\scalebox{0.45}{$0$}}}$, ${\mathcal H}^{a_{\scalebox{0.45}{$L$}}}$ and ${\mathcal H}^{a_{\scalebox{0.45}{$R$}}}$ are minimal in the sense that
\begin{align}
\dim{\mathcal H}^{a_{\scalebox{0.45}{$L$}}}=\max_{j\in J}\dim{\mathcal H}_j^{a_{\scalebox{0.45}{$L$}}},\;\dim{\mathcal H}^{a_{\scalebox{0.45}{$R$}}}=\max_{j\in J}\dim{\mathcal H}_j^{a_{\scalebox{0.45}{$R$}}}\nonumber
\end{align}
and
\begin{align}
\forall j\in J,\:\exists k\text{ s.t. }p_{j|k}>0.\nonumber
\end{align}

\end{enumerate}
\end{thm}
We call $\Gamma$ as the {\it KI isometry on system $A$ with respect to ${\mathfrak S}$}. The KI decomposition and the corresponding KI isometry are uniquely determined from ${\mathfrak S}$, up to trivial changes of the basis (Lemma 7 in \cite{koashi02}). The dimensions of ${\mathcal H}^{a_{\scalebox{0.45}{$0$}}}$, ${\mathcal H}^{a_{\scalebox{0.45}{$L$}}}$ and ${\mathcal H}^{a_{\scalebox{0.45}{$R$}}}$ are at most $d_A$. An algorithm for obtaining the KI decomposition is proposed in \cite{koashi02}.

It is also proved in \cite{koashi02} that the sets of states $\{\rho_{j|k}\}_k$ in (\ref{eq:kidecset}) are {\it irreducible} in the following sense.

\begin{lmm}\label{lmm:irrki} (Corollary of Lemma 6 in \cite{koashi02})
The set of states $\{\rho_{j|k}\}_k$ in (\ref{eq:kidecset}) satisfies the following properties.
\begin{enumerate}
\item If a linear operator $N$ on ${\mathcal H}_j^{a_{\scalebox{0.45}{$R$}}}:={\rm supp}\sum_k\rho_{j|k}$ satisfies $p_{j|k}N\rho_{j|k}=p_{j|k}\rho_{j|k}N$ for all $k$, then $N=cI_j^{a_{\scalebox{0.45}{$R$}}}$ for a complex number $c$, where $I_j^{a_{\scalebox{0.45}{$R$}}}$ is the identity operator on ${\mathcal H}_j^{a_{\scalebox{0.45}{$R$}}}$.
\item If a linear operator $N:{\mathcal H}_j^{a_{\scalebox{0.45}{$R$}}}\rightarrow{\mathcal H}_{j'}^{a_{\scalebox{0.45}{$R$}}}\;(j\neq j')$ satisfies $p_{j|k}N\rho_{j|k}=p_{j'|k}\rho_{j'|k}N$ for all $k$, then $N=0$.
\end{enumerate}
\end{lmm}

Let us now describe an extension of the KI decomposition to a bipartite quantum states, which is introduced in \cite{hayden04}. Associated to any bipartite state $\Psi^{AA'}\!\in{\mathcal S}({\mathcal H}^A\otimes{\mathcal H}^{A'})$, there exists a set of states on $A$ to which system $A$ can be {\it steered} through $\Psi^{AA'}$, i.e., the set of states that can be prepared by performing a measurement on $A'$ on the state $\Psi^{AA'}$ and post-selecting one outcome. The KI decomposition of $A$ with respect to the set is then associated to $\Psi^{AA'}$. It happens that any quantum operation on $A$ which leaves all states in the set invariant also leaves $\Psi^{AA'}$ invariant, and vice versa. Hence the set of operations preserving $\Psi^{AA'}$ is completely characterized by the corresponding KI decomposition. More precisely, we have the following statements.
\begin{dfn}\label{dfn:kibipart}
Consider quantum systems $A$ and $A'$ described by finite dimensional Hilbert spaces ${\mathcal H}^A$ and ${\mathcal H}^{A'}$, respectively. The KI decomposition of system $A$ with respect to a bipartite state $\Psi^{AA'}\!\in{\mathcal S}({\mathcal H}^A\otimes{\mathcal H}^{A'})$ is defined as the KI decomposition of $A$ with respect to the following set ${\mathfrak S}_{\Psi^{A'\rightarrow A}}$ of states, i.e., using ${\mathcal L}({\mathcal H}^{A'})$ to denote the set of linear operators on ${\mathcal H}^{A'}$,
\begin{eqnarray}
&&\!\!\!\!\!\!\!\!\!\!\!\!\!\!\!\!\!{\mathfrak S}_{\Psi^{A'\rightarrow A}}:=\{\rho\in{\mathcal S}({\mathcal H}^A)\:|\:\text{$\exists M \in {\mathcal L}({\mathcal H}^{A'})$ s.t.}\nonumber\\
&&\;\;\;\;\;\;\;\;\;\;\;\;\;\;\;\;\;\;\;\;\;\;\;\;\;\;\;\;\;\text{$\rho={\rm Tr}_{A'}[M^{A'}\!\Psi^{AA'}M^{\dagger A'}]$}\}.\label{eq:steerablestates}
\end{eqnarray}
The KI isometry on system $A$ with respect to $\Psi^{AA'}$ is defined as that with respect to ${\mathfrak S}_{\Psi^{A'\rightarrow A}}$.
\end{dfn}

\begin{lmm}\label{lmm:qqq}(See the proof of Theorem 6 in \cite{hayden04} and Equality (14) therein.)
Let $\Gamma$ be the KI isometry on $A$ with respect to $\Psi^{AA'}$, and define ${\mathcal H}^A_\Psi:={\rm supp}[\Psi^A]\subseteq{\mathcal H}^A$. $\Gamma$ satisfies the following properties.
\begin{enumerate}
\item
$\Gamma$ gives
\begin{eqnarray}
\Psi^{AA'}_{K\!I}:=\Gamma^A\Psi^{AA'}\Gamma^{\dagger A}=\sum_{j\in J}p_j\proj{j}^{a_{\scalebox{0.45}{$0$}}}\otimes\omega_j^{a_{\scalebox{0.45}{$L$}}}\otimes\varphi_j^{a_{\scalebox{0.45}{$R$}}A'}\!\!\!\!\!\!\!\!\!\!\nonumber\\
\label{eq:koashi02ofbistate}
\end{eqnarray}
with some probability distribution $\{p_{j}\}_{j\in J}$, orthonormal basis $\{\ket{j}\}_{j\in J}$ of ${\mathcal H}^{a_{\scalebox{0.45}{$0$}}}$, states $\omega_j\in{\mathcal S}({\mathcal H}^{a_{\scalebox{0.45}{$L$}}})$ and $\varphi_j\in{\mathcal S}({\mathcal H}^{a_{\scalebox{0.45}{$R$}}}\otimes{\mathcal H}^{A'})$.
\item A quantum operation $\mathcal E$ on ${\mathcal S}({\mathcal H}^A_\Psi)$ leaves $\Psi^{AA'}$ invariant only if there exists an isometry $U:{\mathcal H}^A_{\Psi}\rightarrow{\mathcal H}^A_{\Psi}\otimes{\mathcal H}^E$ such that a Stinespring dilation of $\mathcal E$ is given by ${\mathcal E}(\tau)={\rm Tr}_E[U\tau U^{\dagger}]$, and that $U$ is decomposed by $\Gamma$ as (\ref{eq:kidecofsd}), in which case we define ${\mathcal H}_j^{a_{\scalebox{0.45}{$R$}}}:={\rm supp}\varphi_j^{a_{\scalebox{0.45}{$R$}}}$.
\end{enumerate}
\end{lmm}
We call (\ref{eq:koashi02ofbistate}) as the {\it KI decomposition of $\Psi^{AA'}$ on $A$}. The following lemma, regarding the equivalence between the KI isometries of two bipartite states, immediately follows.
\begin{lmm}\label{lmm:equivkidec}
The following conditions are equivalent when ${\mathcal H}^A_{\Psi_1}={\mathcal H}^A_{\Psi_2}={\mathcal H}^A$:
\begin{enumerate}
\item A quantum operation on $A$ leaves a state $\Psi_1^{AA'}$ invariant if and only if it leaves a state $\Psi_2^{AA''}$ invariant. 
\item The KI isometries on $A$ with respect to $\Psi_1^{AA'}$ and $\Psi_2^{AA''}$ are the same.
\end{enumerate}
\end{lmm}
We define the sub-KI isometries as follows.
\begin{dfn}
Consider a bipartite state $\Psi^{AA'}$, three Hilbert spaces ${\mathcal H}^{a_{\scalebox{0.45}{$0$}}}$, ${\mathcal H}^{a_{\scalebox{0.45}{$L$}}}$, ${\mathcal H}^{a_{\scalebox{0.45}{$R$}}}$ and let $\Gamma$ be a linear isometry from ${\mathcal H}^A_\Psi$ to ${\mathcal H}^{a_{\scalebox{0.45}{$0$}}}\otimes{\mathcal H}^{a_{\scalebox{0.45}{$L$}}}\otimes{\mathcal H}^{a_{\scalebox{0.45}{$R$}}}$. We call $\Gamma$ as a sub-KI isometry on system $A$ with respect to $\Psi^{AA'}$ if it satisfies Condition 1) in Lemma \ref{lmm:qqq}.
\end{dfn}

\begin{figure}[t]
\begin{center}
\includegraphics[bb={0 0 396 302}, scale=0.3]{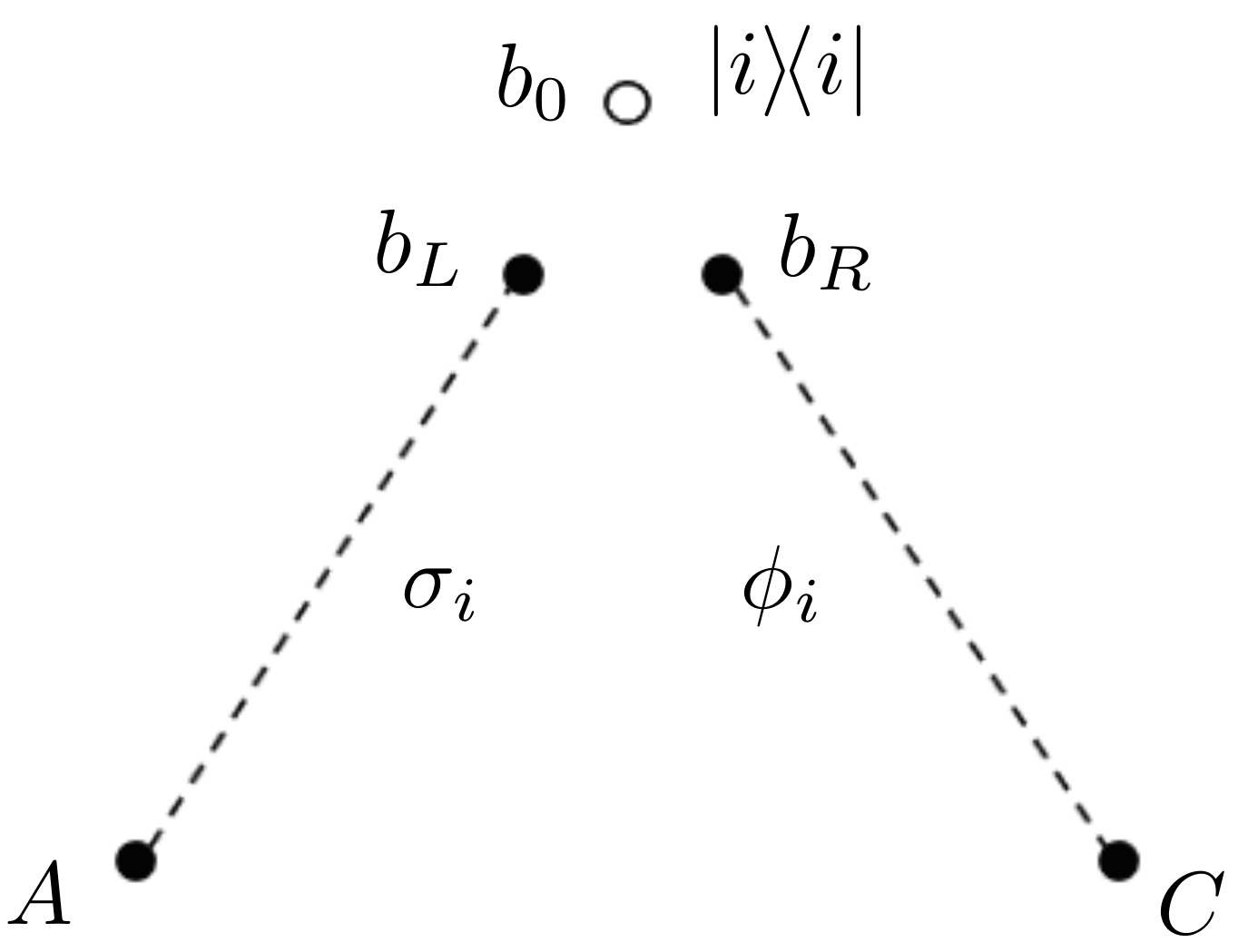}
\end{center}\caption{A graphical representation of the Markov decomposition of the Markov state (\ref{eq:markovdecofstate}).  Each vertex corresponds to a quantum system, and the white circle represents a `classical' system, where the state of the whole system is diagonal with respect to $|{i}\rangle^{b_0}$. The dotted lines represent mixed states. The whole state is the probabilistic mixture of the above state with probability $q_i$, namely, $\sum_{i}q_i\proj{i}^{b_0}\otimes\sigma_i^{Ab_L}\otimes\phi_i^{b_RC}$.}
\label{fig:markovdec}
\end{figure}

\subsection{Markov States}
A tripartite quantum state $\Upsilon^{ABC}$ is called a Markov state conditioned by $B$ if it satisfies $I(A:C|B)_\Upsilon=0$. It is proved in \cite{hayden04} that the structure of Markov states is characterized by the KI decomposition as follows.
\begin{thm}\label{thm:markovdechayden}(See Theorem 6 in \cite{hayden04} and the proof thereof.)
The following three conditions are equivalent:
\begin{enumerate}
\item $\Upsilon^{ABC}$ is a Markov state conditioned by $B$.
\item There exist three Hilbert spaces ${\mathcal H}^{b_{\scalebox{0.45}{$0$}}}$, ${\mathcal H}^{b_{\scalebox{0.45}{$L$}}}$, ${\mathcal H}^{b_{\scalebox{0.45}{$R$}}}$ and a linear isometry $\Gamma$ from ${\mathcal H}^B_\Upsilon:={\rm supp}[\Upsilon^B]$ to ${\mathcal H}^{b_{\scalebox{0.45}{$0$}}}\otimes{\mathcal H}^{b_{\scalebox{0.45}{$L$}}}\otimes{\mathcal H}^{b_{\scalebox{0.45}{$R$}}}$ such that $\Upsilon^{ABC}$ is decomposed by $\Gamma$ as
\begin{eqnarray}
\Gamma^B\Upsilon^{ABC}\Gamma^{\dagger B}=\sum_{i}q_i\proj{i}^{b_{\scalebox{0.45}{$0$}}}\otimes\sigma_i^{Ab_{\scalebox{0.45}{$L$}}}\otimes\phi_i^{b_{\scalebox{0.45}{$R$}}C}
\label{eq:markovdecofstate}
\end{eqnarray}
with some probability distribution $\{q_i\}_{i}$, orthonormal basis $\{\ket{i}\}_i$ of ${\mathcal H}^{b_{\scalebox{0.45}{$0$}}}$, states $\sigma_i\in{\mathcal S}({\mathcal H}^A\otimes{\mathcal H}^{b_{\scalebox{0.45}{$L$}}})$ and $\phi_i\in{\mathcal S}({\mathcal H}^{b_{\scalebox{0.45}{$R$}}}\otimes{\mathcal H}^{C})$.
\item $\Upsilon^{ABC}$ is decomposed in the form of (\ref{eq:markovdecofstate}) with $\Gamma$ being the KI isometry on $B$ with respect to $\Upsilon^{BC}$.
\item  There exist quantum operations $\mathcal R$ from $B$ to $BC$ and ${\mathcal R}'$ from $B$ to $AB$ such that
\begin{eqnarray}
\Upsilon^{ABC}={\mathcal R}(\Upsilon^{AB})={\mathcal R}'(\Upsilon^{BC}).\label{eq:recoverable}
\end{eqnarray}
\end{enumerate}
\end{thm}
We call (\ref{eq:markovdecofstate}) as a {\it Markov decomposition} of a Markov state $\Upsilon^{ABC}$ (Figure \ref{fig:markovdec}).

\begin{rmk}
The KI decomposition is first proved in \cite{koashi02} by an algorithmic construction, and by an algebraic proof in \cite{hayden04} afterward. A similar decomposition is derived in \cite{robin08} and \cite{robin10} in the context of ``information preserving structure''. In these literatures, the decomposition is given in the form of the direct sum of Hilbert spaces as $\bigoplus_j{\mathcal H}_j^L\otimes{\mathcal H}_j^R$. This is equivalent to the decomposition in the form of a tensor product of three Hilbert spaces described in this section, as verified by choosing ${\mathcal H}^{a_{\scalebox{0.45}{$0$}}}$, ${\mathcal H}^{a_{\scalebox{0.45}{$L$}}}$ and ${\mathcal H}^{a_{\scalebox{0.45}{$R$}}}$ such that $\dim{{\mathcal H}^{a_{\scalebox{0.45}{$0$}}}}=|J|$, $\dim{{\mathcal H}^{a_{\scalebox{0.45}{$L$}}}}=\max_{j\in J}{\mathcal H}_j^L$ and $\dim{{\mathcal H}^{a_{\scalebox{0.45}{$R$}}}}=\max_{j\in J}{\mathcal H}_j^R$. The corresponding KI isometry is defined as
\begin{align}
\Gamma:=\sum_{j\in J}|j\rangle^{a_{\scalebox{0.45}{$0$}}}\otimes(\Gamma_{L,j}\otimes\Gamma_{R,j})P_j
\end{align}
where $\Gamma_{L,j}:{\mathcal H}_j^L\rightarrow{\mathcal H}^{a_{\scalebox{0.45}{$L$}}}$ and $\Gamma_{R,j}:{\mathcal H}_j^R\rightarrow{\mathcal H}^{a_{\scalebox{0.45}{$R$}}}$ are linear isometries and $P_j$ is the projection onto ${\mathcal H}_j^L\otimes{\mathcal H}_j^R\in{\mathcal H}$. As stressed in \cite{koashi02}, ${\mathcal H}^{a_{\scalebox{0.45}{$0$}}}$ in (\ref{eq:kidecset}) holds the ``classical'' part of information possessed by $\rho_k$, ${\mathcal H}^{a_{\scalebox{0.45}{$R$}}}$ the ``quantum'' part, and ${\mathcal H}^{a_{\scalebox{0.45}{$L$}}}$ the redundant part.
\end{rmk}

\hfill

\section{Definitions and Main Results}
\label{sec:results}

In this section, we introduce the formal definition of Markovianization, and state the main results on the Markovianizing cost of tripartite pure states. The outlines of proofs are given in Section \ref{sec:outline}. Rigorous proofs will be given in Appendix \ref{app:rigorous} and \ref{app:comp}. 

\begin{dfn}\label{dfn:mofpsi}
A tripartite state $\rho^{ABC}$ is {\it Markovianized} with the randomness cost $R$ on $A$, conditioned by $B$, if the following statement holds. That is, for any $\epsilon>0$, there exists $n_\epsilon$ such that for any $n\geq n_\epsilon$, we find a random unitary operation ${\mathcal V}_n:\tau\mapsto2^{-nR}\sum_{k=1}^{2^{nR}}V_k\tau V_k^{\dagger}$ on $A^n$ and a Markov state $\Upsilon^{A^nB^nC^n}$ conditioned by $B^n$ that satisfy
\begin{eqnarray}
\left\|{\mathcal V}_n^{A^n}(\rho^{\otimes n})-\Upsilon^{A^nB^nC^n}\right\|_1\leq\epsilon.
\label{eq:defmarkovianizing}
\end{eqnarray}
The {\it Markovianizing cost} of $\rho^{ABC}$ is defined as  $M_{A|B}(\rho^{ABC}):=\inf\{R\:|\:\rho^{ABC}$ is Markovianized with the randomness cost $R$ on $A$, conditioned by $B\}$.
\end{dfn}

The following theorem is the main contribution of this work. The outline of the proof is given in the next section.
\begin{thm}\label{thm:strongmarkcostequality}
Let $|\Psi\rangle^{ABC}$ be a pure state, and let
\begin{eqnarray}
\Psi_{K\!I}^{AC}=\sum_{j\in J}p_j\proj{j}^{a_{\scalebox{0.45}{$0$}}}\otimes\omega_j^{a_{\scalebox{0.45}{$L$}}}\otimes\varphi_j^{a_{\scalebox{0.45}{$R$}}C}
\label{eq:kidecofpsi}
\end{eqnarray}
be the KI decomposition of $\Psi^{AC}$ on $A$. Then we have
\begin{eqnarray}
M_{A|B}(\Psi^{ABC})=H(\{p_j\}_{j\in J})+2\sum_{j\in J}p_jS(\varphi_j^{a_{\scalebox{0.45}{$R$}}}).\nonumber
\end{eqnarray}
\end{thm}

Based on this theorem, it is possible to compute the Markovianizing cost of pure states once we obtain the KI decomposition of its bipartite reduced density matrix. However, the algorithm for obtaining the KI decomposition, which is proposed in \cite{koashi02}, involves repeated application of decompositions of the Hilbert space into subspaces, and is difficult to execute in general.

Below we propose an algorithm by which we can compute the Markovianizing cost for a particular class of pure states, without obtaining an explicit form of the KI decomposition. The algorithm is based on the following theorem, which connects the Markovianizing cost of a pure state and the Petz recovery map corresponding to the state. Here, the {\it Petz recovery map} of a tripartite state $\Psi^{ABC}$ from $A$ to $AC$, an idea first introduced in \cite{hayden04}, is defined by
\begin{eqnarray}
\;\;\;{\mathcal R}_\Psi^{A\rightarrow AC}(\tau)=(\Psi^{AC})^{\frac{1}{2}}(\Psi^{A})^{-\frac{1}{2}}\tau(\Psi^{A})^{-\frac{1}{2}}(\Psi^{AC})^{\frac{1}{2}}\nonumber\\
(\forall\tau\in{\mathcal S}({\mathcal H}^A)).\nonumber
\end{eqnarray}
A proof of the theorem will be given in Appendix \ref{app:comp}.

\begin{thm}\label{thm:comp}
Let $|\Psi\rangle^{ABC}$ be a pure state, such that a CPTP map ${\mathcal E}$ on ${\mathcal S}({\mathcal H}^A_{\Psi})$ defined by
\begin{eqnarray}
{\mathcal E}:={\rm Tr}_C\circ{\mathcal R}_\Psi^{A\rightarrow AC}
\label{eq:defmathe}
\end{eqnarray}
is self-adjoint. Define another CPTP map ${\mathcal E}_\infty$ by
\begin{eqnarray}
{\mathcal E}_\infty:=\lim_{N\rightarrow\infty}\frac{1}{N}\sum_{n=1}^N{\mathcal E}^n,
\label{def:einfty}
\end{eqnarray}
and consider the state
\begin{eqnarray}
\Psi_\infty^{ABC}:={\mathcal E}_\infty^A(|\Psi\rangle\!\langle\Psi|^{ABC}).\label{def:einfty2}
\end{eqnarray}
Then we have
\begin{eqnarray}
M_{A|B}(\Psi^{ABC})=S(\Psi_\infty^{ABC}).\label{eq:miss}
\end{eqnarray}
\end{thm} 

Due to this theorem, the Markovianizing cost of pure states can be computed by the following algorithm, based on a matrix representation of CPTP maps. Here, $\{|k\rangle\}_{k=1}^{d_A}$ is an orthonormal basis of ${\mathcal H}^A$, and  $[\cdot]_{kl,mn}$ denotes a matrix element in the $kl$-th row and the $mn$-th column. (See also Remark in Appendix \ref{app:prfmiss}.)
\begin{enumerate}
\item Compute $d_A^2$-dimensional square matrices $\Lambda_1$, $\Lambda_2$ and $\Lambda$ given by
\begin{eqnarray}
\![\Lambda_1]_{kl,mn}\!\!&=&\!\!\langle k|(\Psi^A)^{\frac{1}{2}}|m\rangle\langle n|(\Psi^A)^{\frac{1}{2}}|l\rangle,\nonumber\\
\![\Lambda_2]_{kl,mn}\!\!&=&\!\!{\rm Tr}\left[\langle k|^A(\Psi^{AC})^{\frac{1}{2}}|m\rangle^A\langle n|^A(\Psi^{AC})^{\frac{1}{2}}|l\rangle^A\right]\nonumber
\end{eqnarray}
and ${\Lambda}=\Lambda_2\Lambda_1^{-1}$, where the superscript $-1$ denotes the generalized inverse. 
\item Check the hermiticity of ${\Lambda}$, which is equivalent to the self-adjointness of $\mathcal E$. If it is Hermitian, continue to Step 3. If not, this algorithm is not applicable.
\item Compute a matrix ${\Lambda}_\infty$ corresponding to ${\mathcal E}_\infty$, which is given by the projection onto the eigensubspace of ${\Lambda}$ corresponding to the eigenvalue 1. Then compute ${\tilde\Lambda}_\infty={\Lambda}_\infty\Lambda_1$.
\item Compute $\Omega_{\infty}^{AA'}$ given by
\begin{eqnarray}
\Omega_{\infty}^{AA'}=\sum_{klmn}[{\tilde\Lambda}_\infty]_{kl,mn}|k\rangle\!\langle l|^A\otimes|m\rangle\!\langle n|^{A'}.\nonumber
\end{eqnarray}
\item Compute the Shannon entropy of the eigenvalues of $\Omega_{\infty}^{AA'}$, which is equal to $M_{A|B}(\Psi^{ABC})$.
\end{enumerate}

\begin{figure}[t]
\begin{center}
\includegraphics[bb={0 90 438 410}, scale=0.3]{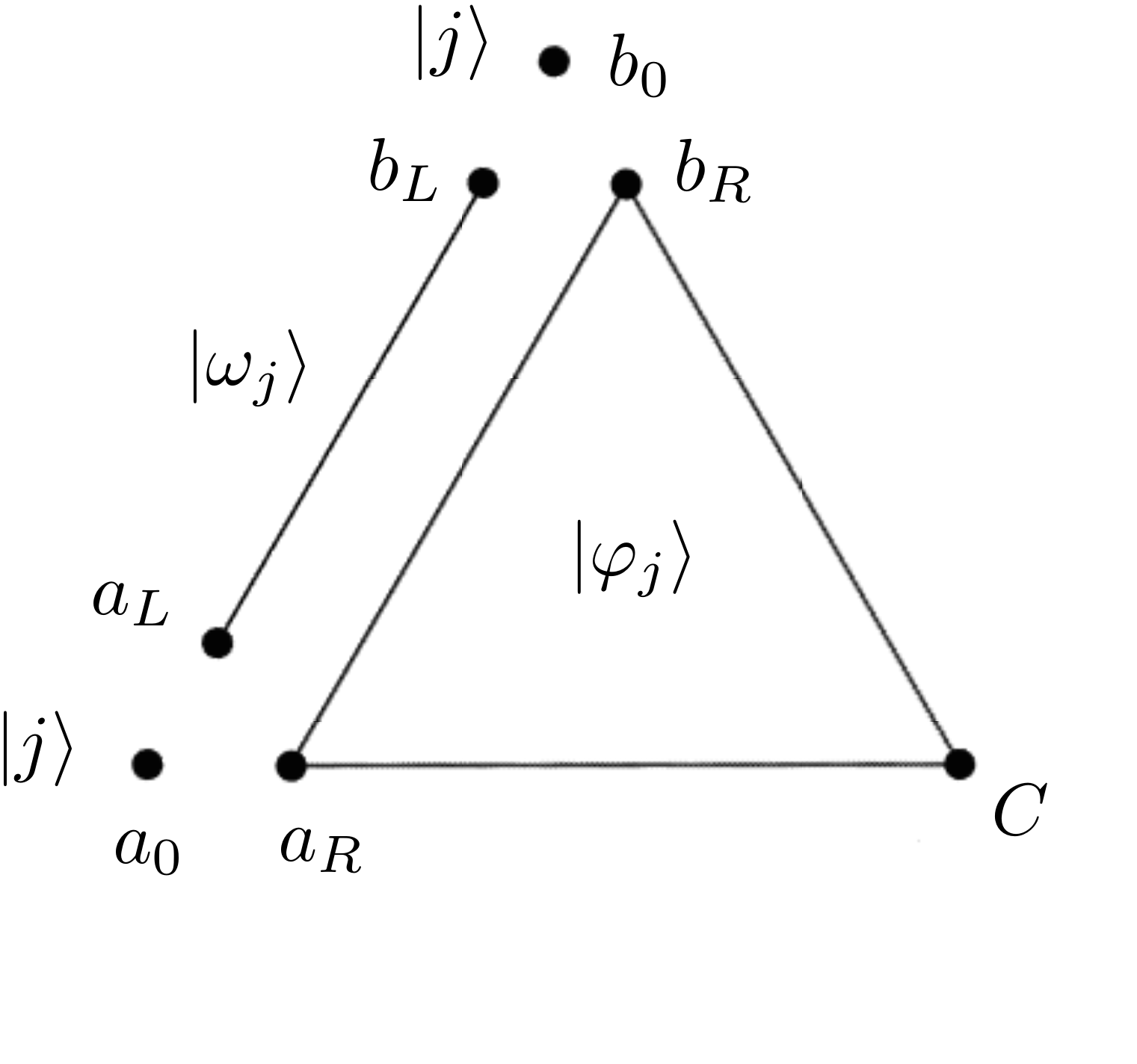}
\end{center}\caption{A graphical representation of the KI decomposition of tripartite pure states (\ref{eq:kidecoftristate}). Each vertex corresponds to a quantum system. The solid lines express pure states. The whole state is the superposition of the above states with the probability amplitude $\sqrt{p_j}$, namely, $\sum_{j\in J}\sqrt{p_j}\ket{j}^{a_0}\ket{j}^{b_0}\ket{\omega_j}^{a_Lb_L}\ket{\varphi_j}^{a_Rb_RC}$.}
\label{fig:kipure}
\end{figure}

\hfill

\section{Outline of Proofs of the Main Theorems}\label{sec:outline}

In this section, we describe the outline, main concepts and technical ingredients for the proofs of Theorem \ref{thm:strongmarkcostequality} and \ref{thm:comp}. Detailed proofs are given in Appendix \ref{app:rigorous} and \ref{app:comp}. 

We first introduce an adaptation of the KI decomposition to {\it tripartite} pure states as follows.
\begin{lmm}\label{lmm:kipure}
Let $\ket{\Psi}^{ABC}$ be a tripartite pure state and suppose that the KI decomposition of ${\Psi}^{AC}$ on $A$ is given by
\begin{eqnarray}
\Gamma^A\Psi^{AC}\Gamma^{\dagger A}=\sum_{j\in J}p_j\proj{j}^{a_{\scalebox{0.45}{$0$}}}\otimes\omega_j^{a_{\scalebox{0.45}{$L$}}}\otimes\varphi_j^{a_{\scalebox{0.45}{$R$}}C}.\nonumber
\end{eqnarray}
There exists a linear isometry $\Gamma':{\mathcal H}_\Psi^B\rightarrow {\mathcal H}^{b_{\scalebox{0.45}{$0$}}}\otimes{\mathcal H}^{b_{\scalebox{0.45}{$L$}}}\otimes{\mathcal H}^{b_{\scalebox{0.45}{$R$}}}$ that decomposes $|\Psi\rangle^{ABC}$ together with $\Gamma$ as
\begin{eqnarray}
(\Gamma^A\otimes\Gamma'^B)|\Psi\rangle^{ABC}=\sum_{j\in J}\sqrt{p_j}\ket{j}^{a_{\scalebox{0.45}{$0$}}}\ket{j}^{b_{\scalebox{0.45}{$0$}}}\ket{\omega_j}^{a_{\scalebox{0.45}{$L$}}b_{\scalebox{0.45}{$L$}}}\ket{\varphi_j}^{a_{\scalebox{0.45}{$R$}}b_{\scalebox{0.45}{$R$}}C},\!\!\!\!\!\!\!\!\!\!\!\!\nonumber\\
\label{eq:kidecoftristate}
\end{eqnarray}
where $\ket{\omega_j}^{a_{\scalebox{0.45}{$L$}}b_{\scalebox{0.6}{$L$}}}$ and $\ket{\varphi_j}^{a_{\scalebox{0.45}{$R$}}b_{\scalebox{0.45}{$R$}}C}$ are purifications of $\omega_j^{a_{\scalebox{0.45}{$L$}}}$ and $\varphi_j^{a_{\scalebox{0.45}{$R$}}C}$, respectively, and $\inpro{j}{j'}^{b_{\scalebox{0.45}{$0$}}}=\delta_{jj'}$. Moreover, $\Gamma'$ is the sub-KI isometry on $B$ with respect to ${\Psi}^{BC}$.
\end{lmm}
\begin{prf}
The existence of $\Gamma'$ follows from Uhlmann's theorem (\!\!\cite{uhlmann}, see Appendix \ref{app:uhlman}). It is straightforward to verify that $\Gamma'$ is the sub-KI isometry, since we have
\begin{eqnarray}
\Gamma'^B\Psi^{BC}\Gamma'^{\dagger B}=\sum_{j\in J}p_j\proj{j}^{b_{\scalebox{0.45}{$0$}}}\otimes\omega_j^{b_{\scalebox{0.45}{$L$}}}\otimes\varphi_j^{b_{\scalebox{0.45}{$R$}}C}\nonumber
\end{eqnarray}
from (\ref{eq:kidecoftristate}).
\hfill$\blacksquare$\end{prf}
We call (\ref{eq:kidecoftristate}) as the {\it KI decomposition of $\ket{\Psi}^{ABC}$ on $A$ and $B$} (Figure \ref{fig:kipure}), and denote it by $|\Psi_{K\!I}\rangle$.
 
\begin{figure}[t]
\begin{center}
\includegraphics[bb={0 70 722 302}, scale=0.36]{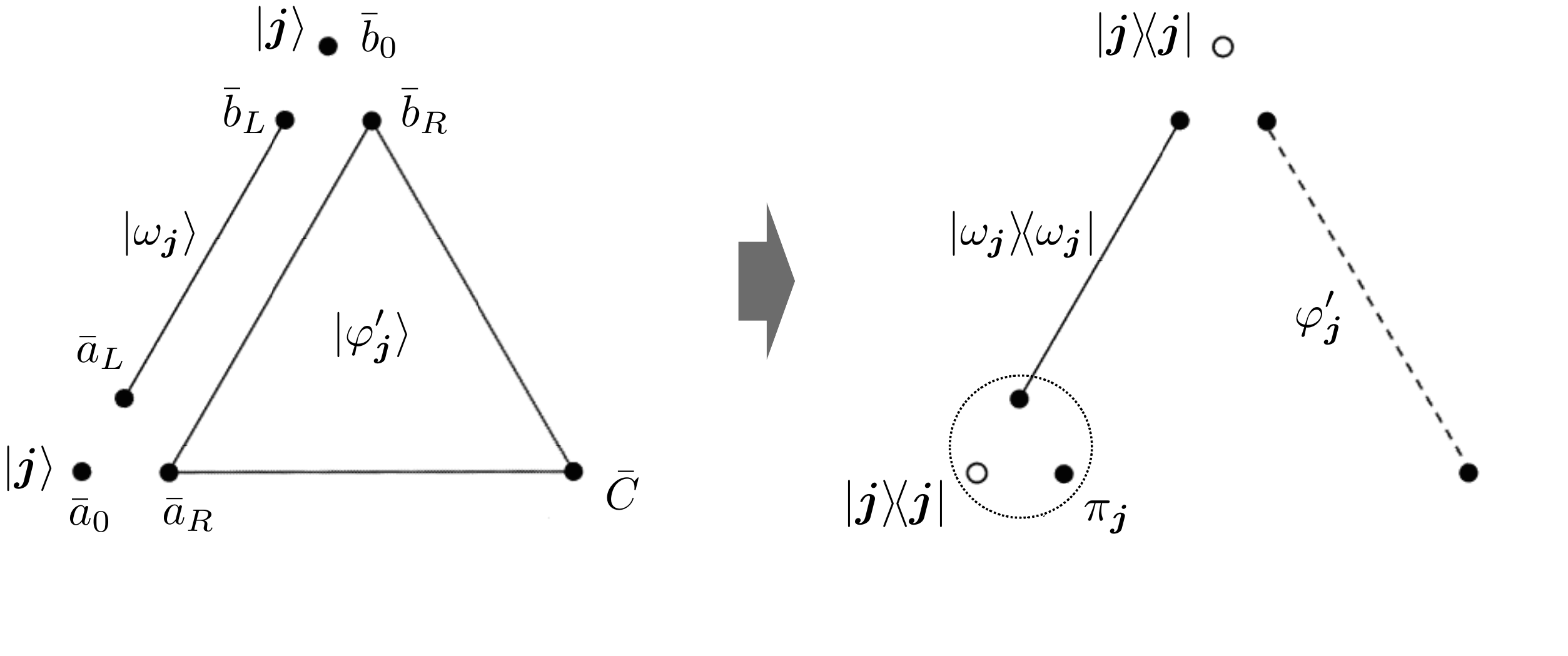}
\end{center}\caption{A graphical representation of the state transformation from $\Psi'_{n,\delta}$ in (\ref{eq:psinepsilon2}) to ${\bar \Psi}_{n,\delta}$ in (\ref{eq:transformmarkov2}) by a random unitary operation given by (\ref{eq:vensemble2})}.
\label{fig:markovunitary}
\end{figure}

\subsection{For Theorem \ref{thm:strongmarkcostequality}: Achievability}\label{sec:prfachieve}

The direct part of Theorem \ref{thm:strongmarkcostequality} is formulated by the following inequality:
\begin{eqnarray}
M_{A|B}(\Psi^{ABC})\leq H(\{p_j\}_{j\in J})+2\sum_{j\in J}p_jS(\varphi_j^{a_{\scalebox{0.45}{$R$}}}).
\label{eq:markoviancost}
\end{eqnarray}
The outline of the proof is as follows. The state $|\Psi^{ABC}\rangle^{\otimes n}$ is local unitarily equivalent to $|\Psi_{K\!I}\rangle^{\otimes n}$, which is almost equal to the state defined by
\begin{eqnarray}
|\Psi'_{n,\delta}\rangle:=\!\!\sum_{{\bm j}\in J_{n,\delta}}\!\!\sqrt{p_{\bm j}}\ket{\bm j}^{{\bar a}_{\scalebox{0.45}{$0$}}}\ket{\bm j}^{{\bar b}_{\scalebox{0.45}{$0$}}}\ket{\omega_{\bm j}}^{{\bar a}_{\scalebox{0.45}{$L$}}{\bar b}_{\scalebox{0.45}{$L$}}}\Pi_{{\bm j}, \delta}^{{\bar a}_{\scalebox{0.45}{$R$}}}\ket{\varphi_{\bm j}}^{{\bar a}_{\scalebox{0.45}{$R$}}{\bar b}_{\scalebox{0.45}{$R$}}{\bar C}}\!\!\!\!\!\!\!\!\!\nonumber\\
\label{eq:psinepsilon2}
\end{eqnarray} 
for sufficiently large $n$. Here, we have introduced notations ${\bm j}=j_1\cdots j_n$, $\varphi_{\bm j}=\varphi_{j_1}\otimes\cdots\otimes\varphi_{j_n}$ and $\omega_{\bm j}=\omega_{j_1}\otimes\cdots\otimes\omega_{j_n}$. $J_{n,\delta}$ is the $\delta$-strongly typical set with respect to the probability distribution $\{p_j\}_j$, and $\Pi_{{\bm j},\delta}^{{\bar a}_{\scalebox{0.45}{$R$}}}$ is the projection onto the conditionally typical subspace of $\varphi_{\bm j}^{{\bar a}_{\scalebox{0.45}{$R$}}}$ conditioned by ${\bm j}$. 
Consider a unitary operation on ${\rm supp}\:[\Psi_{n,\delta}'^{{\bar a}_{\scalebox{0.45}{$0$}}{\bar a}_{\scalebox{0.45}{$L$}}{\bar a}_{\scalebox{0.45}{$R$}}}]$ of the form
\begin{eqnarray}
V:=\sum_{{\bm j}\in J_{n,\delta}}\proj{\bm j}^{{\bar a}_{\scalebox{0.45}{$0$}}}\otimes I_{\bm j}^{{\bar a}_{\scalebox{0.45}{$L$}}}\otimes v_{{\bm j}}^{{\bar a}_{\scalebox{0.45}{$R$}}},
\label{eq:vensemble2}
\end{eqnarray}
where $I_{\bm j}^{{\bar a}_{\scalebox{0.45}{$L$}}}$ is the identity operator on ${\rm supp}\:\omega_{\bm j}^{{\bar a}_{\scalebox{0.45}{$L$}}}$ and $v_{{\bm j}}^{{\bar a}_{\scalebox{0.45}{$R$}}}$ is a unitary on the support of $\Pi_{{\bm j}, \delta}^{{\bar a}_{\scalebox{0.45}{$R$}}}$. We apply $V^{{\bar a}_{\scalebox{0.45}{$0$}}{\bar a}_{\scalebox{0.45}{$L$}}{\bar a}_{\scalebox{0.45}{$R$}}}$ on $\Psi'_{n,\delta}$ by independently choosing $v_{{\bm j}}^{{\bar a}_{\scalebox{0.45}{$R$}}}$ from the Haar distributed random unitary ensemble for each ${\bm j}$. By this random unitary operation, the state (\ref{eq:psinepsilon2}) is transformed to the following state
\begin{eqnarray}
{\bar \Psi}_{n,\delta}:=\!\sum_{{\bm j}\in J_{n,\delta}}\!{p_{\bm j}}\proj{{\bm j}{\bm j}}^{{\bar a}_{\scalebox{0.45}{$0$}}{\bar b}_{\scalebox{0.45}{$0$}}}\!\otimes\proj{\omega_{\bm j}}^{{\bar a}_{\scalebox{0.45}{$L$}}{\bar b}_{\scalebox{0.45}{$L$}}}\!\otimes\pi_{{\bm j}}^{{\bar a}_{\scalebox{0.45}{$R$}}}\!\otimes\varphi'^{{\bar b}_{\scalebox{0.45}{$R$}}{\bar C}}_{\bm j},\!\!\!\!\!\!\!\!\!\!\!\!\nonumber\\
\label{eq:transformmarkov2}
\end{eqnarray}
where $\varphi_{\bm j}':={\rm Tr}_{a_{\scalebox{0.45}{$R$}}}[\Pi_{\bm j, \delta}^{{\bar a}_{\scalebox{0.45}{$R$}}}\proj{\varphi_{\bm j}}]$ and $\pi_{{\bm j}}^{{\bar a}_{\scalebox{0.45}{$R$}}}=\Pi_{\bm j, \delta}^{{\bar a}_{\scalebox{0.45}{$R$}}}/{\rm Tr}[\Pi_{\bm j,\delta}^{{\bar a}_{\scalebox{0.45}{$R$}}}]$. ${\bar \Psi}_{n,\delta}$ is a Markov state conditioned by $B$ (Figure \ref{fig:markovunitary}). 

To Markovianize $|\Psi^{ABC}\rangle^{\otimes n}$, it is sufficient that we approximate the transformation from (\ref{eq:psinepsilon2}) to (\ref{eq:transformmarkov2}) by ${\mathcal V}_n$ with a vanishingly small error, where $V_k$ in ${\mathcal V}_n$ are unitaries which are decomposed by $\Gamma^{\otimes n}$ as (\ref{eq:vensemble2}). By a random coding method and the operator Chernoff bound \cite{berry08}, it is shown that a sufficient number of unitaries in ${\mathcal V}_n$ for this approximation is almost equal to the inverse of the minimum nonzero eigenvalue of (\ref{eq:transformmarkov2}), and is given as $H(\{p_j\}_{j\in J})+2\sum_{j\in J}p_jS(\varphi_j^{a_{\scalebox{0.6}{$R$}}})$ per copy. We note that the error $\epsilon$ converges exponentially with $n$ to zero.

\subsection{For Theorem \ref{thm:strongmarkcostequality}: Optimality}

The converse part of Theorem \ref{thm:strongmarkcostequality} is formulated by the following inequality:
\begin{eqnarray}
M_{A|B}(\Psi^{ABC})\geq H(\{p_j\}_{j\in J})+2\sum_{j\in J}p_jS(\varphi_j^{a_{\scalebox{0.45}{$R$}}}).
\label{eq:lowermarkov}
\end{eqnarray}

Let us first assume tentatively that a Markov decomposition of $\Upsilon^{{\bar A}{\bar B}{\bar C}}$ in (\ref{eq:defmarkovianizing}) is given by
\begin{eqnarray}
\;\;(\Gamma'^B)^{\otimes n}\Upsilon^{{\bar A}{\bar B}{\bar C}}(\Gamma'^B)^{\otimes n}=\sum_{{\bm j}\in J^n}p_{\bm j}'\proj{\bm j}^{b_{\scalebox{0.45}{$0$}}}\otimes\sigma_{\bm j}^{{\bar A}b_{\scalebox{0.45}{$L$}}}\otimes\phi_{\bm j}^{b_{\scalebox{0.45}{$R$}}{\bar C}},\!\!\!\!\!\!\nonumber\\
\label{eq:markovupsilonin}
\end{eqnarray}
with $\Gamma'$ being the KI isometry on $B$ with respect to $\Psi^{BC}$. In this case, it is not difficult to show that the amount of randomness per copy required for transforming $|\Psi^{\otimes n}\rangle^{{\bar A}{\bar B}{\bar C}}$ to $\Upsilon^{{\bar A}{\bar B}{\bar C}}$ is bounded below by the R.H.S. of (\ref{eq:lowermarkov}). Indeed, in order to transform  $|\Psi_{K\!I}\rangle^{\otimes n}$ to a Markov state in the form of (\ref{eq:markovupsilonin}), it is necessary that (i) the off-diagonal terms with respect to $|{\bm j}\rangle$ vanish, and (ii) the correlation between ${\bar a}_{\scalebox{0.6}{$R$}}$ and ${\bar b}_{\scalebox{0.6}{$R$}}{\bar C}$ in the state $|{\varphi_{\bm j}}\rangle^{{\bar a}_{\scalebox{0.45}{$R$}}{\bar b}_{\scalebox{0.45}{$R$}}{\bar C}}$ is destroyed for each ${\bm j}$. An optimal way for satisfying these two conditions is transforming the state (\ref{eq:psinepsilon2}) close to a state of the form (\ref{eq:transformmarkov2}). Since the entropy of the state (\ref{eq:transformmarkov2}) is approximately equal to $n(H(\{p_j\}_{j\in J})+2\sum_{j\in J}p_jS(\varphi_j^{a_{\scalebox{0.45}{$R$}}}))$, the cost of randomness required for this transformation is at least about $H(\{p_j\}_{j\in J})+2\sum_{j\in J}p_jS(\varphi_j^{a_{\scalebox{0.45}{$R$}}})$ bits per copy.

However, it might be possible in general that the amount of randomness can be further reduced by appropriately choosing $\Upsilon^{{\bar A}{\bar B}{\bar C}}$ and the corresponding KI decomposition of ${\bar B}$. We shall see that our choice presented above is indeed optimal. At the core of the proof lies the following lemma.
\begin{lmm}\label{lmm:koashi}
Let $\Psi^{AC}$ be a bipartite quantum state, and let $\Gamma_\Psi:{\mathcal H}_\Psi^{A}\rightarrow{\mathcal H}^{{a}_{\scalebox{0.45}{$0$}}}\otimes{\mathcal H}^{{a}_{\scalebox{0.45}{$L$}}}\otimes{\mathcal H}^{{a}_{\scalebox{0.45}{$R$}}}$ be the KI isometry on $A$ with respect to $\Psi^{AC}$. For any $n$ and $\epsilon>0$, let $\chi^{{\bar A}{\bar C}}$ be a state that satisfies
\begin{eqnarray}
\left\|(\Psi^{\otimes n})^{{\bar A}{\bar C}}-\chi^{{\bar A}{\bar C}}\right\|_1\leq\epsilon,
\label{eq:distpsirho}
\end{eqnarray}
and let $\Gamma_\chi:{\mathcal H}_\chi^{\bar A}\rightarrow{\mathcal H}^{{\hat a}_{\scalebox{0.45}{$0$}}}\otimes{\mathcal H}^{{\hat a}_{\scalebox{0.45}{$L$}}}\otimes{\mathcal H}^{{\hat a}_{\scalebox{0.45}{$R$}}}$ be a sub-KI isometry on $\bar A$ with respect to $\chi^{{\bar A}{\bar C}}$. Denoting the decompositions $\Gamma_\Psi^A\Psi^{AC}\Gamma_\Psi^{\dagger A}$ and $\Gamma_\chi^{\bar A}\chi^{{\bar A}{\bar C}}\Gamma_\chi^{\dagger{\bar A}}$ by $\Psi_{K\!I}^{AC}$ and $\chi_{sK\!I}^{{\bar A}{\bar C}}$, respectively, we have
\begin{eqnarray}
&&\!\!\!\!\!\!\!\!\frac{1}{n}\left(S({\hat a}_{\scalebox{0.6}{$0$}})_{\chi_{sK\!I}}+2S({\hat a}_{\scalebox{0.6}{$R$}}|{\hat a}_{\scalebox{0.6}{$0$}})_{\chi_{sK\!I}}\right)\nonumber\\
&&\geq S({a}_{\scalebox{0.6}{$0$}})_{\Psi_{K\!I}}+2S(a_{\scalebox{0.6}{$R$}}|a_{\scalebox{0.6}{$0$}})_{\Psi_{K\!I}}-\zeta'_{{}_{\Psi}}\!(\epsilon)\log{d_A}.\label{eq:pqcomineq}
\end{eqnarray}
Here, $\zeta'_{{}_{\Psi}}\!(\epsilon)$ is a function of $\epsilon>0$ and $\Psi$,  which does not depend on $n$, and satisfies $\lim_{\epsilon\rightarrow0}\zeta'_{{}_{\Psi}}\!(\epsilon)=0$. See Equality (\ref{eq:zetaprime}) in Appendix \ref{sec:prflmm10} for a rigorous definition.
\end{lmm}

\begin{figure}[t]
\begin{center}
\includegraphics[bb={0 0 638 208}, scale=0.35]{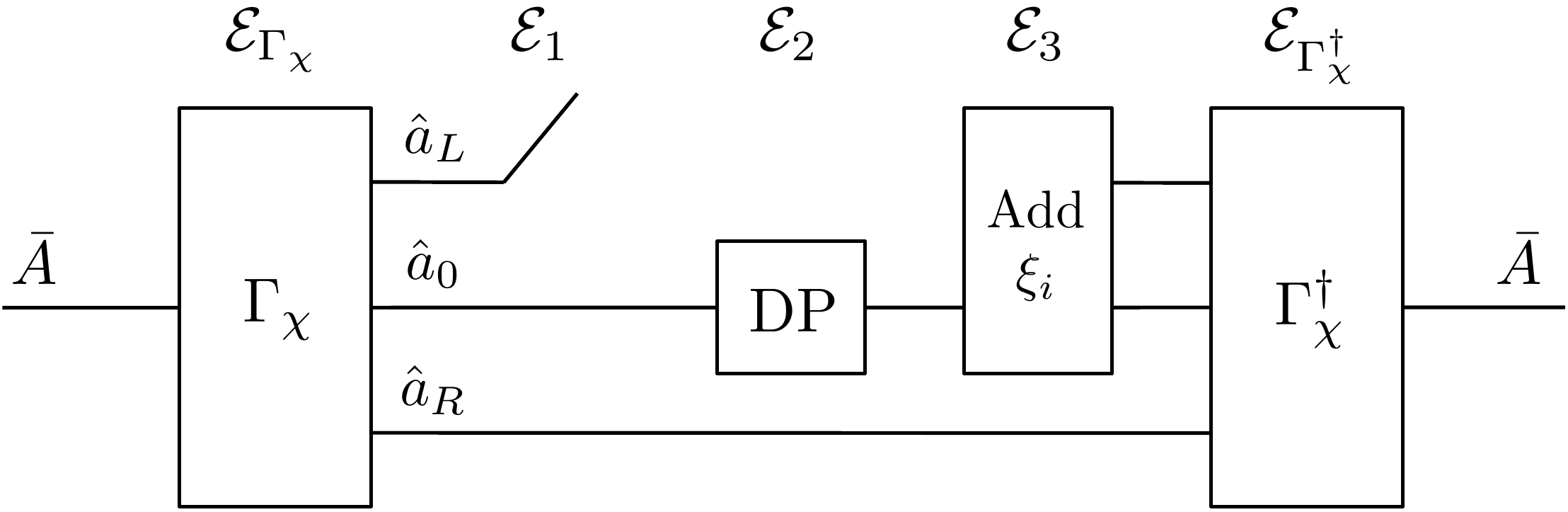}
\end{center}\caption{A graphical representation of the channel ${\mathcal E}_{\chi}$. Due to the completely dephasing channel denoted as DP, the system ${\hat a}_0$ has some capacity to transmit classical information, but has no capacity to transfer entanglement. The system ${\hat a}_R$ has some capacity to transfer entanglement.}
\label{fig:mixcomp1}
\end{figure}

{\it Proof Outline for Lemma \ref{lmm:koashi}:}
Assume here for simplicity that
\begin{eqnarray}
({\mathcal H}_\Psi^A)^{\otimes n}={\mathcal H}_\chi^{\bar A}={\mathcal H}^{\bar A},\label{eq:wondagold}
\end{eqnarray}
and suppose the decompositions of $\Psi^{AC}$ and that of $\chi^{{\bar A}{\bar C}}$ are given by
\begin{eqnarray}
\Psi_{K\!I}^{AC}=\sum_{j\in J}p_j\proj{j}^{a_{\scalebox{0.45}{$0$}}}\otimes\omega_j^{a_{\scalebox{0.45}{$L$}}}\otimes\varphi_j^{a_{\scalebox{0.45}{$R$}}C}\nonumber
\end{eqnarray} 
and
\begin{eqnarray}
\chi^{{\bar A}{\bar C}}_{sK\!I}=\sum_iq_i\proj{i}^{{\hat a}_{\scalebox{0.45}{$0$}}}\otimes\xi_i^{{\hat a}_{\scalebox{0.45}{$L$}}}\otimes\phi_i^{{\hat a}_{\scalebox{0.45}{$R$}}{\bar C}},
\label{eq:kiofrho}
\end{eqnarray}
respectively. Consider a quantum channel ${\mathcal E}_\chi$ on ${\bar A}$ defined by
\begin{eqnarray}
{\mathcal E}_{\chi}(\tau)=\Gamma_{\chi}^{\dagger}\!\left(\sum_i\proj{i}^{{\hat a}_{\scalebox{0.45}{$0$}}}{\rm Tr}_{{\hat a}_{\scalebox{0.45}{$L$}}}[\Gamma_{\chi}\tau\Gamma_{\chi}^{\dagger}]\proj{i}^{{\hat a}_{\scalebox{0.45}{$0$}}}\otimes\xi_i^{{\hat a}_{\scalebox{0.45}{$L$}}}\!\right)\!\Gamma_{\chi},\!\!\!\!\!\!\!\!\!\nonumber\\
\label{eq:defofchannel}
\end{eqnarray} 
which is decomposed as ${\mathcal E}_{\Gamma_{\chi}^{\dagger}}\circ{\mathcal E}_{3}\circ{\mathcal E}_{2}\circ{\mathcal E}_{1}\circ{\mathcal E}_{\Gamma_{\chi}}$. The maps ${\mathcal E}_{\Gamma_{\chi}}$ and ${\mathcal E}_{\Gamma_{\chi}^{\dagger}}$ are isometry channels corresponding to $\Gamma_{\chi}$ and $\Gamma_{\chi}^{\dagger}$, respectively; ${\mathcal E}_{1}$ is discarding of system ${\hat a}_{\scalebox{0.6}{$L$}}$; ${\mathcal E}_{2}$ is the completely dephasing channel on ${\hat a}_{\scalebox{0.6}{$0$}}$ with respect to the basis $\ket{i}$; ${\mathcal E}_{3}$ is appending of the state $\xi_i^{{\hat a}_{\scalebox{0.45}{$L$}}}$, conditioned by ${\hat a}_{\scalebox{0.6}{$0$}}$ (Figure \ref{fig:mixcomp1}). The linearity and the complete positivity of ${\mathcal E}_{\chi}$ immediately follows from (\ref{eq:defofchannel}), and the trace-preserving property results from the fact that $\Gamma_\chi$ satisfies Condition (\ref{eq:imgGamma}). 

The state $\chi^{{\bar A}{\bar C}}$ is invariant under the action of ${\mathcal E}_{\chi}$, and thus $(\Psi^{\otimes n})^{{\bar A}{\bar C}}$ is almost unchanged due to (\ref{eq:distpsirho}). By extending the data compression theorem for quantum mixed-state ensembles\cite{mixcomp1}, it follows that any state of the form $(\psi^{\otimes n})^{{\bar A}{\bar {C'}}}$ is almost unchanged by ${\mathcal E}_\chi$ on average, as long as $\psi^{A}=\Psi^{A}$ holds and the KI isometry on $A$ with respect to $\psi^{{A}{C'}}$ is equal to $\Gamma_{\Psi}$. We consider $\psi^{A{C'}}$ such that its KI decomposition on $A$ is, up to an additional decomposition on $C'$, given by
\begin{eqnarray}
\psi_{K\!I}^{AC'}&=&\sum_{j\in J}p_j\proj{j}^{a_{\scalebox{0.45}{$0$}}}\otimes\omega_j^{a_{\scalebox{0.45}{$L$}}}\otimes|\tilde{\varphi}_j\rangle\!\langle\tilde{\varphi}_j|^{a_{\scalebox{0.45}{$R$}}c_{\scalebox{0.45}{$R$}}'}\otimes\proj{j}^{c_{\scalebox{0.45}{$0$}}'},\nonumber
\end{eqnarray}
where $|\tilde{\varphi}_j\rangle^{a_{\scalebox{0.45}{$R$}}c_{\scalebox{0.45}{$R$}}'}$ is a purification of ${\varphi}_j^{a_{\scalebox{0.45}{$R$}}}$. The correlation between $\bar A$ and $\bar{C'}$ in the state $\psi^{\otimes n}$, measured by QMI, is equal to
\begin{eqnarray}
nI(A:C')_{\psi}&=&nI(a_{\scalebox{0.6}{$0$}}a_{\scalebox{0.6}{$L$}}a_{\scalebox{0.6}{$R$}}:c_{\scalebox{0.6}{$0$}}'c_{\scalebox{0.6}{$R$}}')_{\psi_{K\!I}}\nonumber\\
&=&n\left(S(a_{\scalebox{0.6}{$0$}})_{\psi_{K\!I}}+2S(a_{\scalebox{0.6}{$R$}}|a_{\scalebox{0.6}{$0$}})_{\psi_{K\!I}}\right).
\label{eq:ament}
\end{eqnarray}
It can be shown that this amount of correlation is almost conserved under ${\mathcal E}_{\chi}$. 

\begin{figure}[t]
\begin{center}
\includegraphics[bb={0 0 843 316}, scale=0.294]{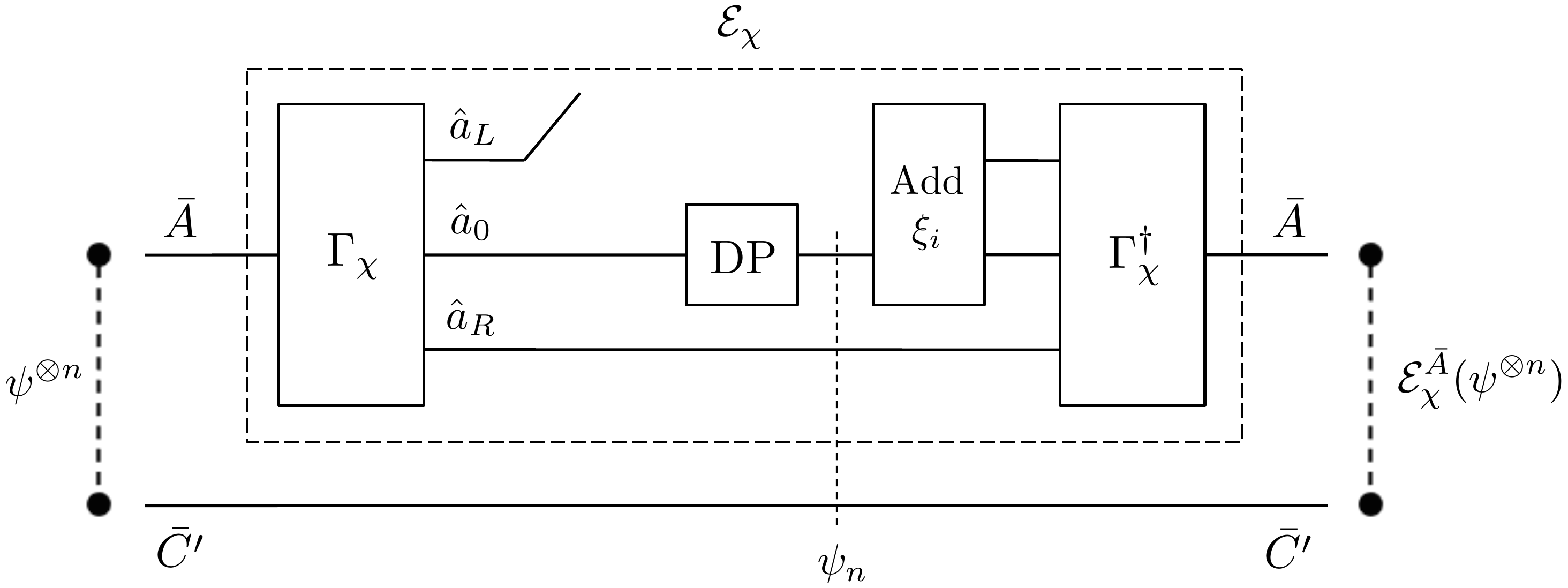}
\end{center}\caption{A graphical representation of the state transformation of $\psi^{\otimes n}$ under ${\mathcal E}_{\chi}$. The channel ${\mathcal E}_{\chi}$ has little effect on the state $\psi^{\otimes n}$ on average. In particular, it almost conserves the correlation that $\psi^{\otimes n}$ initially has. Thus the intermediate state $\psi_n$ has the same amount of correlation due to the monotonicity.}
\label{fig:mixcomp2}
\end{figure}

Due to the monotonicity of QMI, it follows that the correlation between ${\hat a}_{\scalebox{0.6}{$0$}}{\hat a}_{\scalebox{0.6}{$L$}}{\hat a}_{\scalebox{0.6}{$R$}}$ and $\bar{C'}$ is approximately equal to (\ref{eq:ament}) at any intermediate step of ${\mathcal E}_\chi$ (see Figure \ref{fig:mixcomp1}). After the action of the completely dephasing channel ${\mathcal E}_{2}$, the system ${\hat a}_{\scalebox{0.6}{$0$}}$ holds no quantum correlation with other systems, and thus the correlation between ${\hat a}_{\scalebox{0.6}{$0$}}{\hat a}_{\scalebox{0.6}{$R$}}$ and $\bar{C'}$ is bound to be at most $S({\hat a}_{\scalebox{0.6}{$0$}})+2S({\hat a}_{\scalebox{0.6}{$R$}}|{\hat a}_{\scalebox{0.6}{$0$}})$ (Figure \ref{fig:mixcomp2}). Moreover, the state on ${\hat a}_{\scalebox{0.6}{$0$}}{\hat a}_{\scalebox{0.6}{$R$}}$ after ${\mathcal E}_{2}$ is almost equal to $\chi_{sK\!I}^{{\hat a}_{\scalebox{0.6}{$0$}}{\hat a}_{\scalebox{0.6}{$R$}}}$ due to (\ref{eq:distpsirho}). A more detailed argument reveals that
\begin{eqnarray}
&&S({\hat a}_{\scalebox{0.6}{$0$}})_{\chi_{sK\!I}}+2S({\hat a}_{\scalebox{0.6}{$R$}}|{\hat a}_{\scalebox{0.6}{$0$}})_{\chi_{sK\!I}}\nonumber\\
&&\;\;\;\;\;\;\;\;\;\;\;\;\;\;\gtrsim n\left(S(a_{\scalebox{0.6}{$0$}})_{\psi_{K\!I}}+2S(a_{\scalebox{0.6}{$R$}}|a_{\scalebox{0.6}{$0$}})_{\psi_{K\!I}}\right),\nonumber
\end{eqnarray}
and consequently proving (\ref{eq:pqcomineq}).

\subsection{For Theorem \ref{thm:comp}}
Let us first express (\ref{eq:defmathe}), (\ref{def:einfty}) and (\ref{def:einfty2}) in terms of the ``decomposed'' Hilbert space ${\mathcal H}^{a_{\scalebox{0.45}{$0$}}}\otimes{\mathcal H}^{a_{\scalebox{0.45}{$L$}}}\otimes{\mathcal H}^{a_{\scalebox{0.45}{$R$}}}$. A Kraus representation of a map $\mathcal E$ defined by (\ref{eq:defmathe}) is given by ${\mathcal E}(\cdot)=\sum_{kl}E_{kl}(\cdot)E_{kl}^\dagger$, with the Kraus operators
\begin{eqnarray}
E_{kl}:=\langle k|^C(\Psi^{AC})^{\frac{1}{2}}|l\rangle^C(\Psi^A)^{-\frac{1}{2}}.\label{eq:krause}
\end{eqnarray}
Let $\Gamma$ be the KI isometry on $A$ with respect to $\Psi^{AC}$, and suppose the KI decomposition of $\Psi^{AC}$ is given by (\ref{eq:kidecofpsi}). For each $E_{kl}$, we have
\begin{eqnarray}
{\hat E}_{kl}:=\Gamma E_{kl}\Gamma^\dagger=\sum_{j\in J}\proj{j}^{a_{\scalebox{0.45}{$0$}}}\otimes I_j^{a_{\scalebox{0.45}{$L$}}}\otimes e_{j,kl}^{a_{\scalebox{0.45}{$R$}}},
\label{eq:hatekl}
\end{eqnarray}
where
\begin{eqnarray}
e_{j,kl}:=\langle k|^C(\varphi_j^{a_{\scalebox{0.45}{$R$}}C})^{\frac{1}{2}}|l\rangle^C(\varphi_j^{a_{\scalebox{0.45}{$R$}}})^{-\frac{1}{2}}.\label{eq:defejkl}
\end{eqnarray}
By an extension of Lemma \ref{lmm:irrki}, it follows that $\{e_{j,kl}\}_{kl}$ is irreducible in the sense that it satisfies Property 1) and 2). It is straightforward to verify from (\ref{eq:defejkl}) that maps ${\mathcal E}_j^{a_{\scalebox{0.45}{$R$}}}$ on ${\mathcal S}({\mathcal H}_j^{a_{\scalebox{0.45}{$R$}}})$, defined by ${\mathcal E}_j^{a_{\scalebox{0.45}{$R$}}}(\cdot):=\sum_{k,l}e_{j,kl}(\cdot)e_{j,kl}^\dagger\;(j=1,\cdots,|J|)$, are trace-preserving. Representations of $\mathcal E$ and ${\mathcal E}_\infty$ in the decomposed Hilbert space are given by
\begin{eqnarray}
\hat{\mathcal E}(\cdot)=\sum_{kl}{\hat E}_{kl}(\cdot){\hat E}_{kl}^\dagger\label{eq:mathcalhate}
\end{eqnarray}
and $\hat{\mathcal E}_\infty:=\lim_{n\rightarrow\infty}(1/N)\sum_{n=1}^N{\hat{\mathcal E}}^n$, respectively, and that of $\Psi_\infty^{ABC}$ is given by $\hat{\mathcal E}_\infty(|\Psi_{K\!I}\rangle\!\langle\Psi_{K\!I}|)$.

Due to ${\mathcal E}\circ{\mathcal E}_\infty={\mathcal E}_\infty$ and the irreducibility of $\{e_{j,kl}\}_{kl}$, we have
\begin{eqnarray}
\hat{\mathcal E}_\infty(|\Psi_{K\!I}\rangle\!\langle\Psi_{K\!I}|)=\sum_{j\in J}p_j'\proj{j}^{a_{\scalebox{0.45}{$0$}}}\otimes{\tilde\omega}_j\otimes \pi_{j}^{a_{\scalebox{0.45}{$R$}}},\label{eq:mathhateki}
\end{eqnarray}
where $\{p_j'\}_{j\in J}$ is a probability distribution and ${\tilde\omega}_j\:(j\in J)$ are states on ${a_{\scalebox{0.6}{$L$}}b_{\scalebox{0.6}{$0$}}b_{\scalebox{0.6}{$L$}}b_{\scalebox{0.6}{$R$}}C}$. Explicit forms of ${\tilde\omega}_j$ and $p_j'$ are obtained as follows. First, from (\ref{eq:hatekl}), (\ref{eq:mathcalhate}) and the trace-preserving property of ${\mathcal E}_j$, we have
\begin{eqnarray}
{\rm Tr}[\langle j|^{a_{\scalebox{0.45}{$0$}}}\hat{\mathcal E}({\hat\tau})|j\rangle^{a_{\scalebox{0.45}{$0$}}}]&=&{\rm Tr}[({\rm id}^{a_{\scalebox{0.45}{$L$}}}\otimes{\mathcal E}_j^{a_{\scalebox{0.45}{$R$}}})(\langle j|^{a_{\scalebox{0.45}{$0$}}}{\hat\tau}|j\rangle^{a_{\scalebox{0.45}{$0$}}})]\nonumber\\
&=&{\rm Tr}[\langle j|^{a_{\scalebox{0.45}{$0$}}}{\hat\tau}|j\rangle^{a_{\scalebox{0.45}{$0$}}}]\nonumber
\end{eqnarray}
for any $j$ and ${\hat\tau}=\Gamma\tau\Gamma^\dagger$ ($\tau\in{\mathcal S}({\mathcal H}_\Psi^A)$). This implies that the probability amplitude with respect to the basis $\{|j\rangle\}_j$ is conserved by $\hat{\mathcal E}$, as well as by $\hat{\mathcal E}_\infty$. Thus we have $p_j'=p_j$. Observe from (\ref{eq:hatekl}) that $\hat{\mathcal E}$ and $\hat{\mathcal E}_\infty$ do not affect the system $a_{\scalebox{0.6}{$L$}}$, which implies
\begin{eqnarray}
{\tilde\omega}_j=\proj{\omega_j}^{a_{\scalebox{0.45}{$L$}}b_{\scalebox{0.45}{$L$}}}\otimes\proj{j}^{b_{\scalebox{0.45}{$0$}}}\otimes\varphi_{j}^{b_{\scalebox{0.45}{$R$}}C}.\nonumber
\end{eqnarray}
Hence the von Neumann entropy of $\Psi_\infty^{ABC}$, which is equal to that of (\ref{eq:mathhateki}), is given by
\begin{eqnarray}
S(\Psi_\infty^{ABC})=H(\{p_j\}_j)+\sum_jp_j\left(S(\varphi_{j}^{b_{\scalebox{0.45}{$R$}}C})+S(\pi_{j}^{a_{\scalebox{0.45}{$R$}}})\right).\nonumber
\end{eqnarray}
In addition, that the self-adjointness of $\mathcal E$ implies $\varphi_{j}^{a_{\scalebox{0.45}{$R$}}}=\pi_{j}^{a_{\scalebox{0.45}{$R$}}}$. Since $|\varphi_{j}\rangle^{a_{\scalebox{0.45}{$R$}}b_{\scalebox{0.45}{$R$}}C}$ is a purification of $\varphi_{j}^{b_{\scalebox{0.45}{$R$}}C}$, we finally obtain that
\begin{eqnarray}
S(\Psi_\infty^{ABC})=H(\{p_j\}_j)+2\sum_jp_jS(\varphi_{j}^{a_{\scalebox{0.45}{$R$}}})=M_{A|B}(\Psi^{ABC}).
\nonumber
\end{eqnarray}

\hfill

\section{Properties}\label{sec:prop}
In this section, we describe properties of the Markovianizing cost of tripartite quantum states. We first consider arbitrary (possibly mixed) states, and then focus on the case of pure states.

\subsection{General Properties}
Let $\rho^{ABC}$ be an arbitrary tripartite state on finite dimensional quantum systems $A$, $B$ and $C$. The Markovianizing cost of $\rho^{ABC}$ satisfies
\begin{eqnarray}
I(A:C|B)_\rho\leq M_{A|B}(\rho^{ABC})\leq I(A:BC)_\rho.\label{eq:markovqcmi}
\end{eqnarray}
The second inequality directly follows from the fact that decoupling $A$ from $BC$ is sufficient for converting the state to a Markov state and that the cost of randomness for decoupling bipartite states is asymptotically given by QMI \cite{berry08}. The first inequality is proved in Appendix \ref{app:prfmarkovqcmi}. Consequently, the Markovianizing cost is equal to zero only for Markov states.

The Markovianizing cost satisfies a kind of the data processing inequality, namely, that
\begin{eqnarray}
M_{A|B}({\mathcal E}^C(\rho^{ABC}))\leq M_{A|B}(\rho^{ABC})\nonumber
\end{eqnarray}
under any quantum operation $\mathcal E$ on $C$. This is because any random unitary operation ${\mathcal V}_n$ on $A^n$ satisfying (\ref{eq:defmarkovianizing}) also satisfies
\begin{eqnarray}
\left\|{\mathcal V}_n^{A^n}([{\mathcal E}^C(\rho)]^{\otimes n})-({\mathcal E}^C)^{\otimes n}(\Upsilon^{A^nB^nC^n})\right\|_1\leq\epsilon,\nonumber
\end{eqnarray}
and $({\mathcal E}^C)^{\otimes n}(\Upsilon^{A^nB^nC^n})$ is a Markov state conditioned by $B^n$. As a consequence, an upper bound on the Markovianizing cost of a mixed state is obtained as
\begin{eqnarray}
M_{A|B}(\rho^{ABC})\leq M_{A|B}(\psi_\rho^{ABC'}),\nonumber
\end{eqnarray}
where $|\psi_\rho\rangle^{ABC'}$ is a purification of $\rho^{AB}$. This is because there always exists a quantum operation ${\mathcal F}_\rho:C'\rightarrow C$ such that ${\mathcal F}_\rho(\psi_\rho^{ABC'})=\rho^{ABC}$.

\subsection{Pure States}\label{sec:pureprop}

\begin{figure}[t]
\begin{center}
\includegraphics[bb={0 0 428 341}, scale=0.3]{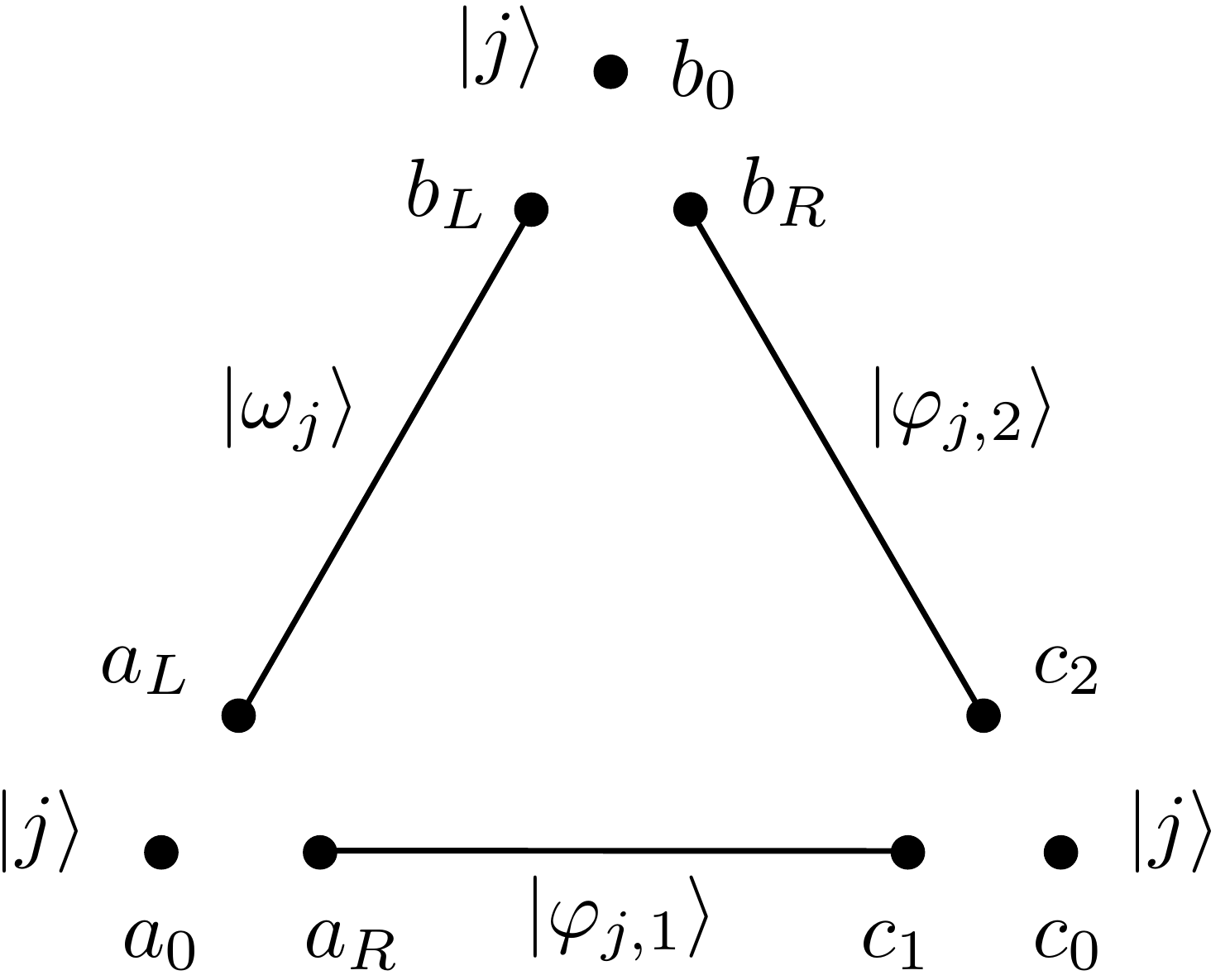}
\end{center}\caption{A graphical representation of a decomposition of tripartite pure states for which the Markovianizing cost is equal to QCMI. The whole state is the superposition of the above states with the probability amplitude $\sqrt{p_j}$, namely, $\sum_{j\in J}\sqrt{p_j}\ket{j}^{a_0}\ket{j}^{b_0}\ket{j}^{c_0}\ket{\omega_j}^{a_Lb_L}\ket{\varphi_{j,1}}^{a_Rc_1}\ket{\varphi_{j,2}}^{b_Rc_2}$.}
\label{fig:mkisqcmi}
\end{figure}

Let us now consider pure states, based on the result presented in Section \ref{sec:results}. First, we see that the Markovianizing cost $M_{A|B}$ of two pure states $\Psi_1$ and $\Psi_2$ are equal if there exist $\lambda\;(0<\lambda\leq1)$ and $\sigma^C\in{\mathcal S}({\mathcal H}^C)$ such that
\begin{eqnarray}
\Psi_2^{AC}=\lambda\Psi_1^{AC}+(1-\lambda)\Psi_1^A\otimes\sigma^C.
\label{eq:psi1and2}
\end{eqnarray}
This is because $\Psi_1^{A}=\Psi_2^{A}$ and the KI decompositions of $A$ with respect to $\Psi_1^{AC}$ and $\Psi_2^{AC}$ are equal, the latter of which follows from Lemma \ref{lmm:equivkidec}. Indeed, any quantum operation on $A$ which keeps $\Psi_1^{AC}$ invariant also keeps $\Psi_2^{AC}$ invariant and vice versa, as can be seen by observing from (\ref{eq:psi1and2}) that we have
\begin{eqnarray}
\Psi_1^{AC}=\frac{1}{\lambda}\Psi_2^{AC}-\frac{1-\lambda}{\lambda}\Psi_2^A\otimes\sigma^C.\nonumber
\end{eqnarray}
Second, we obtain a necessary and sufficient condition for the Markovianizing cost of pure states to be equal to QCMI as follows (Figure \ref{fig:mkisqcmi}).
\begin{thm}\label{thm:mareqmi}
Let $|\Psi\rangle^{ABC}$ be a pure state, and let
\begin{eqnarray}
|\Psi_{K\!I}\rangle^{ABC}=\sum_{j\in J}\sqrt{p_j}\ket{j}^{a_{\scalebox{0.45}{$0$}}}\ket{j}^{b_{\scalebox{0.45}{$0$}}}\ket{\omega_j}^{a_{\scalebox{0.45}{$L$}}b_{\scalebox{0.45}{$L$}}}\ket{\varphi_j}^{a_{\scalebox{0.45}{$R$}}b_{\scalebox{0.45}{$R$}}C},\!\!\!\!\!\!\!\!\!\!\!\!\nonumber
\end{eqnarray}
be its KI decomposition on $A$ and $B$ (see Lemma \ref{lmm:kipure}). Then we have $M_{A|B}(\Psi^{ABC})=I(A:C|B)_\Psi$ if and only if there exists an isometry $\Gamma'':{\mathcal H}^C\rightarrow {\mathcal H}^{c_{\scalebox{0.45}{$0$}}}\otimes{\mathcal H}^{c_{\scalebox{0.45}{$1$}}}\otimes{\mathcal H}^{c_{\scalebox{0.45}{$2$}}}$ such that $\varphi_j$ is decomposed as
\begin{eqnarray}
\Gamma''^C\ket{\varphi_j}^{a_{\scalebox{0.45}{$R$}}b_{\scalebox{0.45}{$R$}}C}=\ket{j}^{c_{\scalebox{0.45}{$0$}}}\ket{\varphi_{j,1}}^{a_{\scalebox{0.45}{$R$}}c_1}\ket{\varphi_{j,2}}^{b_{\scalebox{0.45}{$R$}}c_2},\!\!\!\!\!\!\!\!\!\!\!\!\nonumber
\end{eqnarray}
where $\inpro{j}{j'}=\delta_{j,j'}$.
\end{thm}

\begin{prf}
We have
\begin{eqnarray}
&&I(A:C|B)_\Psi=I(A:C)_\Psi=I(a_{\scalebox{0.6}{$0$}}a_{\scalebox{0.6}{$L$}}a_{\scalebox{0.6}{$R$}}:C)_{\Psi_{\!K\!I}}\nonumber\\
&=&I(a_{\scalebox{0.6}{$0$}}:C)_{\Psi_{\!K\!I}}+I(a_{\scalebox{0.6}{$L$}}a_{\scalebox{0.6}{$R$}}:C|a_{\scalebox{0.6}{$0$}})_{\Psi_{\!K\!I}}\nonumber\\
&=&I(a_{\scalebox{0.6}{$0$}}:C)_{\Psi_{\!K\!I}}+\sum_jp_jI(a_{\scalebox{0.6}{$R$}}:C)_{\varphi_j},\nonumber
\end{eqnarray}
as well as
\begin{eqnarray}
M_{A|B}(\Psi^{ABC})=H(a_{\scalebox{0.6}{$0$}})+2\sum_jp_jS(a_{\scalebox{0.6}{$R$}})_{\varphi_j}.\nonumber
\end{eqnarray}
Since $\Psi_{\!K\!I}^{a_{\scalebox{0.45}{$0$}}C}$ is a classical-quantum state, we have $I(a_{\scalebox{0.6}{$0$}}:C)_{\Psi_{\!K\!I}}\leq H(a_{\scalebox{0.6}{$0$}})$ with equality if and only if $\{{\rm supp}[\varphi_j^C]\}_j$ is mutually orthogonal. We also have $2S(a_{\scalebox{0.6}{$R$}})_{\varphi_j}\geq I(a_{\scalebox{0.6}{$R$}}:C)_{\varphi_j}$, which is saturated if and only if $2S(a_{\scalebox{0.6}{$R$}})_{\varphi_j}-I(a_{\scalebox{0.6}{$R$}}:C)_{\varphi_j}=I(a_{\scalebox{0.6}{$R$}}:b_{\scalebox{0.6}{$R$}})_{\varphi_j}=0$ (see Inequality (\ref{eq:ileq2s}) in Appendix \ref{app:entandmutual}). Hence we have $M_{A|B}(\Psi^{ABC})=I(A:C|B)_\Psi$ if and only if 
\begin{eqnarray}
\Psi_{K\!I}^{AB}=\sum_{j\in J}{p_j}\proj{j,j}^{a_{\scalebox{0.45}{$0$}}b_{\scalebox{0.45}{$0$}}}\otimes\proj{\omega_j}^{a_{\scalebox{0.45}{$L$}}b_{\scalebox{0.45}{$L$}}}\otimes\varphi_j^{a_{\scalebox{0.45}{$R$}}}\otimes\varphi_j^{b_{\scalebox{0.45}{$R$}}},\!\!\!\!\!\!\!\!\!\!\!\!\nonumber
\end{eqnarray}
which concludes the proof due to Uhlmann's theorem (\!\!\cite{uhlmann}, see Appendix \ref{app:uhlman}).
\hfill${\blacksquare}$
\end{prf}
An example of states that satisfy the above conditions is given in Section \ref{sec:exghz}. 

\hfill

\section{Examples}
\label{sec:examples}

In this section, we consider examples of pure states to illustrate discontinuity and asymmetry of the Markovianizing cost. We also give an example of states for which the Markovianizing cost is equal to QCMI.

\subsection{Discontinuity}
We consider tripartite pure states that are expressed as
\begin{eqnarray}
|\Psi_\lambda\rangle&=&\sqrt{\frac{d^2\lambda-1}{d^2-1}}|00\rangle^B|\Phi_d\rangle^{AC}\nonumber\\
&&\;\;\;\;+\sqrt{\frac{1-\lambda}{d^2-1}}\sum_{k,l=1}^d|kl\rangle^B|k\rangle^A|l\rangle^C,\nonumber
\end{eqnarray}
where $d={\rm dim}{\mathcal H}^A={\rm dim}{\mathcal H}^C$, ${\rm dim}{\mathcal H}^B=d^2+1$, $1/d^2\leq\lambda\leq1$, and $\Phi_d$ is a maximally entangled state of Schmidt rank $d$, defined as
\begin{eqnarray}
|\Phi_d\rangle=\frac{1}{\sqrt{d}}\sum_{k=1}^d|k\rangle|k\rangle.
\label{eq:maxent}
\end{eqnarray}
For this state, we have
\begin{eqnarray}
I(A:C|B)_{\Psi_\lambda}&=&2\log{d}-h(\lambda)-(1-\lambda)\log{(d^2-1)},\!\nonumber\\
\label{eq:distmarkpsilambda}
\end{eqnarray}
where $h$ denotes the binary entropy defined by $h(\lambda)=-\lambda\log{\lambda}-(1-\lambda)\log{(1-\lambda)}$. The state is a Markov state if and only if $\lambda=1/d^2$. The distance to the closest Markov state is bounded from above by
\begin{eqnarray}
\left\|\Psi_\lambda^{ABC}-\Psi_{1/d^2}^{ABC}\right\|_1&=&2\sqrt{1-|\langle\Psi_\lambda|\Psi_{1/d^2}\rangle|^2}\nonumber\\
&=&2\sqrt{\frac{d^2\lambda-1}{d^2-1}}.\label{eq:distmarkpsilambda2}
\end{eqnarray}
The reduced state on $AC$ is given by
\begin{eqnarray}
\Psi_\lambda^{AC}&=&\frac{d^2\lambda-1}{d^2-1}\:|\Phi_d\rangle\!\langle\Phi_d|^{AC}+\frac{1-\lambda}{d^2-1}\:I^A\otimes I^C\nonumber\\
&=&\lambda'|\Phi_d\rangle\!\langle\Phi_d|^{AC}+(1-\lambda')\:\pi^A\otimes \pi^C,\nonumber
\end{eqnarray}
where $\pi$ is the $d$-dimensional maximally mixed state and $\lambda'=(d^2\lambda-1)/(d^2-1)$. Hence the Markovianizing cost does not depend on $\lambda'$ when $\lambda'>0$, as we proved in Section \ref{sec:pureprop}. As directly verified by considering the case of $\lambda'=1$, the Markovianizing cost is equal to $2\log{d}$ for $\lambda'>0$. Taking the symmetry of $\Psi_\lambda$ between $A$ and $C$ into account, we obtain
\begin{eqnarray}
M_{A|B}(\Psi_\lambda)=M_{C|B}(\Psi_\lambda)=\begin{cases}
2\log{d}&(\lambda>1/d^2)\\
0&(\lambda=1/d^2)\end{cases}.\label{eq:markdisc}
\end{eqnarray}
Hence the Markovianizing cost is not a continuous function of states. In a particular case where $\lambda=2/d^2$, the Markovianizing cost grows logarithmically with respect to the dimension of the system, whereas QCMI, as well as the distance to the closest Markov state, approaches zero as indicated by (\ref{eq:distmarkpsilambda}) and (\ref{eq:distmarkpsilambda2}).

 We note that $\Psi_\lambda^{ABC}$ is ``approximately recoverable'' if $\lambda$ is close to $1/d^2$, i.e., it satisfies Equalities (\ref{eq:recoverable}) approximately if $\lambda\approx1/d^2$. Indeed, since $\Psi_{1/d^2}$ is a Markov state, there exist quantum operations ${\mathcal R}:B\rightarrow BC$ and ${\mathcal R}':B\rightarrow AB$ such that
\begin{eqnarray}
\Psi_{1/d^2}^{ABC}={\mathcal R}(\Psi_{1/d^2}^{AB})={\mathcal R}'(\Psi_{1/d^2}^{BC}).\nonumber
\end{eqnarray}
Due to the triangle inequality and the monotonicity of the trace distance (see Appendix \ref{app:uhlman}), we have
\begin{eqnarray}
&&\left\|\Psi_\lambda^{ABC}-{\mathcal R}(\Psi_\lambda^{AB})\right\|_1\nonumber\\
&\leq&\left\|\Psi_\lambda^{ABC}-\Psi_{1/d^2}^{ABC}\right\|_1+\left\|\Psi_{1/d^2}^{ABC}-{\mathcal R}(\Psi_{1/d^2}^{AB})\right\|_1\nonumber\\
&&\quad+\left\|{\mathcal R}(\Psi_{1/d^2}^{AB})-{\mathcal R}(\Psi_\lambda^{AB})\right\|_1\nonumber\\
&\leq&2\left\|\Psi_\lambda^{ABC}-\Psi_{1/d^2}^{ABC}\right\|_1\nonumber
\end{eqnarray}
as well as
\begin{eqnarray}
&&\left\|\Psi_\lambda^{ABC}-{\mathcal R}'(\Psi_\lambda^{BC})\right\|_1\leq2\left\|\Psi_\lambda^{ABC}-\Psi_{1/d^2}^{ABC}\right\|_1.\nonumber
\end{eqnarray}
Thus Equality (\ref{eq:distmarkpsilambda2}) implies
\begin{align}
\left\|\Psi_\lambda^{ABC}-{\mathcal R}(\Psi_\lambda^{AB})\right\|_1,\:\left\|\Psi_\lambda^{ABC}-{\mathcal R}'(\Psi_\lambda^{BC})\right\|_1\nonumber\\
\leq4\sqrt{\frac{d^2\lambda-1}{d^2-1}}.\nonumber
\end{align}

\subsection{Asymmetry}

We consider tripartite pure states that are expressed as
\begin{eqnarray}
|\Psi_\lambda\rangle=\sqrt{\lambda}|0\rangle^B|\Phi_d\rangle^{AC}+\sqrt{1-\lambda}|0\rangle^A|\Phi_d\rangle^{BC},\nonumber
\end{eqnarray}
where $d={\rm dim}{\mathcal H}^C$, ${\rm dim}{\mathcal H}^A={\rm dim}{\mathcal H}^B=d+1$, $0\leq\lambda\leq1$ and $\Phi_d$ is a maximally entangled state defined by (\ref{eq:maxent}). The reduced state on $AC$ is given by
\begin{eqnarray}
\Psi_\lambda^{AC}=\lambda|\Phi_d\rangle\!\langle\Phi_d|^{AC}+(1-\lambda)\: |0\rangle\!\langle0|^A\otimes\pi^C.
\label{eq:redpsiac2}\end{eqnarray}
Note that $|\Psi_d\rangle^{AC}$ does not have any $|0\rangle^A|0\rangle^C$ component. Hence the CPTP maps on $A$ defined as (\ref{eq:defmathe}) and (\ref{def:einfty}) are given by
\begin{eqnarray}
{\mathcal E}(\tau)&=&{\rm Tr}[P_1\tau]\cdot\pi_1+{\rm Tr}[P_0\tau]\cdot|0\rangle\!\langle0|\nonumber\\
&&+\frac{1}{d}P_1\tau P_0+\frac{1}{d}P_0\tau P_1\nonumber
\end{eqnarray}
and
\begin{eqnarray}
{\mathcal E}_\infty(\tau)={\rm Tr}[P_1\tau]\cdot\pi_1+{\rm Tr}[P_0\tau]\cdot|0\rangle\!\langle0|,\nonumber
\end{eqnarray}
respectively, where $P_0=|0\rangle\!\langle0|$, $P_1=I-P_0$ and $\pi_1=P_1/d$. It is straightforward to verify that ${\mathcal E}$ is self-adjoint. By applying ${\mathcal E}_\infty$ to $|\Psi_\lambda\rangle$ on $A$, we obtain that
\begin{eqnarray}
\Psi_\infty^{ABC}&=&\lambda\:\pi_1^A\otimes|0\rangle\!\langle0|^B\otimes\pi_1^{C}\nonumber\\
&&+(1-\lambda)\:|0\rangle\!\langle0|^A\otimes|\Phi_d\rangle\!\langle\Phi_d|^{BC}.\nonumber
\end{eqnarray}
Therefore, due to Theorem \ref{thm:comp}, the Markovianizing cost is given by
\begin{eqnarray}
M_{A|B}(\Psi_\lambda)=h(\lambda)+2\lambda\log{d}.\nonumber
\end{eqnarray}

On the other hand, from (\ref{eq:redpsiac2}), the Markovianizing costs $M_{C|B}$ of $\Psi_\lambda$ does not depend on $\lambda$ when $\lambda>0$ as proved in Section \ref{sec:pureprop}. Thus we have
\begin{eqnarray}
M_{C|B}(\Psi_\lambda)=\begin{cases}
2\log{d}&(\lambda>0)\\
0&(\lambda=0)\end{cases}\nonumber
\end{eqnarray}
in the same way as (\ref{eq:markdisc}). Hence the Markovianizing cost is not symmetric in $A$ and $C$, as opposed to QCMI, which satisfies $I(A:C|B)=I(C:A|B)$.

\subsection{States for which the Markovianizing cost coincides QCMI}\label{sec:exghz}

We consider states that are expressed as
\begin{eqnarray}
|\Psi_{\{\!\lambda_k\!\}}\rangle:=\sum_{k=1}^d\sqrt{\lambda_k}|k\rangle^A|k\rangle^B|k\rangle^C,\nonumber
\end{eqnarray}
where $\lambda_k\geq0$ and $\sum_{k=1}^d\lambda_k=1$. These states satisfy conditions in Theorem \ref{thm:mareqmi}, thus the Markovianizing cost is given by
\begin{eqnarray}
M_{A|B}(\Psi_{\{\!\lambda_k\!\}})=I(A:C|B)_{\Psi_{\{\!\lambda_k\!\}}}=H(\{\lambda_k\}_k).\nonumber
\end{eqnarray}

\hfill

\section{Conclusions and Discussions}
\label{sec:conclusion}
We have introduced the task of Markovianization, and derived a single-letter formula for the minimum cost of randomness required for Markovianizing tripartite pure states. We have also proposed an algorithm to compute the Markovianizing cost of a class of pure states without obtaining an explicit form of the Koashi-Imoto decomposition. We then have computed the Markovianizing cost for certain pure states, and revealed its discontinuity and asymmetry. Our results have an application in analyzing optimal costs of resources for simulating a bipartite unitary gate by local operations and classical communication\cite{waka15_2}. Some open questions are generalization to mixed states, formulation of a classical analog of Markovianization, in addition to finding an alternative formulation of Markovianization for which we obtain QCMI as the cost function.

In \cite{waka15_rec}, we have introduced and analyzed an alternative formulation of Markivianization and the Markovianizing cost. Instead of requiring Condition (\ref{eq:defmarkovianizing}), we require that the state after a random unitary operation is ``approximately recoverable'', i.e., it satisfies Equalities (\ref{eq:recoverable}) approximately. For pure states, we have proved that the Markovianizing cost in that case is equal to the one obtained in this paper.


\hfill

\section*{Acknowledgment}

The authors thank Tomohiro Ogawa and Masato Koashi for useful discussions. 

\hfill
\bibliographystyle{IEEEtran}
\bibliography{markov}

\hfill

\appendices

\section{Mathematical Preliminaries}

In this appendix, we summarize frequently used facts and technical tools used when studying quantum Shannon theory and also in the following appendices. Readers who are familiar with the material may skip this section. For the references, see e.g. \cite{nielsentext,hayashitext,wildetext}.

\subsection{Trace Distance and Uhlmann's Theorem}\label{app:uhlman}

The trace distance between two quantum states $\rho,\sigma\in{\mathcal S}({\mathcal H})$ is defined by
\begin{eqnarray}
d(\rho,\sigma):=\frac{1}{2}\|\rho-\sigma\|_1=\frac{1}{2}{\rm Tr}\left[\sqrt{(\rho-\sigma)^2}\right].\nonumber
\end{eqnarray}
In the following, we omit the coefficient $1/2$ for simplicity. For pure states $|\psi\rangle,|\phi\rangle\in{\mathcal H}$, the trace distance takes a simple form of
\begin{eqnarray}
\left\||\psi\rangle\!\langle\psi|-|\phi\rangle\!\langle\phi|\right\|_1=2\sqrt{1-|\langle\psi|\phi\rangle|^2}.\nonumber
\end{eqnarray}
For $\rho,\sigma,\tau\in{\mathcal S}({\mathcal H})$, we have
\begin{eqnarray}
\left\|\rho-\tau\right\|_1\leq\left\|\rho-\sigma\right\|_1+\left\|\sigma-\tau\right\|_1,\nonumber
\end{eqnarray}
which is called the {\it triangle inequality}. The trace distance is monotonically nonincreasing under quantum operations, i.e., it satisfies
\begin{eqnarray}
\|\rho-\sigma\|_1\geq\|{\mathcal E}(\rho)-{\mathcal E}(\sigma)\|_1\nonumber
\end{eqnarray}
for any linear CPTP map ${\mathcal E}:{\mathcal S}({\mathcal H})\rightarrow{\mathcal S}({\mathcal H}')$. As a particular case, the trace distance between two states on a composite system is nonincreasing under taking the partial trace, that is, for $\rho,\sigma\in{\mathcal S}({\mathcal H}^A\otimes{\mathcal H}^B)$ we have
\begin{eqnarray}
\|\rho^{AB}-\sigma^{AB}\|_1\geq\|\rho^A-\sigma^A\|_1.\nonumber
\end{eqnarray}

Consider two states $\rho,\sigma\in{\mathcal S}({\mathcal H}^A)$ satisfying $\|\rho-\sigma\|_1\leq\epsilon$, and let $|\psi_\rho\rangle^{AB}$ and $|\phi_\sigma\rangle^{AB'}$ be purifications of the two states, respectively. If $d_B\leq d_{B'}$, there exists an embedding of ${\mathcal H}^B$ into ${\mathcal H}^{B'}$, represented by an isometry from ${\mathcal H}^B$ to ${\mathcal H}^{B'}$, such that
\begin{eqnarray}
\left\|\proj{\psi_\rho}^{AB'}-\proj{\psi_\sigma}^{AB'}\right\|_1\leq2\sqrt{\epsilon}.\nonumber
\end{eqnarray}
This relation is referred to as {\it Uhlmann's theorem} (\!\cite{uhlmann}, see also Lemma 2.2 in \cite{deve08}). In the case of $\epsilon=0$, the above statement implies that all purifications are equivalent up to a local isometry.

The {\it gentle measurement lemma} (Lemma 9.4.1 in \cite{wildetext}) states that for any $\rho\in{\mathcal S}({\mathcal H})$, $X\in{\mathcal L}({\mathcal H})$ and $\epsilon\geq0$ such that $0\leq X\leq I$ and ${\rm Tr}[\rho X]\geq1-\epsilon$, we have
\begin{eqnarray}
\left\|\rho-\frac{\sqrt{X}\rho\sqrt{X}}{{\rm Tr}[\rho X]}\right\|_1\leq2\sqrt{\epsilon}.\label{eq:gentlemeasurement2}
\end{eqnarray}
As a corollary, when two bipartite states $\rho\in{\mathcal S}({\mathcal H}^A\otimes{\mathcal H}^B)$ and $\sigma\in{\mathcal S}({\mathcal H}^A\otimes{\mathcal H}^{B'})$ satisfies $\|\rho^A-\sigma^A\|_1\leq\epsilon$, and $\Pi_\sigma$ is the projection onto ${\rm supp}[\sigma^A]\subseteq{\mathcal H}^A$, we have
\begin{eqnarray}
\left\|\rho^{AB}-\frac{\Pi_\sigma^A\rho^{AB}\Pi_\sigma^A}{{\rm Tr}[\rho^A\Pi_\sigma^A]}\right\|_1\leq2\sqrt{\epsilon}.\label{eq:namacha}
\end{eqnarray}
This is because we have
\begin{eqnarray}
\epsilon&\geq&\left\|\Pi_\sigma\rho^{A}\Pi_\sigma+\Pi_\sigma^\perp\rho^{A}\Pi_\sigma^\perp-\sigma^A\right\|_1\nonumber\\
&=&\left\|\sigma^A-\Pi_\sigma\rho^{A}\Pi_\sigma\right\|_1+{\rm Tr}[\rho^A\Pi_\sigma^\perp],\nonumber
\end{eqnarray}
where $\Pi_\sigma^\perp$ denotes the projection onto the orthogonal complement of ${\rm supp}[\sigma^A]\subseteq{\mathcal H}^A$, and thus have
\begin{eqnarray}
{\rm Tr}[\rho^{AB}\Pi_\sigma^A]={\rm Tr}[\rho^A\Pi_\sigma^A]=1-{\rm Tr}[\rho^A\Pi_\sigma^\perp]\geq1-\epsilon.\nonumber
\end{eqnarray}

\subsection{Quantum Entropies and Mutual Informations}\label{app:entandmutual}

The Shannon entropy of a probability distribution $\{p_x\}_{x\in{\mathcal X}}$ is defined as
\begin{eqnarray}
H(\{p_x\}_{x\in{\mathcal X}}):=-\sum_{x\in{\mathcal X}}p_x\log{p_x}.\nonumber
\end{eqnarray}
The von Neumann entropy of a quantum state $\rho^A\in {\mathcal S}({\mathcal H}^A)$ is defined as
\begin{eqnarray}
S(\rho^A)=S(A)_\rho:=-{\rm Tr}[\rho^A\log{\rho^A}].\nonumber
\end{eqnarray}
If $\rho^A$ is a probabilistic mixture of pure states as $\rho^A=\sum_jp_j|\phi_j\rangle\!\langle\phi_j|$, we have $S(\rho^A)\leq H(\{p_j\}_j)$ with equality if and only if $\{|\phi_j\rangle\}_j$ is mutually orthogonal. The von Neumann entropy is monotonically nondecreasing under random unitary operations, that is, we have $S(A)_\rho\leq S(A)_{{\mathcal V}(\rho)}$ for any random unitary operation $\mathcal V$ on $A$. For a bipartite pure state $|\psi\rangle^{AA'}$, we have
\begin{eqnarray}
S(\psi^A)=S(\psi^{A'}).\label{eq:vNpure}
\end{eqnarray}

For a bipartite state $\rho\in{\mathcal S}({\mathcal H}^A\otimes{\mathcal H}^B)$, the quantum conditional entropy and the quantum mutual information (QMI) are defined as
\begin{eqnarray}
&&S(A|B)_\rho=S(AB)_\rho-S(B)_\rho,\nonumber\\
&&I(A:B)_\rho=S(A)_\rho+S(B)_\rho-S(AB)_\rho,\nonumber
\end{eqnarray}
respectively. The von Neumann entropy satisfies the {\it subadditivity}, expressed as
\begin{eqnarray}
S(A)_\rho+S(B)_\rho\geq S(AB)_\rho,\label{eq:subadditivity}
\end{eqnarray}
which guarantees the nonnegativity of QMI. The equality holds if and only if $\rho^{AB}=\rho^A\otimes\rho^B$. Applying (\ref{eq:subadditivity}) to $|\psi_\sigma\rangle^{ABC}$, which is a purification of $\sigma^{AC}$, and by using (\ref{eq:vNpure}), we obtain
\begin{eqnarray}
S( C)_\sigma-S(A)_\sigma\leq S(AC)_\sigma.\label{eq:subadditivity2}
\end{eqnarray}
Hence QMI is bounded above as
\begin{eqnarray}
I(A:C)_\rho\leq\min\{2S(A)_\rho,2S(C)_\rho\}.\label{eq:ileq2s}
\end{eqnarray}
For any $\rho\in{\mathcal S}({\mathcal H}^A\otimes{\mathcal H}^B)$ and quantum operation $\mathcal E$ on $B$, we have
\begin{eqnarray}
S(A|B)_\rho\leq S(A|B)_{{\mathcal E}(\rho)},\;\;I(A:B)_\rho\geq I(A:B)_{{\mathcal E}(\rho)}.\label{eq:dpi}
\end{eqnarray}
Inequalities (\ref{eq:dpi}) are called the {\it data processing inequality}.

For a tripartite state $\rho\in{\mathcal S}({\mathcal H}^A\otimes{\mathcal H}^B\otimes{\mathcal H}^C)$, the quantum conditional mutual information (QCMI) is defined as
\begin{eqnarray}
I(A:C|B)_\rho=S(AB)_\rho+S(BC)_\rho-S(B)_\rho-S(ABC)_\rho.\nonumber
\end{eqnarray}
QCMI is nonnegative because of the {\it strong subadditivity} of the von Neumann entropy\cite{lieb73}, which is also equivalent to the data processing inequality. QMI and QCMI are related by a simple relation as
\begin{eqnarray}
I(A:BC)_\rho=I(A:B)_\rho+I(A:C|B)_\rho,\nonumber
\end{eqnarray}
which is called the {\it chain rule}.

For a class of states called the {\it classical-quantum states}, the quantum conditional entropy and QCMI take simple forms. That is, for states $\rho\in{\mathcal S}({\mathcal H}^X\otimes{\mathcal H}^A)$ and $\sigma\in{\mathcal S}({\mathcal H}^X\otimes{\mathcal H}^A\otimes{\mathcal H}^B)$, given as
\begin{eqnarray}
\rho^{XA}=\sum_ip_i\proj{i}^X\otimes\rho_i^A,\nonumber\\
\sigma^{XAB}=\sum_iq_i\proj{i}^X\otimes\sigma_i^{AB},\nonumber
\end{eqnarray}
where $\{|i\rangle\}_i$ is an orthonormal basis of ${\mathcal H}^X$, we have
\begin{eqnarray}
S(A|X)_\rho&=&\sum_ip_iS(\rho_i^A),\nonumber\\
I(A:B|X)_\sigma&=&\sum_iq_iI(A:B)_{\sigma_i}.\nonumber
\end{eqnarray}
QMI of a classical-quantum state takes the form of
\begin{eqnarray}
I(X:A)_\rho=S({\bar\rho}^A)-\sum_ip_iS(\rho_i^A),\nonumber
\end{eqnarray}
where ${\bar\rho}^A=\sum_ip_i\rho_i^A$. This quantity is equal to the {\it Holevo information}\cite{holevo73}, and satisfies
\begin{eqnarray}
I(X:A)_\rho\leq S(X)_\rho=H(\{p_i\}_i) \nonumber
\end{eqnarray}
with equality if and only if $\{{\rm supp}[\rho_i^A]\}_i$ is mutually orthogonal.

\subsection{Continuity of Quantum Entropies}
Define
\begin{eqnarray}
\eta_0(x):=\begin{cases}
-x\log{x}&(x\leq 1/e)\\
\frac{1}{e}&(x\geq 1/e)
\end{cases},\nonumber
\end{eqnarray}
$\eta(x)=x+\eta_0(x)$ and $h(x):=\eta_0(x)+\eta_0(1-x)$, where $e$ is the base of the natural logarithm. For two states $\rho$ and $\sigma$ in a $d$-dimensional quantum system ($d<\infty$) such that $\|\rho-\sigma\|_1\leq\epsilon$, we have
\begin{eqnarray}
|S(\rho)-S(\sigma)|\leq\epsilon\log{d}+\eta_0(\epsilon)\leq\eta(\epsilon)\log{d},\label{eq:fannes}
\end{eqnarray}
which is called the {\it Fannes inequality}\cite{fannes73}. It follows that for two bipartite states $\rho,\sigma\in{\mathcal S}({\mathcal H}^A\otimes{\mathcal H}^B)$ such that  $\|\rho-\sigma\|_1\leq\epsilon$, we have
\begin{eqnarray}
|S(A|B)_\rho-S(A|B)_\sigma|\leq\eta(\epsilon)\log{(d_Ad_B^2)}\label{eq:alicki}
\end{eqnarray}
and
\begin{eqnarray}
|I(A:B)_\rho-I(A:B)_\sigma|\leq2\eta(\epsilon)\log{(d_Ad_B)}.\label{eq:contmi}
\end{eqnarray}

\subsection{Typical Sequences and Subspaces (\!\!\cite{cover05,schumacher95}, see also Appendices in \cite{abey09} for further details.)}

Let $X$ be a discrete random variable with finite alphabet $\mathcal X$ and probability distribution $p_x={\rm Pr}\{X=x\}$ where $x\in{\mathcal X}$. A sequence ${\bm x}=(x_1,\cdots,x_n)\in{\mathcal X}^n$ is said to be {\it$\delta$-weakly typical with respect to} $\{p_x\}_{x\in{\mathcal X}}$ if it satisfies
\begin{eqnarray}
2^{-n(H(X)+\delta)}\leq\prod_{i=1}^Np_{x_i}\leq2^{-n(H(X)-\delta)}.\nonumber
\end{eqnarray}
The set of all $\delta$-weakly typical sequences is called the {\it$\delta$-weakly typical set}, and is denoted by ${\mathcal T}_{n,\delta}$ in the following. Denoting $\prod_{i=1}^Np_{x_i}$ by $p_{\bm x}$, we have
\begin{eqnarray}
1=\sum_{{\bm x}\in{\mathcal X}^n}p_{\bm x}\geq\sum_{{\bm x}\in{\mathcal T}_{n,\delta}}p_{\bm x}\geq|{\mathcal T}_{n,\delta}|\cdot2^{-n(H(X)+\delta)},\nonumber
\end{eqnarray}
which implies that
\begin{eqnarray}
|{\mathcal T}_{n,\delta}|\leq2^{n(H(X)+\delta)}.
\label{eq:cardtyp}\end{eqnarray}

A sequence ${\bm x}=(x_1,\cdots,x_n)\in{\mathcal X}^n$ is called {\it$\delta$-strongly typical with respect to} $\{p_x\}_{x\in{\mathcal X}}$ if it satisfies
\begin{eqnarray}
\left|\frac{1}{n}N_{x|{\bm x}}-p_x\right|<\frac{\delta}{|{\mathcal X}|}\nonumber
\end{eqnarray}
for all $x\in{\mathcal X}$ and $N_{x|{\bm x}}=0$ if $p_x=0$. Here, $N_{x|{\bm x}}$ is the number of occurrences of the symbol $x$ in the sequence $\bm x$. The set of all $\delta$-strongly typical sequences is called the {\it$\delta$-strongly typical set}, and denoted by ${\mathcal T}_{n,\delta}^*$ in the following. From the weak law of large numbers, we have that for any $\epsilon,\delta>0$ and sufficiently large $n$,
\begin{eqnarray}
&&{\rm Pr}\{(X_1,\cdots,X_n)\in{\mathcal T}_{n,\delta}\}\geq1-\epsilon,\label{eq:weaktyp}\\
&&{\rm Pr}\{(X_1,\cdots,X_n)\in{\mathcal T}_{n,\delta}^*\}\geq1-\epsilon.\label{eq:strongtyp}
\end{eqnarray}

Suppose the spectral decomposition of $\rho\in{\mathcal S}({\mathcal H})$ is given by $\rho=\sum_{x}p_x\proj{x}$. The $\delta$-weakly typical subspace ${\mathcal H}_{n,\delta}\subset{\mathcal H}^{\otimes n}$ with respect to $\rho$ is defined as
\begin{eqnarray}
{\mathcal H}_{n,\delta}:={\rm span}\{\ket{x_1}\cdots\ket{x_n}\in{\mathcal H}^{\otimes n}|(x_1,\cdots,x_n)\in{\mathcal T}_{n,\delta}\},\!\!\!\!\!\!\!\!\!\nonumber
\end{eqnarray}
where ${\mathcal T}_{n,\delta}$ is the $\delta$-weakly typical set with respect to $p_x$. 
Similarly, the $\delta$-strongly typical subspace ${\mathcal H}_{n,\delta}^*\subset{\mathcal H}^{\otimes n}$ with respect to $\rho$ is defined as
\begin{eqnarray}
{\mathcal H}_{n,\delta}^*:={\rm span}\{\ket{x_1}\cdots\ket{x_n}\in{\mathcal H}^{\otimes n}|(x_1,\cdots,x_n)\in{\mathcal T}_{n,\delta}^*\}.\!\!\!\!\!\!\!\!\!\nonumber
\end{eqnarray}

Suppose the Schmidt decomposition of $|\psi\rangle^{AB}\in{\mathcal H}^A\otimes{\mathcal H}^B$ is given by $|\psi\rangle^{AB}=\sum_x\sqrt{p_x}|x\rangle^A|x\rangle^B$. For any $\delta>0$ and $n$, let  ${\mathcal H}_{n,\delta}$ and ${\mathcal H}_{n,\delta}^*$ be the $\delta$-weakly and strongly typical subspace of $({\mathcal H}^A)^{\otimes n}$ with respect to $\psi^A={\rm Tr}_B[|\psi\rangle\!\langle\psi|^{AB}]$, and let $\Pi_{n,\delta}$ and $\Pi_{n,\delta}^*$ be the projection onto those subspaces, respectively. From (\ref{eq:cardtyp}), we have
\begin{eqnarray}
{\rm rank}\:\Pi_{n,\delta}={\rm dim}\:{\mathcal H}_{n,\delta}\leq2^{n(H(A)_\psi+\delta)}.\nonumber
\end{eqnarray}
From (\ref{eq:weaktyp}) and (\ref{eq:strongtyp}), we have
\begin{eqnarray}
&&{\rm Tr}[\Pi_{n,\delta}^{\bar A}(|\psi\rangle\!\langle\psi|^{AB})^{\otimes n}]=\sum_{{\bm x}\in{\mathcal T}_{n,\delta}}p_{\bm x}\geq1-\epsilon,\label{eq:aaaaaa}\\
&&{\rm Tr}[\Pi_{n,\delta}^{*{\bar A}}(|\psi\rangle\!\langle\psi|^{AB})^{\otimes n}]=\sum_{{\bm x}\in{\mathcal T}_{n,\delta}^*}p_{\bm x}\geq1-\epsilon\label{eq:aaaaaaa}
\end{eqnarray}
for any $\epsilon,\delta>0$ and sufficiently large $n$.

\hfill

\section{Proof of Theorem \ref{thm:strongmarkcostequality}}\label{app:rigorous}

In this Appendix, we show a detailed proof of Theorem \ref{thm:strongmarkcostequality}. In the following, we informally denote the composite systems $a_{\scalebox{0.6}{$0$}}a_{\scalebox{0.6}{$L$}}a_{\scalebox{0.6}{$R$}}$ by $A$ and $b_{\scalebox{0.6}{$0$}}b_{\scalebox{0.6}{$L$}}b_{\scalebox{0.6}{$R$}}$ by $B$, when there is no fear of confusion.

\subsection{Proof of Achievability (Inequality (\ref{eq:markoviancost}))}\label{app:achieve}
Fix arbitrary $n$ and $\delta\in(0,1]$. Let $J_{n,\delta}\subset J^n$ be the $\delta$-strongly typical set with respect to $\{p_j\}_{j\in J}$. For each $j\in J$ and ${\bm j}=j_1\cdots j_n\in J_{n,\delta}$, define ${\mathfrak L}_{j,{\bm j}}:=\{l|j_l=j,1\leq l\leq n\}$. The number of elements in the set is bounded as
\begin{eqnarray}
n\left(p_j-\frac{\delta}{|J|}\right)\leq|{\mathfrak L}_{j,{\bm j}}|\leq n\left(p_j+\frac{\delta}{|J|}\right).\nonumber
\end{eqnarray}
For each ${\bm j}\in J_{n,\delta}$, we sort $({\mathcal H}^{a_{\scalebox{0.45}{$R$}}})^{\otimes n}={\mathcal H}^{a_{\scalebox{0.45}{$R$}_1}}\otimes\cdots\otimes{\mathcal H}^{a_{\scalebox{0.45}{$R$}_n}}$ as 
\begin{eqnarray}
({\mathcal H}^{a_{\scalebox{0.45}{$R$}}})^{\otimes n}=\bigotimes_{j\in J}\left(\bigotimes_{l\in{\mathfrak L}_{j,{\bm j}}}{\mathcal H}^{a_{\scalebox{0.45}{$R$}_l}}\right).\nonumber
\end{eqnarray}
For each $j$ and ${\bm j}$, let ${\mathcal H}_{j,{\bm j},\delta}$ be the $\delta$-weakly typical subspace of $\bigotimes_{l\in{\mathfrak L}_{j,{\bm j}}}{\mathcal H}^{a_{\scalebox{0.45}{$R$}_l}}$ with respect to $\varphi_j^{a_{\scalebox{0.45}{$R$}}}$, $\Pi_{j,{\bm j},\delta}$ be the projection onto ${\mathcal H}_{j,{\bm j},\delta}$, and let $\Pi_{{\bm j},\delta}^{{\bar a}_{\scalebox{0.45}{$R$}}}:=\bigotimes_{j\in J}\Pi_{j,{\bm j},\delta}$. Define 
\begin{eqnarray}
\Pi^{\bar A}_\delta:=\sum_{{\bm j}\in J_{n,\delta}}\proj{\bm j}^{{\bar a}_{\scalebox{0.45}{$0$}}}\otimes I_{\bm j}^{{\bar a}_{\scalebox{0.45}{$L$}}}\otimes\Pi_{{\bm j},\delta}^{{\bar a}_{\scalebox{0.45}{$R$}}}\label{eq:pindelta}
\end{eqnarray}
and
\begin{eqnarray}
&&\!\!\!\!\!\!\!\!\!\!\!\!\!\!\!\!|\Psi'_{n,\delta}\rangle^{{\bar A}{\bar B}{\bar C}}:=\Pi^{\bar A}_\delta|\Psi_{K\!I}^{\otimes n}\rangle^{{\bar A}{\bar B}{\bar C}}\nonumber\\
&&\!\!\!\!\!=\sum_{{\bm j}\in J_{n,\delta}}\sqrt{p_{\bm j}}\ket{\bm j}^{{\bar a}_{\scalebox{0.45}{$0$}}}\ket{\bm j}^{{\bar b}_{\scalebox{0.45}{$0$}}}\ket{\omega_{\bm j}}^{{\bar a}_{\scalebox{0.45}{$L$}}{\bar b}_{\scalebox{0.45}{$L$}}}\Pi_{{\bm j},\delta}^{{\bar a}_{\scalebox{0.45}{$R$}}}\ket{\varphi_{\bm j}}^{{\bar a}_{\scalebox{0.45}{$R$}}{\bar b}_{\scalebox{0.45}{$R$}}{\bar C}},\label{eq:psindelta}
\end{eqnarray}
where we introduced notations $p_{\bm j}=p_{j_1}\!\times\cdots\times p_{j_n}$, $\varphi_{\bm j}=\varphi_{j_1}\otimes\cdots\otimes\varphi_{j_n}$ and $\omega_{\bm j}=\omega_{j_1}\otimes\cdots\otimes\omega_{j_n}$.

Let $v_{{\bm j}}$ be any unitary acting on $\bigotimes_{j\in J}{\mathcal H}_{j,{\bm j}}$, and define a unitary on $\bigotimes_{{\bm j}\in J_{n,\delta}}(\bigotimes_{j\in J}{\mathcal H}_{j,{\bm j}})$ by
\begin{eqnarray}
V^{\bar A}:=\sum_{{\bm j}\in J_{n,\delta}}\proj{\bm j}^{{\bar a}_{\scalebox{0.45}{$0$}}}\otimes I_{\bm j}^{{\bar a}_{\scalebox{0.45}{$L$}}}\otimes v_{{\bm j}}^{{\bar a}_{\scalebox{0.45}{$R$}}},
\label{eq:vensemble}
\end{eqnarray}
 as (\ref{eq:vensemble2}). We have
\begin{eqnarray}
&&\!\!\!\!\!\!\!\!\!\!|\Psi'_{n,\delta}(V)\rangle^{{\bar A}{\bar B}{\bar C}}:=V^{\bar A}|\Psi'_{n,\delta}\rangle^{{\bar A}{\bar B}{\bar C}}\nonumber\\
&&=\sum_{{\bm j}\in J_{n,\delta}}\sqrt{p_{\bm j}}\ket{\bm j}^{{\bar a}_{\scalebox{0.45}{$0$}}}\ket{\bm j}^{{\bar b}_{\scalebox{0.45}{$0$}}}\ket{\omega_{\bm j}}^{{\bar a}_{\scalebox{0.45}{$L$}}{\bar b}_{\scalebox{0.45}{$L$}}}v_{{\bm j}}^{{\bar a}_{\scalebox{0.45}{$R$}}}|{\varphi'_{\bm j}}\rangle^{{\bar a}_{\scalebox{0.45}{$R$}}{\bar b}_{\scalebox{0.45}{$R$}}{\bar C}},\;\;\;\;
\label{eq:vvv}
\end{eqnarray}
where $|\varphi'_{\bm j}\rangle:=\Pi_{{\bm j}}^{{\bar a}_{\scalebox{0.45}{$R$}}}\ket{\varphi_{\bm j}}$.
 
Let $\{p(dV),V\}$ be the ensemble of unitaries generated by choosing $v_{\bm j}$ randomly and independently according to the Haar measure for each ${\bm j}$ in (\ref{eq:vensemble}). Due to Schur's lemma, as an ensemble average we have
\begin{eqnarray}
{\mathbb E}\left[v_{{\bm j}}^{{\bar a}_{\scalebox{0.45}{$R$}}}|\varphi'_{\bm j}\rangle\!\langle\varphi'_{\bm j}|v_{{\bm j}}^{\dagger {\bar a}_{\scalebox{0.45}{$R$}}}\right]=\pi_{{\bm j}}^{{\bar a}_{\scalebox{0.45}{$R$}}}\otimes\varphi'^{{\bar b}_{\scalebox{0.45}{$R$}}{\bar C}}_{\bm j},\nonumber
\end{eqnarray}
where $\pi_{{\bm j}}^{{\bar a}_{\scalebox{0.45}{$R$}}}={\Pi_{{\bm j}, \delta}^{{\bar a}_{\scalebox{0.45}{$R$}}}}/{{\rm Tr}\Pi_{{\bm j},\delta}^{{\bar a}_{\scalebox{0.45}{$R$}}}}$, and
\begin{eqnarray}
{\mathbb E}\left[v_{{\bm j}}^{{\bar a}_{\scalebox{0.45}{$R$}}}|\varphi'_{\bm j}\rangle\!\langle\varphi'_{{\bm j}'}|v_{{\bm j}'}^{\dagger {\bar a}_{\scalebox{0.45}{$R$}}}\right]=0\nonumber
\end{eqnarray}
for ${\bm j}\neq{\bm j}'$. Thus the average state of (\ref{eq:vvv}) is given by
\begin{eqnarray}
&&\!\!\!\!\!\!\!\!\!\!\!{\bar \Psi}_{n,\delta}:={\mathbb E}\left[|\Psi'_{n,\delta}(V)\rangle\!\langle\Psi'_{n,\delta}(V)|^{{\bar A}{\bar B}{\bar C}}\right]\nonumber\\
&&\!\!\!\!\!\!\!\!\!\!=\sum_{{\bm j}\in J_{n,\delta}}\!\!{p_{\bm j}}\proj{{\bm j}{\bm j}}^{{\bar a}_{\scalebox{0.45}{$0$}}{\bar b}_{\scalebox{0.45}{$0$}}}\!\!\otimes\proj{\omega_{\bm j}}^{{\bar a}_{\scalebox{0.45}{$L$}}{\bar b}_{\scalebox{0.45}{$L$}}}\otimes\pi_{{\bm j}}^{{\bar a}_{\scalebox{0.45}{$R$}}}\otimes\varphi'^{{\bar b}_{\scalebox{0.45}{$R$}}{\bar C}}_{\bm j}\nonumber\\
&&\!\!\!\!\!\!\!\!\!\!=\sum_{{\bm j}\in J_{n,\delta}}\!\!{p_{\bm j}}\proj{{\bm j}}^{{\bar b}_{\scalebox{0.45}{$0$}}}\!\!\otimes\left(\pi_{{\bm j}}^{{\bar a}_{\scalebox{0.45}{$R$}}}\!\otimes\proj{{\bm j},\omega_{\bm j}}^{{\bar a}_{\scalebox{0.45}{$0$}}{\bar a}_{\scalebox{0.45}{$L$}}{\bar b}_{\scalebox{0.45}{$L$}}}\right)\otimes\varphi'^{{\bar b}_{\scalebox{0.45}{$R$}}{\bar C}}_{\bm j},\nonumber\\
\label{eq:submarkov}
\end{eqnarray}
which is a subnormalized Markov state conditioned by $\bar B$ corresponding to (\ref{eq:transformmarkov2}) (see Figure \ref{fig:markovunitary}).

The minimum nonzero eigenvalue of ${\bar \Psi}_{n,\delta}$ is calculated as follows. First, due to the definition of $J_{n,\delta}$, we have
\begin{eqnarray}
{p_{\bm j}}\geq \prod_{j\in J}p_j^{n(p_j+\delta/|J|)}=2^{-n\left(H(\{p_j\}_j)+\delta H'(\{p_j\}_j)\right)},\nonumber
\end{eqnarray}
where
\begin{eqnarray}
H'(\{p_j\}_j):=\frac{1}{|J|}\sum_j\log{p_j}>-\infty.\nonumber
\end{eqnarray}
Second, since the spectrums of $\varphi'^{{\bar a}_{\scalebox{0.45}{$R$}}}_{\bm j}$ and $\varphi'^{{\bar b}_{\scalebox{0.45}{$R$}}{\bar C}}_{\bm j}$ are the same, the minimum nonzero eigenvalue $\mu_{\bm j}$ of $\varphi'^{{\bar b}_{\scalebox{0.45}{$R$}}{\bar C}}_{\bm j}$ is bounded from below as
\begin{eqnarray}
\mu_{\bm j}&\geq&\prod_{j\in J}2^{-{\mathfrak L}_{j,{\bm j}}(S(\varphi_j^{a_{\scalebox{0.45}{$R$}}})+\delta)}\nonumber\\
&\geq&\prod_{j\in J}2^{-n(p_j+\delta/|J|)(S(\varphi_j^{a_{\scalebox{0.45}{$R$}}})+\delta)}\nonumber\\
&\geq&2^{-n\left(\sum_{j}p_jS(\varphi_j^{a_{\scalebox{0.45}{$R$}}})+\delta\log{(4d_A)}\right)},\nonumber
\end{eqnarray}
where the last line follows from
\begin{eqnarray}
&&\sum_{j}\left(p_j+\frac{\delta}{|J|}\right)\left(S(\varphi_j^{a_{\scalebox{0.45}{$R$}}})+\delta\right)\nonumber\\
&=&\sum_{j}p_jS(\varphi_j^{a_{\scalebox{0.45}{$R$}}})+\delta\left(1+\frac{1}{|J|}\sum_{j}S(\varphi_j^{a_{\scalebox{0.45}{$R$}}})+\delta\right)\nonumber\\
&\leq&\sum_{j}p_jS(\varphi_j^{a_{\scalebox{0.45}{$R$}}})+\delta\left(2+\log{d_A}\right)\nonumber\\
&=&\sum_{j}p_jS(\varphi_j^{a_{\scalebox{0.45}{$R$}}})+\delta\log{(4d_A)}.\nonumber
\end{eqnarray}
Third, we have 
\begin{eqnarray}
{\rm rank}\:\Pi_{j,{\bm j},\delta}\leq2^{|{\mathfrak L}_{j,{\bm j}}|(S(\varphi_j^{a_{\scalebox{0.45}{$R$}}})+\delta)}\leq2^{n(p_j+\delta/|J|)(S(\varphi_j^{a_{\scalebox{0.45}{$R$}}})+\delta)}\nonumber
\end{eqnarray}
and
\begin{eqnarray}
{\rm rank}\:\Pi_{\bm j,\delta}^{{\bar a}_{\scalebox{0.45}{$R$}}}=\prod_{j\in J}{\rm rank}\:\Pi_{j,{\bm j}, \delta}\leq\prod_{j\in J}2^{n(p_j+\delta/|J|)(S(\varphi_j^{a_{\scalebox{0.45}{$R$}}})+\delta)}.\!\!\!\nonumber
\end{eqnarray}
Thus the nonzero eigenvalue $\nu_{\bm j}$ of $\pi_{\bm j}^{{\bar a}_{\scalebox{0.45}{$R$}}}$ is, in the same way as $\mu_{\bm j}$, bounded from below as
\begin{eqnarray}
\nu_{\bm j}\geq\prod_{j\in J}2^{-n(p_j+\delta/|J|)(S(\varphi_j^{a_{\scalebox{0.45}{$R$}}})+\delta)}.\nonumber
\end{eqnarray}
All in all, the minimum nonzero eigenvalue $\lambda$ of ${\bar \Psi}_{n,\delta}$ is bounded as
\begin{eqnarray}
\lambda&=&{p_{\bm j}}\mu_{\bm j}\nu_{\bm j}\nonumber\\
&\geq&2^{-n\left[H(\{p_j\}_{j})+2\sum_{j}p_jS(\varphi_j^{a_{\scalebox{0.45}{$R$}}})+\delta\left(H'(\{p_j\}_j)+2\log{(4d_A)}\right)\right]}.\nonumber
\end{eqnarray}
We also have
\begin{eqnarray}
{\rm rank}\:{\bar \Psi}_{n,\delta}\leq |J_{n,\delta}|\times{\rm rank}\:{\pi^{{\bar a}_{\scalebox{0.45}{$R$}}}_{\bm j}}\times{\rm rank}\:{\varphi'^{{\bar a}_{\scalebox{0.45}{$R$}}}_{\bm j}}\leq d_A^{3n}.\nonumber
\end{eqnarray}

Suppose $V_1,\cdots,V_N$ are unitaries that are randomly and independently chosen from the ensemble $\{p(dV),V\}$. Due to the operator Chernoff bound (Lemma 3 in \cite{berry08}), we have
\begin{eqnarray}
&&{\rm Pr}\left\{\frac{1}{N}\sum_{i=1}^N\Psi'_{n,\delta}(V_i)\notin[(1-\epsilon_{{}_1}){\bar \Psi}_{n,\delta},(1+\epsilon_{{}_1}){\bar \Psi}_{n,\delta}]\right\}\nonumber\\
&&\leq2d_A^{3n}\exp{\left(-\frac{N\lambda\epsilon_{{}_1}^2}{2}\right)}\nonumber
\end{eqnarray}
for any $\epsilon_{{}_1}\in(0,1]$, which implies that
\begin{eqnarray}
&&{\rm Pr}\left\{\left\|\frac{1}{2^{nR}}\sum_{i=1}^{2^{nR}}\Psi'_{n,\delta}(V_i)-{\bar \Psi}_{n,\delta}\right\|_1\leq2\epsilon_{{}_1}\right\}\nonumber\\
&&\geq1-2d_A^{3n}\exp{\left(-\frac{2^{nR}\lambda\epsilon_{{}_1}^2}{2}\right)}\label{eq:chercher}
\end{eqnarray}
for an arbitrary $R>0$. Therefore, if $R$ satisfies
\begin{eqnarray}
&&R>H(\{p_j\}_{j})+2\sum_{j}p_jS(\varphi_j^{a_{\scalebox{0.45}{$R$}}})\nonumber\\
&&\;\;\;\;\;\;\;\;\;\;\;+\delta\left(H'(\{p_j\}_j)+2\log{(4d_A)}\right),\label{eq:misalove}
\end{eqnarray}
and if $n$ is sufficiently large so that the R.H.S. in (\ref{eq:chercher}) is greater than $0$, there exists a set of unitaries $\{V_i\}_{i=1}^{2^{nR}}$ such that
\begin{eqnarray}
\left\|\frac{1}{2^{nR}}\sum_{i=1}^{2^{nR}}\Psi'_{n,\delta}(V_i)-{\bar \Psi}_{n,\delta}\right\|_1\leq2\epsilon_{{}_1}.\label{eq:misonolove}
\end{eqnarray}
Using unitaries in the set, construct a random unitary operation ${\mathcal V}_n$ on ${\hat a}_{\scalebox{0.6}{$0$}}{\hat a}_{\scalebox{0.6}{$L$}}{\hat a}_{\scalebox{0.6}{$R$}}$ as ${\mathcal V}_n(\cdot)=2^{-nR}\sum_{k=1}^{2^{nR}}V_k(\cdot)V_k^{\dagger}$.

Let us evaluate the total error. First, from (\ref{eq:aaaaaa}), (\ref{eq:aaaaaaa}), (\ref{eq:pindelta}) and (\ref{eq:psindelta}), we have
\begin{eqnarray}
D_{n,\delta}:={\rm Tr}[\Pi^{\bar A}_{\delta}|\Psi_{K\!I}^{\otimes n}\rangle\!\langle\Psi_{K\!I}^{\otimes n}|]=\langle\Psi'_{n,\delta}|\Psi'_{n,\delta}\rangle\geq1-\epsilon_{{}_2}\label{eq:defddeltan}
\end{eqnarray}
for any $\epsilon_{{}_2}>0$ and sufficiently large $n$. Thus, by the gentle measurement lemma (\ref{eq:gentlemeasurement2}), we have
\begin{eqnarray}
\left\|\frac{|\Psi'_{n,\delta}\rangle\!\langle\Psi'_{n,\delta}|}{D_{n,\delta}}-|\Psi_{K\!I}^{\otimes n}\rangle\!\langle\Psi_{K\!I}^{\otimes n}|\right\|_1\leq2\sqrt{\epsilon_{{}_2}},\nonumber
\end{eqnarray}
which leads to
\begin{eqnarray}
\left\|{\mathcal V}_n\left(\frac{|{\Psi'_{n,\delta}}\rangle\!\langle{\Psi'_{n,\delta}}|}{D_{n,\delta}}\right)-{\mathcal V}_n\left(|\Psi_{K\!I}^{\otimes n}\rangle\!\langle\Psi_{K\!I}^{\otimes n}|\right)\right\|_1\leq2\sqrt{\epsilon_{{}_2}}.\nonumber
\end{eqnarray}
Second, from (\ref{eq:misonolove}) and (\ref{eq:defddeltan}), we have
\begin{eqnarray}
\left\|{\mathcal V}_n\left(\frac{|{\Psi'_{n,\delta}}\rangle\!\langle{\Psi'_{n,\delta}}|}{D_{n,\delta}}\right)-\frac{{\bar \Psi}_{n,\delta}}{D_{n,\delta}}\right\|_1\leq\frac{2\epsilon_{{}_1}}{D_{n,\delta}}\leq\frac{2\epsilon_{{}_1}}{1-\epsilon_{{}_2}}.\nonumber
\end{eqnarray}
Therefore, by the triangle inequality, we obtain
\begin{eqnarray}
\left\|{\mathcal V}_n\left(|\Psi_{K\!I}^{\otimes n}\rangle\!\langle\Psi_{K\!I}^{\otimes n}|\right)-\frac{{\bar \Psi}_{n,\delta}}{D_{n,\delta}}\right\|_1\leq2\sqrt{\epsilon_{{}_2}}+\frac{2\epsilon_{{}_1}}{1-\epsilon_{{}_2}}.\label{eq:erikalove}
\end{eqnarray}
From (\ref{eq:vvv}) and (\ref{eq:submarkov}), we have ${\rm Tr}[{\bar \Psi}_{n,\delta}]=D_{n,\delta}$, which implies that ${\bar \Psi}_{n,\delta}/D_{n,\delta}$ is a normalized Markov state conditioned by $B^n$. Since the relation (\ref{eq:erikalove}) holds for any $\epsilon_{{}_1},\epsilon_{{}_2}>0$, $R>H(\{p_j\}_{j})+2\sum_{j}p_jS(\varphi_j^{a_{\scalebox{0.45}{$R$}}})$, any $\delta\in(0,1]$ that satisfies (\ref{eq:misalove}) and sufficiently large $n$, we obtain (\ref{eq:markoviancost}). \hfill$\blacksquare$

\subsection{Convergence Speed of the Error}\label{app:convspeed}
We prove that, in the direct part of Theorem \ref{thm:strongmarkcostequality}, the error $\epsilon$ vanishes exponentially in the asymptotic limit of $n\rightarrow\infty$. More precisely, we prove the following theorem.
\begin{thm}
There exists a constant $c_{{}_\Psi}>0$ such that for any $R>M_{A|B}(\Psi^{ABC})$, sufficiently small $\delta>0$ and any sufficiently large $n$, we find a random unitary operation ${\mathcal V}_n:\tau\mapsto2^{-nR}\sum_{k=1}^{2^{nR}}V_k\tau V_k^{\dagger}$ on $A^n$ and a Markov state $\Upsilon^{A^nB^nC^n}$ conditioned by $B^n$ that satisfy
\begin{eqnarray}
\left\|{\mathcal V}_n(\rho^{\otimes n})-\Upsilon^{A^nB^nC^n}\right\|_1\leq6\exp{\left(-\frac{c_{{}_\Psi}\delta^2n}{2}\right)}.\nonumber
\end{eqnarray}
\end{thm}

\begin{prf}
Let $X_1,\cdots,X_n$ be a sequence of i.i.d. random variables obeying a probability distribution ${\{p_x\}_x}$.
It is proved in \cite{ahlswede80} that there exists a constant $c>0$, which depends on ${\{p_x\}_x}$, such that for any $\delta>0$ and $n$, we have
\begin{eqnarray}
&&{\rm Pr}\{(X_1,\cdots,X_n)\in{\mathcal T}_{n,\delta}\}\geq1-\exp{(-c\delta^2n)},\nonumber\\
&&{\rm Pr}\{(X_1,\cdots,X_n)\in{\mathcal T}_{n,\delta}^*\}\geq1-\exp{(-c\delta^2n)}.\nonumber
\end{eqnarray}
As a consequence, there exists a constant $c_\psi>0$ such that we have
\begin{eqnarray}
&&\!\!\!\!{\rm Tr}[\Pi_{n,\delta}(|\psi\rangle\!\langle\psi|^{AB})^{\otimes n}]=\sum_{{\bm x}\in{\mathcal T}_{n,\delta}}p_{\bm x}\geq1-\exp{(-c_\psi\delta^2n)}\nonumber\\
&&\!\!\!\!{\rm Tr}[\Pi_{n,\delta}^*(|\psi\rangle\!\langle\psi|^{AB})^{\otimes n}]=\sum_{{\bm x}\in{\mathcal T}_{n,\delta}^*}p_{\bm x}\geq1-\exp{(-c_\psi\delta^2n)}\nonumber
\end{eqnarray}
for any $\delta>0$ and $n$, corresponding to (\ref{eq:aaaaaa}) and (\ref{eq:aaaaaaa}). Thus, for any $\delta>0$, $n$ and $D_{n,\delta}$ defined by (\ref{eq:defddeltan}), we obtain
\begin{eqnarray}
D_{n,\delta}\geq1-\exp{(-c_{{}_\Psi}\delta^2n)},\nonumber
\end{eqnarray}
where $c_{{}_\Psi}>0$ is a constant. Hence we have
\begin{eqnarray}
&&\left\|{\mathcal V}_n\left(|\Psi_{K\!I}^{\otimes n}\rangle\!\langle\Psi_{K\!I}^{\otimes n}|\right)-\frac{{\bar \Psi}_{n,\delta}}{D_{n,\delta}}\right\|_1\nonumber\\
&&\leq2\exp{\left(-\frac{c_{{}_\Psi}\delta^2n}{2}\right)}+\frac{2\epsilon_{{}_1}}{1-\exp{(-c_{{}_\Psi}\delta^2n)}}\nonumber
\end{eqnarray}
for {\it any} $\delta,\epsilon_{{}_1}>0$ and $n$, corresponding to (\ref{eq:erikalove}). Substituting $\exp{(-c_{{}_\Psi}\delta^2n/2)}$ into $\epsilon_1$, we obtain
\begin{eqnarray}
\left\|{\mathcal V}_n\left(|\Psi_{K\!I}^{\otimes n}\rangle\!\langle\Psi_{K\!I}^{\otimes n}|\right)-\frac{{\bar \Psi}_{n,\delta}}{D_{n,\delta}}\right\|_1\leq6\exp{\left(-\frac{c_{{}_\Psi}\delta^2n}{2}\right)}\label{eq:erikaloveee}
\end{eqnarray}
for any $\delta>0$ and $n\geq(\ln{2})/c_{{}_\Psi}\delta^2$.

For an arbitrary $R$, choose sufficiently small $\delta>0$ such that
\begin{eqnarray}
&&R>H(\{p_j\}_{j})+2\sum_{j}p_jS(\varphi_j^{a_{\scalebox{0.45}{$R$}}})\nonumber\\
&&\;\;\;\;\;\;\;\;\;\;\;+\delta\left(H'(\{p_j\}_j)+2\log{(4d_A)}+c_{{}_\Psi}\delta\right).\nonumber
\end{eqnarray}
Inequality (\ref{eq:erikaloveee}) then holds for sufficiently large $n$, while keeping the R.H.S. in (\ref{eq:chercher}) strictly greater than $0$. This completes the proof.
\hfill$\blacksquare$
\end{prf}

\subsection{Proof of Optimality (Inequality (\ref{eq:lowermarkov}))}

We assume, without loss of generality, that $d_A\geq d_Bd_C$. This condition is always satisfied by associating a sufficiently large Hilbert space ${\mathcal H}^A$ to system $A$.

Take an arbitrary $R>M_{A|B}(\Psi^{ABC})$. By definition, for any $\epsilon\in(0,1]$ and sufficiently large $n$, there exist a random unitary operation ${\mathcal V}_n:\tau\mapsto2^{-nR}\sum_{k=1}^{2^{nR}}V_k\tau V_k^{\dagger}$ on $\bar A$ and a Markov state $\Upsilon^{{\bar A}{\bar B}{\bar C}}$ conditioned by $\bar B$ such that
\begin{eqnarray}
\left\|{\mathcal V}_n(\Psi^{\otimes n})-\Upsilon^{{\bar A}{\bar B}{\bar C}}\right\|_1\leq\epsilon.
\label{eq:markovdistance}
\end{eqnarray}
By tracing out $A^n$, we have
\begin{eqnarray}
\left\|(\Psi^{\otimes n})^{{\bar B}{\bar C}}-\Upsilon^{{\bar B}{\bar C}}\right\|_1\leq\epsilon.\label{eq:markovdistanceee}
\end{eqnarray}
Due to Uhlmann's theorem (\!\!\cite{uhlmann}, see Appendix \ref{app:uhlman}), there exists a purification $|\chi\rangle^{{\bar A}{\bar B}{\bar C}}$ of $\Upsilon^{{\bar B}{\bar C}}$ such that we have
\begin{eqnarray}
\left\|(\Psi^{\otimes n})^{{\bar A}{\bar B}{\bar C}}-\chi^{{\bar A}{\bar B}{\bar C}}\right\|_1\leq2\sqrt{\epsilon}.
\label{eq:markovdistance3}
\end{eqnarray}

Let ${\tilde\Gamma}':{\mathcal H}_\chi^{\bar B}\rightarrow{\mathcal H}^{{\hat b}_{\scalebox{0.45}{$0$}}}\otimes{\mathcal H}^{{\hat b}_{\scalebox{0.45}{$L$}}}\otimes{\mathcal H}^{{\hat b}_{\scalebox{0.45}{$R$}}}$ be the KI isometry on ${\bar B}$ with respect to $\chi^{{\bar B}{\bar C}}$. From Lemma \ref{lmm:kipure}, there exists a sub-KI isometry ${\tilde\Gamma}:{\mathcal H}_\chi^{{\bar A}}\rightarrow{\mathcal H}^{{\hat a}_{\scalebox{0.45}{$0$}}}\otimes{\mathcal H}^{{\hat a}_{\scalebox{0.45}{$L$}}}\otimes{\mathcal H}^{{\hat a}_{\scalebox{0.45}{$R$}}}$ such that the KI decomposition of $|\chi\rangle$ on $\bar B$ and ${\bar A}$ is given by
\begin{eqnarray}
\ket{\chi_{{}_{K\!I}}}&:=&({\tilde\Gamma}^{\bar A}\otimes{\tilde\Gamma}'^{{\bar B}})|\chi\rangle\nonumber\\
&=&\sum_{i}\sqrt{q_i}\ket{i}^{{\hat a}_{\scalebox{0.45}{$0$}}}\ket{i}^{{\hat b}_{\scalebox{0.45}{$0$}}}\ket{\xi_i}^{{\hat a}_{\scalebox{0.45}{$L$}}{\hat b}_{\scalebox{0.45}{$L$}}}\ket{\phi_i}^{{\hat a}_{\scalebox{0.45}{$R$}}{\hat b}_{\scalebox{0.45}{$R$}}{\bar C}}.
\label{eq:purechi}
\end{eqnarray}
From Theorem \ref{thm:markovdechayden} and $\chi^{{\bar B}{\bar C}}=\Upsilon^{{\bar B}{\bar C}}$, a Markov decomposition of $\Upsilon^{{\bar A}{\bar B}{\bar C}}$ is obtained by ${\tilde\Gamma}'$ as
\begin{eqnarray}
\Upsilon^{{\bar A}{\bar B}{\bar C}}_{M\!k}\!:=\!{\tilde\Gamma}'^{{\bar B}}\Upsilon'^{{\bar A}{\bar B}{\bar C}}{\tilde\Gamma}'^{\dagger{\bar B}}\!=\!\sum_iq_i\proj{i}^{{\hat b}_{\scalebox{0.45}{$0$}}}\otimes\sigma_i^{{\bar A}{\hat b}_{\scalebox{0.45}{$L$}}}\otimes\phi_i^{{\hat b}_{\scalebox{0.45}{$R$}}{\bar C}}.
\label{eq:upsilonmarkovbar}
\end{eqnarray}

Due to (\ref{eq:markovdistance3}) and the monotonicity of the trace distance, we have
\begin{eqnarray}
\left\|{\mathcal V}_n(\Psi^{\otimes n})-{\mathcal V}_n(\chi^{{\bar A}{\bar B}{\bar C}})\right\|_1\leq2\sqrt{\epsilon}.\nonumber
\end{eqnarray}
Thus from (\ref{eq:markovdistance}) and the triangle inequality, we obtain
\begin{eqnarray}
\left\|{\mathcal V}_n(\chi^{{\bar A}{\bar B}{\bar C}})-\Upsilon^{{\bar A}{\bar B}{\bar C}}\right\|_1\leq2\sqrt{\epsilon}+\epsilon<3\sqrt{\epsilon}.\nonumber
\end{eqnarray}
Applying ${\tilde\Gamma}'^{\bar B}$ yields
\begin{eqnarray}
\left\|{\mathcal V}_n\left({\tilde\Gamma}'^{\bar B}\chi^{{\bar A}{\bar B}{\bar C}}{\tilde\Gamma}'^{\dagger{\bar B}}\right)-\Upsilon^{{\bar A}{\bar B}{\bar C}}_{M\!k}\right\|_1\leq3\sqrt{\epsilon},\nonumber
\end{eqnarray}
due to (\ref{eq:upsilonmarkovbar}). Hence we obtain from (\ref{eq:purechi}) that
\begin{eqnarray}
\left\|{\mathcal V}_n\left({\tilde\Gamma}^{\dagger{\bar A}}\chi_{{}_{K\!I}}^{{\bar A}{\bar B}{\bar C}}{\tilde\Gamma}^{\bar A}\right)-\Upsilon^{{\bar A}{\bar B}{\bar C}}_{M\!k}\right\|_1\leq3\sqrt{\epsilon}.
\label{eq:markovdistance36}
\end{eqnarray}
Let ${\mathcal D}^{{\hat b}_{\scalebox{0.45}{$0$}}}$ be the completely dephasing operation on ${\hat b}_{\scalebox{0.6}{$0$}}$ with respect to the basis $\{\ket{i}^{{\hat b}_{\scalebox{0.45}{$0$}}}\}_i$. From (\ref{eq:upsilonmarkovbar}), we have ${\mathcal D}^{{\hat b}_{\scalebox{0.45}{$0$}}}(\Upsilon^{{\bar A}{\bar B}{\bar C}}_{M\!k})=\Upsilon^{{\bar A}{\bar B}{\bar C}}_{M\!k}$. Thus we obtain from (\ref{eq:markovdistance36}) that
\begin{eqnarray}
\left\|({\mathcal T}'_n\otimes{\mathcal D}^{{\hat b}_{\scalebox{0.45}{$0$}}})(|\chi_{{}_{K\!I}}\rangle\!\langle\chi_{{}_{K\!I}}|)-\Upsilon_{M\!k}^{{\bar A}{\bar B}{\bar C}}\right\|_1\leq3\sqrt{\epsilon}.
\label{eq:ximarkovianized}
\end{eqnarray}
Here, we defined a random isometry operation ${\mathcal T}'_n:={\mathcal V}_n\circ{\mathcal E}_{{\tilde\Gamma}^{\dagger}}$, where ${\mathcal E}_{{\tilde\Gamma}^{\dagger}}$ is an isometry operation corresponding to ${{\tilde\Gamma}^{\dagger}}$.

Due to (\ref{eq:purechi}), we have
\begin{eqnarray}
&&\!\!\!\!\!\!\!\!{\mathcal D}^{{\hat b}_{\scalebox{0.45}{$0$}}}(\proj{\chi_{{}_{K\!I}}})\nonumber\\
&&\!\!\!\!\!=\sum_{i}{q_i}\proj{i}^{{\hat a}_{\scalebox{0.45}{$0$}}}\otimes\proj{i}^{{\hat b}_{\scalebox{0.45}{$0$}}}\otimes\proj{\xi_i}^{{\hat a}_{\scalebox{0.45}{$L$}}{\hat b}_{\scalebox{0.45}{$L$}}}\otimes\proj{\phi_i}^{{\hat a}_{\scalebox{0.45}{$R$}}{\hat b}_{\scalebox{0.45}{$R$}}{\bar C}},\;\;\;\;\;\;\nonumber
\end{eqnarray}
which leads to
\begin{eqnarray}
&&\!\!\!{\rm Tr}_{{\hat b}_{\scalebox{0.45}{$R$}}{\bar C}}\left[{\mathcal D}^{{\hat b}_{\scalebox{0.45}{$0$}}}(\proj{\chi_{{}_{K\!I}}})\right]\nonumber\\
&&=\sum_{i}{q_i}\proj{i}^{{\hat b}_{\scalebox{0.45}{$0$}}}\otimes\proj{i,\xi_i}^{{\hat a}_{\scalebox{0.45}{$0$}}{\hat a}_{\scalebox{0.45}{$L$}}{\hat b}_{\scalebox{0.45}{$L$}}}\otimes\phi_i^{{\hat a}_{\scalebox{0.45}{$R$}}}.\;\;\;\;\;\;\nonumber
\end{eqnarray}
Hence we have
\begin{eqnarray}
{\rm Tr}_{{\hat b}_{\scalebox{0.45}{$R$}}{\bar C}}\left[({\mathcal T}'_n\otimes{\mathcal D}^{{\hat b}_{\scalebox{0.45}{$0$}}})(\proj{\chi_{{}_{K\!I}}})\right]=\sum_{i}{q_i}\proj{i}^{{\hat b}_{\scalebox{0.45}{$0$}}}\otimes\phi_{i,{\mathcal T}_n'}^{{\bar A}{\hat b}_{\scalebox{0.45}{$L$}}},\nonumber
\end{eqnarray}
where we define 
\begin{eqnarray}
\phi_{i,{\mathcal T}_n'}^{{\bar A}{\hat b}_{\scalebox{0.45}{$L$}}}:={\mathcal T}'_n(\proj{i,\xi_i}^{{\hat a}_{\scalebox{0.45}{$0$}}{\hat a}_{\scalebox{0.45}{$L$}}{\hat b}_{\scalebox{0.45}{$L$}}}\otimes\phi_i^{{\hat a}_{\scalebox{0.45}{$R$}}}).\label{eq:phitn}
\end{eqnarray}
From (\ref{eq:upsilonmarkovbar}) we have
\begin{eqnarray}
{\rm Tr}_{{\hat b}_{\scalebox{0.45}{$R$}}{\bar C}}\left[\Upsilon^{{\bar A}{\bar B}{\bar C}}_{M\!k}\right]=\sum_iq_i\proj{i}^{{\hat b}_{\scalebox{0.45}{$0$}}}\otimes\sigma_i^{{\bar A}{\hat b}_{\scalebox{0.45}{$L$}}}.\nonumber
\end{eqnarray}
Therefore, by tracing out ${\hat b}_{\scalebox{0.6}{$R$}}{\bar C}$ in (\ref{eq:ximarkovianized}), we obtain
\begin{eqnarray}
&&\left\|\sum_{i}{q_i}\proj{i}^{{\hat b}_{\scalebox{0.45}{$0$}}}\otimes\phi_{i,{\mathcal T}_n'}^{{\bar A}{\hat b}_{\scalebox{0.45}{$L$}}}-\sum_iq_i\proj{i}^{{\hat b}_{\scalebox{0.45}{$0$}}}\otimes\sigma_i^{{\bar A}{\hat b}_{\scalebox{0.45}{$L$}}}\right\|_1\leq3\sqrt{\epsilon}.\nonumber
\end{eqnarray}
Thus, by Inequality (\ref{eq:fannes}), we have
\begin{eqnarray}
&&H(\{q_i\}_i)+\sum_iq_iS(\sigma_i^{{\bar A}{\hat b}_{\scalebox{0.45}{$L$}}})\nonumber\\
&&\geq H(\{q_i\}_i)+\sum_iq_iS(\phi_{i,{\mathcal T}_n'}^{{\bar A}{\hat b}_{\scalebox{0.45}{$L$}}})-\eta(3\sqrt{\epsilon})\log(d_{\bar A}d_{\bar B}).\nonumber
\end{eqnarray}
Since the von Neumann entropy is nondecreasing under random unitary operations, we have $S(\phi_{i,{\mathcal T}_n'}^{{\bar A}{\hat b}_{\scalebox{0.45}{$L$}}})\geq S(\phi_i^{{\hat a}_{\scalebox{0.45}{$R$}}})$ for each $j$ from (\ref{eq:phitn}). Hence we obtain
\begin{eqnarray}
\sum_iq_iS(\sigma_i^{{\bar A}{\hat b}_{\scalebox{0.45}{$L$}}})\geq\sum_iq_iS(\phi_i^{{\hat a}_{\scalebox{0.45}{$R$}}})-\eta(3\sqrt{\epsilon})\log(d_{\bar A}d_{\bar B}).\label{eq:jamesbond}
\end{eqnarray}

The von Neumann entropy of the state $\Upsilon^{{\bar A}{\bar B}{\bar C}}$ is then bounded below as
\begin{eqnarray}
&&\!\!\!\!\!\!\!\!S({\bar A}{\bar B}{\bar C})_{\Upsilon}=S({\bar A}{\hat b}_{\scalebox{0.6}{$0$}}{\hat b}_{\scalebox{0.6}{$L$}}{\hat b}_{\scalebox{0.6}{$R$}}{\bar C})_{\Upsilon_{M\!k}}\nonumber\\
&&\!\!\!\!\!\!\!\!=S({\hat b}_{\scalebox{0.6}{$0$}})_{\Upsilon_{M\!k}}+S({\bar A}{\hat b}_{\scalebox{0.6}{$L$}}{\hat b}_{\scalebox{0.6}{$R$}}{\bar C}|{\hat b}_{\scalebox{0.6}{$0$}})_{\Upsilon_{M\!k}}\nonumber\\
&&\!\!\!\!\!\!\!\!=H(\{q_i\}_{i})+\sum_iq_i(S(\sigma_i^{{\bar A}{\hat b}_{\scalebox{0.45}{$L$}}})+S(\phi_i^{{\hat b}_{\scalebox{0.45}{$R$}}{\bar C}}))\nonumber\\
&&\!\!\!\!\!\!\!\!=H(\{q_i\}_{i})+\sum_iq_i(S(\sigma_i^{{\bar A}{\hat b}_{\scalebox{0.45}{$L$}}})+S(\phi_i^{{\hat a}_{\scalebox{0.45}{$R$}}}))\nonumber\\
&&\!\!\!\!\!\!\!\!\geq H(\{q_i\}_{i})+2\sum_iq_iS(\phi_i^{{\hat a}_{\scalebox{0.45}{$R$}}})-n\eta(3\sqrt{\epsilon})\log(d_Ad_Bd_C)\nonumber\\
&&\!\!\!\!\!\!\!\!=S({\hat a}_{\scalebox{0.6}{$0$}})_{\chi_{K\!I}}+2S({\hat a}_{\scalebox{0.6}{$R$}}|{\hat a}_{\scalebox{0.6}{$0$}})_{\chi_{K\!I}}-n\eta(3\sqrt{\epsilon})\log(d_Ad_Bd_C).\;\;\;\;\;\;\;
\label{eq:totalentropybound}
\end{eqnarray}
Here, the third line follows from (\ref{eq:upsilonmarkovbar}); the fourth line because of $\phi_i^{{\hat a}_{\scalebox{0.45}{$R$}}{\hat b}_{\scalebox{0.45}{$R$}}{\bar C}}$ being a pure state; the fifth line by Inequality (\ref{eq:jamesbond}); and the sixth line from (\ref{eq:purechi}). From Lemma \ref{lmm:koashi}, (\ref{eq:markovdistance3}) implies
\begin{eqnarray}
&&\!\!\!\!\!\!\!\!S({\hat a}_{\scalebox{0.6}{$0$}})_{\chi_{K\!I}}+2S({\hat a}_{\scalebox{0.6}{$R$}}|{\hat a}_{\scalebox{0.6}{$0$}})_{\chi_{K\!I}}\nonumber\\
&&\!\!\!\!\!\!\!\!\geq n\left(S(a_{\scalebox{0.6}{$0$}})_{\Psi_{K\!I}}+2S(a_{\scalebox{0.6}{$R$}}|a_{\scalebox{0.6}{$0$}})_{\Psi_{K\!I}}-\zeta'_{{}_{\Psi}}\!(2\sqrt{\epsilon})\log{d_A}\right),\;\;
\label{eq:markoventbound}
\end{eqnarray}
where $\zeta'_{{}_{\Psi}}\!(\epsilon)$ is a function defined by (\ref{eq:zetaprime}) in Appendix \ref{sec:prflmm10}. Putting together (\ref{eq:totalentropybound}) and (\ref{eq:markoventbound}), we obtain
\begin{eqnarray}
\frac{1}{n}S({\bar A}{\bar B}{\bar C})_{\Upsilon}\!&\geq&\!S(a_{\scalebox{0.6}{$0$}})_{\Psi_{K\!I}}+2S(a_{\scalebox{0.6}{$R$}}|a_{\scalebox{0.6}{$0$}})_{\Psi_{K\!I}}\nonumber\\
&&\;-\left(2\eta(3\sqrt{\epsilon})+\zeta'_{{}_{\Psi}}\!(2\sqrt{\epsilon})\right)\log(d_Ad_Bd_C).\nonumber
\end{eqnarray} 
Noting that ${\mathcal V}_n(\Psi^{\otimes n})$ is a mixture of $2^{nR}$ (not necessarily orthogonal) pure states,
from (\ref{eq:markovdistance}), we finally obtain
\begin{eqnarray}
R&\geq&\frac{1}{n}S({\bar A}{\bar B}{\bar C})_{{\mathcal V}_n(\Psi^{\otimes n})}\nonumber\\
&\geq&\frac{1}{n}S({\bar A}{\bar B}{\bar C})_{\Upsilon}-\eta(\epsilon)\log(d_Ad_Bd_C)\nonumber\\
&\geq&S(a_{\scalebox{0.6}{$0$}})_{\Psi_{K\!I}}+2S(a_{\scalebox{0.6}{$R$}}|a_{\scalebox{0.6}{$0$}})_{\Psi_{K\!I}}\nonumber\\
&&\;\;\;\;-\left(3\eta(3\sqrt{\epsilon})+\zeta'_{{}_{\Psi}}\!(2\sqrt{\epsilon})\right)\log(d_Ad_Bd_C)\nonumber\\
&=&H(\{p_j\}_{j\in J})+2\sum_{j\in J}p_jS(\varphi_j^{a_{\scalebox{0.45}{$R$}}})\nonumber\\
&&\;\;\;\;-\left(3\eta(3\sqrt{\epsilon})+\zeta'_{{}_{\Psi}}\!(2\sqrt{\epsilon})\right)\log(d_Ad_Bd_C),\nonumber
\end{eqnarray} 
which implies (\ref{eq:lowermarkov}) by taking the limit of $\epsilon\rightarrow0$.\hfill$\blacksquare$

\subsection{Proof of Lemma \ref{lmm:koashi}}\label{sec:prflmm10}

The key idea for the proof of Lemma \ref{lmm:koashi} is similar to the one used in \cite{mixcomp1}. Let $\psi^{A{C'}}$ be a state such that the KI isometry on $A$ with respect to $\psi^{A{C'}}$ is the same as that with respect to $\Psi^{AC}$, and that it is decomposed as
\begin{eqnarray}
\psi_{K\!I}^{A{C'}}&:=&(\Gamma_{\Psi}^A\otimes\Gamma_{\psi}^{C'})\psi^{A{C'}}(\Gamma_{\Psi}^A\otimes\Gamma_{\psi}^{C'})^{\dagger}\nonumber\\
&=&\sum_{j\in J}p_j\proj{j}^{a_{\scalebox{0.45}{$0$}}}\otimes\omega_j^{a_{\scalebox{0.45}{$L$}}}\otimes|\tilde{\varphi}_j\rangle\!\langle\tilde{\varphi}_j|^{a_{\scalebox{0.45}{$R$}}c_{\scalebox{0.45}{$R$}}'}\otimes\proj{j}^{c_{\scalebox{0.45}{$0$}}'},\nonumber\\
\label{eq:psikidef}
\end{eqnarray}
where $\Gamma_{\psi}:{\mathcal H}^{C'}\rightarrow{\mathcal H}^{c_{\scalebox{0.45}{$0$}}'}\otimes{\mathcal H}^{c_{\scalebox{0.45}{$R$}}'}$ is an isometry and $|\tilde{\varphi}_j\rangle^{a_{\scalebox{0.45}{$R$}}c_{\scalebox{0.45}{$R$}}'}$ is a purification of ${\varphi}_j^{a_{\scalebox{0.45}{$R$}}}$.  The state satisfies $\psi^A=\Psi^A$. Note that we have
\begin{eqnarray}
d_{C'}=\sum_j{\rm rank}\:\tilde{\varphi}_j^{c_{\scalebox{0.45}{$R$}}'}=\sum_j{\rm rank}\:\tilde{\varphi}_j^{a_{\scalebox{0.45}{$R$}}}\leq d_A.\nonumber
\end{eqnarray}
Let ${\mathfrak E}$ be the set of all linear CPTP maps on ${\mathcal S}({\mathcal H}^A)$, and define two functions $f,g:{\mathfrak E}\rightarrow {\mathbb R}$ by
\begin{eqnarray}
&&f({\mathcal E})=\left\|{\mathcal E}(\Psi^{AC})-\Psi^{AC}\right\|_1,\nonumber\\
&&g({\mathcal E})=\left\|{\mathcal E}(\psi^{A{C'}})-\psi^{A{C'}}\right\|_1.\nonumber
\end{eqnarray}
Since the KI decomposition of $A$ with respect to $\Psi^{AC}$ and that with respect to $\psi^{A{C'}}$ are the same, $f({\mathcal E})=0$ if and only if $g({\mathcal E})=0$ (see Lemma \ref{lmm:equivkidec}). Define
\begin{eqnarray}
\zeta_{{}_{\Psi}}\!(\epsilon):=\sup_{{\mathcal E}\in{\mathfrak E}}{\{g({\mathcal E})|f({\mathcal E})\leq\epsilon\}}.
\label{def:zetapsi}
\end{eqnarray}
This is a monotonically nondecreasing function of $\epsilon$ by definition, and satisfies $\lim_{\epsilon\rightarrow0}\zeta_{{}_{\Psi}}\!(\epsilon)=0$ as we prove in Appendix \ref{app:compact}.

We consider a general situation in which the relation (\ref{eq:wondagold}) does not necessarily hold. Let $\Pi_\chi$ be the projection onto ${\mathcal H}_\chi^{\bar A}\subseteq{\mathcal H}^{\bar A}$, and $\Pi_\chi^\perp$ be that onto its orthogonal complement. Using a quantum channel ${\mathcal E}_{\chi}$ on ${\mathcal S}({\mathcal H}_\chi^{\bar A})$ defined by (\ref{eq:defofchannel}), construct another quantum channel ${\mathcal E}^*_{\chi}$ on ${\mathcal S}({\mathcal H}^{\bar A})$ by
\begin{eqnarray}
{\mathcal E}_{\chi}^*(\tau)={\mathcal E}_{\chi}(\Pi_\chi\tau\Pi_\chi)+\Pi_\chi^\perp\tau\Pi_\chi^\perp\;\;\;\;(\forall\tau\in{\mathcal S}({\mathcal H}^{\bar A})).\!\!\label{eq:ooiocha}
\end{eqnarray} 
Define quantum channels ${\mathcal E}_l$ on $A_l$ ($1\leq l\leq n$) by
\begin{eqnarray}
&&{\mathcal E}_l(\tau^{A_l})={\rm Tr}_{{\bar A}\setminus A_l}\left[{\mathcal E}_\chi^*\left(\Psi^{A_1}\otimes\cdots\otimes\Psi^{A_{l-1}}\otimes\tau^{A_l}\right.\right.\;\;\;\;\nonumber\\
&&\;\;\;\;\;\;\;\;\;\;\;\;\;\;\;\;\;\;\;\;\;\;\;\;\;\;\;\;\;\;\;\;\left.\left.\otimes\Psi^{A_{l+1}}\otimes\cdots\otimes\Psi^{A_{n}}\right)\right],\nonumber
\end{eqnarray}
where ${\rm Tr}_{{\bar A}\setminus A_l}$ denotes the partial trace over $A_1\cdots A_{l-1}A_{l+1}\cdots A_n$. From (\ref{eq:kiofrho}), we have ${\mathcal E}_{\chi}^*(\chi^{{\bar A}{\bar C}})=\:{\mathcal E}_{\chi}(\chi^{{\bar A}{\bar C}})=\chi^{{\bar A}{\bar C}}$, and thus from (\ref{eq:distpsirho}) and the triangle inequality, we have
\begin{eqnarray} 
\left\|{\mathcal E}_{\chi}^*(\Psi^{\otimes n})^{{\bar A}{\bar C}}-(\Psi^{\otimes n})^{{\bar A}{\bar C}}\right\|_1\leq2\epsilon,\nonumber
\end{eqnarray}
which implies
\begin{eqnarray}
\left\|{\mathcal E}_{l}(\Psi^{{A_l}{C_l}})-\Psi^{{A_l}{C_l}}\right\|_1\leq2\epsilon\nonumber
\end{eqnarray}
by taking the partial trace. Thus we have
\begin{eqnarray}
\left\|{\mathcal E}_{l}(\psi^{{A_l}{C_l'}})-\psi^{{A_l}{C_l'}}\right\|_1\leq\zeta_{{}_{\Psi}}\!(2\epsilon)\nonumber
\end{eqnarray}
for any $1\leq l\leq n$. By Inequality (\ref{eq:contmi}) and $d_{C'}\leq d_A$, it follows that
\begin{eqnarray}
I(A:C')_{\psi}-I(A_l:C_l')_{{\mathcal E}_l(\psi)}\leq 4\eta(\zeta_{{}_{\Psi}}\!(2\epsilon))\log{d_A},\nonumber
\end{eqnarray}
and consequently, that
\begin{eqnarray}
nI(A:C')_{\psi}-\sum_{l=1}^nI(A_l:C_l')_{{\mathcal E}_l(\psi)}\leq 4n\eta(\zeta_{{}_{\Psi}}\!(2\epsilon))\log{d_A}.\!\!\!\!\!\!\!\nonumber\\
\label{eq:condentdif}
\end{eqnarray}
We also have
\begin{eqnarray}
&&I({\bar A}:{\bar C'})_{{\mathcal E}_{\chi}^*(\psi^{\otimes n})}=S({\bar C'})_{{\mathcal E}_{\chi}^*(\psi^{\otimes n})}-S({\bar C'}|{\bar A})_{{\mathcal E}_{\chi}^*(\psi^{\otimes n})}\nonumber\\
&&=S({\bar C'})_{\psi^{\otimes n}}-\sum_{l=1}^nS(C'_l|A_1\cdots A_nC'_1\cdots C'_{l-1})_{{\mathcal E}_{\chi}^*(\psi^{\otimes n})}\nonumber\\
&&\geq \sum_{l=1}^nS(C_l')_{\psi}-\sum_{l=1}^nS(C'_l|A_l)_{{\mathcal E}_{\chi}^*(\psi^{\otimes n})}\nonumber\\
&&=\sum_{l=1}^nS(C_l')_{{\mathcal E}_l(\psi)}-\sum_{l=1}^nS(C_l'|A_l)_{{\mathcal E}_l(\psi)}\nonumber\\
&&=\sum_{l=1}^nI(A_l:C_l')_{{\mathcal E}_l(\psi)}.
\label{eq:condentdif2}
\end{eqnarray}
Here, we used the fact that ${\mathcal E}_\chi^*$ on $\bar A$ does not change the reduced state on $\bar C'$, and that 
\begin{eqnarray}
{\rm Tr}_{{\bar A}\setminus A_l,{\bar C'}\setminus C_l'}\left[{\mathcal E}_\chi^*\left(\psi^{\otimes{n}}\right)\right]={\mathcal E}_l(\psi^{A_lC_l'}),\nonumber
\end{eqnarray}
because of $\Psi^A_{l'}=\psi^A_{l'}$. Combining (\ref{eq:condentdif}) and (\ref{eq:condentdif2}), we obtain
\begin{eqnarray}
nI(A:C')_{\psi}\leq I({\bar A}:{\bar C'})_{{\mathcal E}_{\chi}^*(\psi^{\otimes n})}+4n\eta(\zeta_{{}_{\Psi}}\!(2\epsilon))\log{d_A}.
\label{eq:condentdif3}
\end{eqnarray}

Define
\begin{eqnarray}
\psi_{n,\chi}^{{\bar A}{\bar C'}}:=\frac{\Pi_\chi^A(\psi^{AC'})^{\otimes n}\Pi_\chi^A}{{\rm Tr}[\Pi_\chi(\psi^A)^{\otimes n}]}\nonumber
\end{eqnarray}
and
\begin{eqnarray}
\psi_{n}^{{\hat a}_{\scalebox{0.45}{$0$}}{\hat a}_{\scalebox{0.45}{$R$}}{\bar C'}}:=({\mathcal E}_{2}\circ{\mathcal E}_{1}\circ{\mathcal E}_{\Gamma_{\chi}})(\psi_{n,\chi}^{{\bar A}{\bar C'}})\nonumber
\end{eqnarray}
as depicted in Figure \ref{fig:mixcomp2}. From Condition (\ref{eq:distpsirho}) and $\psi^A=\Psi^A$, we have $\|(\psi^{\otimes n})^{\bar A}-\chi^{\bar A}\|_1\leq\epsilon$. Thus, due to (\ref{eq:ooiocha}) and Inequality (\ref{eq:namacha}), we have
\begin{align}
&\left\|{\mathcal E}_{\chi}(\psi_{n,\chi}^{{\bar A}{\bar C'}})-{\mathcal E}_{\chi}^*(\psi^{\otimes n})^{{\bar A}{\bar C'}}\right\|_1\nonumber\\
&\leq\left\|\psi_{n,\chi}^{{\bar A}{\bar C'}}-(\psi^{AC'})^{\otimes n}\right\|_1\leq2\sqrt{\epsilon},\nonumber
\end{align}
which leads to
\begin{eqnarray}
I({\bar A}:{\bar C'})_{{\mathcal E}^*_{\mathcal\chi}(\psi^{\otimes n})}\leq I({\bar A}:{\bar C'})_{{\mathcal E}_{\mathcal\chi}(\psi_{n,\chi})}+4n\eta(2\sqrt{\epsilon})\log{d_A}\!\!\!\!\!\!\!\!\nonumber\\\label{eq:aquarious}
\end{eqnarray}
by Inequality (\ref{eq:contmi}) and $d_{C'}\leq d_A$. By the data processing inequality, we also have
\begin{eqnarray}
I({\bar A}:{\bar C'})_{{\mathcal E}_{\mathcal\chi}(\psi_{n,\chi})}\leq I({\hat a}_{\scalebox{0.6}{$0$}}{\hat a}_{\scalebox{0.6}{$R$}}:{\bar C'})_{\psi_{n}}.
\label{eq:condentdif4}
\end{eqnarray}
Consequently, we obtain from (\ref{eq:condentdif3}), (\ref{eq:aquarious}) and (\ref{eq:condentdif4}) that
\begin{eqnarray}
nI(A:C')_{\psi}&\leq&I({\hat a}_{\scalebox{0.6}{$0$}}{\hat a}_{\scalebox{0.6}{$R$}}:{\bar C'})_{\psi_{n}}\nonumber\\
&&+4n\left(\eta(2\sqrt{\epsilon})+\eta(\zeta_{{}_{\Psi}}\!(2\epsilon))\right)\log{d_A}.\;\;\;\;\;
\label{eq:condentdif55}
\end{eqnarray}

The QMIs in (\ref{eq:condentdif55}) are calculated as follows. First, from (\ref{eq:kiofrho}), we have
\begin{eqnarray}
({\mathcal E}_{2}\circ{\mathcal E}_{1}\circ{\mathcal E}_{\Gamma_{\chi}})(\chi^{\bar A})=\chi_{sK\!I}^{{\hat a}_{\scalebox{0.45}{$0$}}{\hat a}_{\scalebox{0.45}{$R$}}}=\sum_iq_i\proj{i}^{{\hat a}_{\scalebox{0.45}{$0$}}}\otimes\phi_i^{{\hat a}_{\scalebox{0.45}{$R$}}}.\label{eq:chia0ar}
\end{eqnarray}
Therefore, from the monotonicity of the trace distance under ${\mathcal E}_{2}\circ{\mathcal E}_{1}\circ{\mathcal E}_{\Gamma_{\chi}}$, Equality (\ref{eq:distpsirho}) and $\psi^A=\Psi^A$, we obtain
\begin{eqnarray}
\left\|\psi_{n}^{{\hat a}_{\scalebox{0.45}{$0$}}{\hat a}_{\scalebox{0.45}{$R$}}}-\chi_{sK\!I}^{{\hat a}_{\scalebox{0.45}{$0$}}{\hat a}_{\scalebox{0.45}{$R$}}}\right\|_1\leq\left\|(\psi^{\otimes n})^{\bar A}-\chi^{\bar A}\right\|_1\leq\epsilon.\label{eq:oraora}
\end{eqnarray}
Due to (\ref{eq:fannes}), (\ref{eq:alicki}) and
\begin{eqnarray}
{\rm dim}{\mathcal H}^{{\hat a}_{\scalebox{0.45}{$0$}}},{\rm dim}{\mathcal H}^{{\hat a}_{\scalebox{0.45}{$R$}}}\leq{\rm dim}{\mathcal H}^{\bar A}=d_A^n,\nonumber
\end{eqnarray}
(\ref{eq:oraora}) implies that
\begin{eqnarray}
&&\left|S({\hat a}_{\scalebox{0.6}{$0$}})_{\chi_{sK\!I}}-S({\hat a}_{\scalebox{0.6}{$0$}})_{\psi_n}\right|\leq n\eta(\epsilon)\log{d_A},\nonumber\\
&&\left|S({\hat a}_{\scalebox{0.6}{$R$}}|{\hat a}_{\scalebox{0.6}{$0$}})_{\chi_{sK\!I}}-S({\hat a}_{\scalebox{0.6}{$R$}}|{\hat a}_{\scalebox{0.6}{$0$}})_{\psi_n}\right|\leq 3n\eta(\epsilon)\log{d_A},\nonumber
\end{eqnarray}
and consequently, that
\begin{eqnarray}
&&S({\hat a}_{\scalebox{0.6}{$0$}})_{\psi_{n}}+2S({\hat a}_{\scalebox{0.6}{$R$}}|{\hat a}_{\scalebox{0.6}{$0$}})_{\psi_{n}}\nonumber\\
&&\leq S({\hat a}_{\scalebox{0.6}{$0$}})_{\chi_{sK\!I}}+2S({\hat a}_{\scalebox{0.6}{$R$}}|{\hat a}_{\scalebox{0.6}{$0$}})_{\chi_{sK\!I}}+7n\eta(\epsilon)\log{d_A}.\;\;\;\;\nonumber
\end{eqnarray}
Since $\psi_{n}^{{\hat a}_{\scalebox{0.45}{$0$}}{\hat a}_{\scalebox{0.45}{$R$}}{\bar C'}}$ is a classical-quantum state between ${\hat a}_{\scalebox{0.6}{$0$}}$ and ${\hat a}_{\scalebox{0.6}{$R$}}{\bar C'}$, we obtain
\begin{eqnarray}
&&I({\hat a}_{\scalebox{0.6}{$0$}}{\hat a}_{\scalebox{0.6}{$R$}}:{\bar C'})_{\psi_{n}}\nonumber\\
&=&I({\hat a}_{\scalebox{0.6}{$0$}}:{\bar C'})_{\psi_{n}}+I({\hat a}_{\scalebox{0.6}{$R$}}:{\bar C'}|{\hat a}_{\scalebox{0.6}{$0$}})_{\psi_{n}}\nonumber\\
&\leq& S({\hat a}_{\scalebox{0.6}{$0$}})_{\psi_{n}}+2S({\hat a}_{\scalebox{0.6}{$R$}}|{\hat a}_{\scalebox{0.6}{$0$}})_{\psi_{n}}\nonumber\\
&\leq& S({\hat a}_{\scalebox{0.6}{$0$}})_{\chi_{sK\!I}}+2S({\hat a}_{\scalebox{0.6}{$R$}}|{\hat a}_{\scalebox{0.6}{$0$}})_{\chi_{sK\!I}}+7n\eta(\epsilon)\log{d_A}\nonumber\\
&=&H(\{q_i\}_i)+2\sum_{i}q_iS(\phi_i^{{\hat a}_{\scalebox{0.45}{$R$}}})+7n\eta(\epsilon)\log{d_A},
\label{eq:condentdif5}
\end{eqnarray}
where the last equality follows from (\ref{eq:chia0ar}).

It is straightforward to obtain from (\ref{eq:psikidef}) that
\begin{eqnarray}
I(A:C')_{\psi}&=&I(a_{\scalebox{0.6}{$0$}}a_{\scalebox{0.6}{$L$}}a_{\scalebox{0.6}{$R$}}:c_{\scalebox{0.6}{$0$}}'c_{\scalebox{0.6}{$R$}}')_{\psi_{K\!I}}\nonumber\\
&=&H(\{p_j\}_j)+2\sum_{j}p_jS(\varphi_j^{{a}_{\scalebox{0.45}{$R$}}}).
\label{eq:iacc}
\end{eqnarray}
Combining (\ref{eq:condentdif55}), (\ref{eq:condentdif5}) (\ref{eq:iacc}), we obtain
\begin{eqnarray}
&&n\left(H(\{p_j\}_j)+2\sum_{j}p_jS(\varphi_j^{{a}_{\scalebox{0.45}{$R$}}})\right)\nonumber\\
&&\leq H(\{q_i\}_i)+2\sum_{i}q_iS(\phi_i^{{\hat a}_{\scalebox{0.45}{$R$}}})+n\zeta'_{{}_{\Psi}}\!(\epsilon)\log{d_A},\nonumber
\end{eqnarray}
where
\begin{eqnarray}
\zeta'_{{}_{\Psi}}\!(\epsilon)=11\eta(2\sqrt{\epsilon})+4\eta(\zeta_{{}_{\Psi}}\!(2\epsilon))\label{eq:zetaprime}
\end{eqnarray}
and $\zeta_{{}_{\Psi}}$ is a function defined by (\ref{def:zetapsi}). Thus we finally arrive at (\ref{eq:pqcomineq}).
\hfill$\blacksquare$


\subsection{Convergence of $\zeta_{{}_{\Psi}}$}
\label{app:compact}

We prove that $\zeta_{{}_{\Psi}}\!(\epsilon)$ defined by (\ref{def:zetapsi}) satisfies $\lim_{\epsilon\rightarrow0}\zeta_{{}_{\Psi}}\!(\epsilon)=0$, based on an idea used in \cite{mixcomp1}. Due to the Choi-Jamiolkowski isomorphism, ${\mathfrak E}$ can be identified with ${\mathcal S}({\mathcal H}^A\otimes{\mathcal H}^A)$. Hence ${\mathfrak E}$ is compact, which implies that the supremum in (\ref{def:zetapsi}) can actually be the maximum:
\begin{eqnarray}
\zeta_{{}_{\Psi}}\!(\epsilon)=\max_{{\mathcal E}\in{\mathfrak E}}{\{g({\mathcal E})|f({\mathcal E})\leq\epsilon\}}.\nonumber
\end{eqnarray} 
Hence we have that
\begin{eqnarray}
\forall\epsilon>0,\:\exists{\mathcal E}\in{\mathfrak E}\;:\;g({\mathcal E})=\zeta_{{}_{\Psi}}\!(\epsilon),\:f({\mathcal E})\leq\epsilon.\nonumber
\end{eqnarray}
Define $\alpha:=\lim_{\epsilon\rightarrow0}\zeta_{{}_{\Psi}}\!(\epsilon)$. Due to the monotonicity, we have $\zeta_{{}_{\Psi}}\!(\epsilon)\geq\alpha$ for all $\epsilon>0$. Consequently, we have that
\begin{eqnarray}
\forall\epsilon>0,\:\exists{\mathcal E}\in{\mathfrak E}\;:\;g({\mathcal E})\geq\alpha,\:f({\mathcal E})\leq\epsilon.
\label{exsalpha}
\end{eqnarray}
Define ${\mathfrak E}_\alpha:=\{{\mathcal E}\in{\mathfrak E}\:|\:g({\mathcal E})\geq\alpha\}$. Due to the continuity of $g$, ${\mathfrak E}_\alpha$ is a closed subset of ${\mathfrak E}$. Hence
\begin{eqnarray}
\beta:=\min_{{\mathcal E}\in{\mathfrak E}_\alpha}f({\mathcal E})\nonumber
\end{eqnarray} 
exists due to the continuity of $f$. By definition, we have that
\begin{eqnarray}
\forall{\mathcal E}\in{\mathfrak E}:\:g({\mathcal E})\geq\alpha\Rightarrow f({\mathcal E})\geq\beta.
\label{exsbeta}
\end{eqnarray} 
Suppose now that $\alpha>0$. We have $f({\mathcal E})>0$ for all ${\mathcal E}\in{\mathfrak E}_\alpha$ due to Lemma \ref{lmm:equivkidec}. Thus we have $\beta>0$, in which case (\ref{exsbeta}) contradicts with (\ref{exsalpha}) because $\epsilon$ can be arbitrarily small.
\hfill$\blacksquare$

\hfill

\section{Proof of Theorem \ref{thm:comp}}\label{app:comp}

In this Appendix, we prove Theorem \ref{thm:comp} based on irreducibility of the KI decomposition.

\subsection{Irreducibility of the KI decomposition}

Similarly to the irreducibility of the KI decomposition of a set of states presented in Lemma \ref{lmm:irrki}, the KI decomposition of a bipartite state defined by Definition \ref{dfn:kibipart} also has a property of irreducibility as follows.

\begin{lmm}\label{lmm:iii}
Suppose the KI decomposition of $\Psi^{AC}$ on $A$ is given by
\begin{eqnarray}
\Psi_{K\!I}^{AC}=\sum_{j\in J}p_j\proj{j}^{a_{\scalebox{0.45}{$0$}}}\otimes\omega_j^{a_{\scalebox{0.45}{$L$}}}\otimes\varphi_j^{a_{\scalebox{0.45}{$R$}}C},\nonumber
\end{eqnarray}
and define $\varphi_{j,kl}:=\langle k|^C\varphi_j^{a_{\scalebox{0.45}{$R$}}C}|l\rangle^C$, where $\{|k\rangle\}_k$ is an orthonormal basis of ${\mathcal H}^C$. Then the following two properties hold.
\begin{enumerate}
\item If a linear operator $N$ on ${\mathcal H}_j^{a_{\scalebox{0.45}{$R$}}}:={\rm supp}[\varphi_j^{a_{\scalebox{0.45}{$R$}}}]$ satisfies $p_{j}N\varphi_{j,kl}=p_{j}\varphi_{j,kl}N$ for all $k$ and $l$, then $N=cI_j^{a_{\scalebox{0.45}{$R$}}}$ for a complex number $c$, where $I_j^{a_{\scalebox{0.45}{$R$}}}$ is the identity operator on ${\mathcal H}_j^{a_{\scalebox{0.45}{$R$}}}$.
\item If a linear operator $N:{\mathcal H}_j^{a_{\scalebox{0.45}{$R$}}}\rightarrow{\mathcal H}_{j'}^{a_{\scalebox{0.45}{$R$}}}\;(j\neq j')$ satisfies $p_{j}N\varphi_{j,kl}=p_{j'}\varphi_{j'\!,kl}N$ for all $k$ and $l$, then $N=0$.
\end{enumerate}
\end{lmm}

\begin{prf}
Define
\begin{eqnarray}
&&p_M:={\rm Tr}[M^C\Psi^{AC}M^{\dagger C}],\nonumber\\
&&\Psi_M:=p_M^{-1}{\rm Tr}_C[M^C\Psi^{AC}M^{\dagger C}],\nonumber\\
&&\varphi_{j,M}^{a_{\scalebox{0.45}{$R$}}}:=p_M^{-1}{\rm Tr}_C[M^C\varphi_j^{a_{\scalebox{0.45}{$R$}}C}M^{\dagger C}]\nonumber
\end{eqnarray}
for $M\in{\mathcal L}({\mathcal H}^{C})$. The set of steerable states corresponding to (\ref{eq:steerablestates}) is given by ${\mathfrak S}_{\varphi^{C\rightarrow A}}:=\{\Psi_M\}_{M\in{\mathcal L}({\mathcal H}^{C})}$. Hence the KI isometry $\Gamma$ on $A$ with respect to $\Psi^{AC}$ is equal to that with respect to ${\mathfrak S}_{\varphi^{C\rightarrow A}}$ by Definition \ref{dfn:kibipart}. Thus $\Psi_M$ is decomposed as
\begin{eqnarray}
\Gamma\Psi_M\Gamma^\dagger=\sum_{j\in J}p_j\proj{j}^{a_{\scalebox{0.45}{$0$}}}\otimes\omega_j^{a_{\scalebox{0.45}{$L$}}}\otimes\varphi_{j,M}^{a_{\scalebox{0.45}{$R$}}},\nonumber
\end{eqnarray}
where $\{\varphi_{j,M}^{a_{\scalebox{0.45}{$R$}}}\}_{M\in{\mathcal L}({\mathcal H}^{C})}$ is a set of states which is irreducible in the sense of Lemma \ref{lmm:irrki}.

To prove Property 1), suppose that $N\in{\mathcal L}({\mathcal H}_j^{a_{\scalebox{0.45}{$R$}}})$ satisfies $p_{j}N\varphi_{j,kl}=p_{j}\varphi_{j,kl}N$ for all $k$ and $l$. Since $\varphi_{j,M}$ is decomposed as $\varphi_{j,M}=\sum_{k,l}\langle l| M^\dagger M|k\rangle\varphi_{j,kl}$, it follows that $p_{j}N\varphi_{j,M}=p_{j}\varphi_{j,M}N$ for all $M\in{\mathcal L}({\mathcal H}^{C})$. Hence we obtain Property 1) due to the irreducibility of $\{\varphi_{j,M}^{a_{\scalebox{0.45}{$R$}}}\}_{M\in{\mathcal L}({\mathcal H}^{C})}$. Property 2) is proved in a similar vein. \hfill$\blacksquare$
\end{prf}

\subsection{Proof of Theorem \ref{thm:comp}}\label{app:prfmiss}

Let us first adduce a useful lemma regarding fixed points of the adjoint map of a linear CPTP map.
\begin{lmm}\label{lmm:kovacs}(See Lemma 11 in \cite{hayden04}.)
Let $\mathcal E$ be a linear CPTP map on ${\mathcal S}({\mathcal H})$, the Kraus representation of which is given by ${\mathcal E}(\cdot)=\sum_kE_k(\cdot)E_k^\dagger$. Let ${\mathcal E}^*$ be the adjoint map of $\mathcal E$ defined by ${\mathcal E}^*(\cdot)=\sum_kE_k^\dagger(\cdot)E_k$. Then $X\in{\mathcal L}({\mathcal H})$ satisfies ${\mathcal E}^*(X)=X$ if and only if $[E_k,X]=[E_k^\dagger,X]=0$ for all $k$. 
\end{lmm}

The proof of Theorem \ref{thm:comp} proceeds as follows. Let $\Gamma$ be the KI isometry on $A$ with respect to $\Psi^{AC}$, and let
\begin{eqnarray}
{\hat\Psi}^{AC}:=\Gamma^A\Psi^{AC}\Gamma^{\dagger A}=\sum_{j\in J}p_j\proj{j}^{a_{\scalebox{0.45}{$0$}}}\otimes\omega_j^{a_{\scalebox{0.45}{$L$}}}\otimes\varphi_j^{a_{\scalebox{0.45}{$R$}}C}\nonumber
\end{eqnarray}
be the KI decomposition of $\Psi^{AC}$ on $A$. We have
\begin{eqnarray}
({\hat\Psi}^{AC})^{\frac{1}{2}}&=&\sum_{j\in J}\sqrt{p_j}\proj{j}^{a_{\scalebox{0.45}{$0$}}}\otimes(\omega_j^{a_{\scalebox{0.45}{$L$}}})^{\frac{1}{2}}\otimes(\varphi_j^{a_{\scalebox{0.45}{$R$}}C})^{\frac{1}{2}},\nonumber\\
({\hat\Psi}^{A})^{-\frac{1}{2}}&=&\sum_{j\in J}\frac{1}{\sqrt{p_j}}\proj{j}^{a_{\scalebox{0.45}{$0$}}}\otimes(\omega_j^{a_{\scalebox{0.45}{$L$}}})^{-\frac{1}{2}}\otimes(\varphi_j^{a_{\scalebox{0.45}{$R$}}})^{-\frac{1}{2}}\nonumber
\end{eqnarray}
and
\begin{eqnarray}
({\hat\Psi}^{AC})^{\frac{1}{2}}({\hat\Psi}^{A})^{-\frac{1}{2}}=\sum_{j\in J}\proj{j}^{a_{\scalebox{0.45}{$0$}}}\otimes I_j^{a_{\scalebox{0.45}{$L$}}}\otimes(\varphi_j^{a_{\scalebox{0.45}{$R$}}C})^{\frac{1}{2}}(\varphi_j^{a_{\scalebox{0.45}{$R$}}})^{-\frac{1}{2}}.\nonumber
\end{eqnarray}
Hence the Kraus operators of $\mathcal E$ defined by (\ref{eq:krause}) is decomposed as (\ref{eq:hatekl}).

It follows from (\ref{eq:defmathe}) that ${\mathcal E}(\Psi^A)=\Psi^A$. Due to Lemma \ref{lmm:kovacs}, we have
\begin{eqnarray}
\forall k,l\:;\;\;\;[E_{kl},\Psi^A]=0,\nonumber
\end{eqnarray}
or equivalently, have
\begin{eqnarray}
\forall k,l\:;\;\;\;[{\hat E}_{kl},{\hat \Psi}^A]=0,\nonumber
\end{eqnarray}
from which it follows that
\begin{eqnarray}
\forall j,k,l\:;\;\;\;[{e}_{j,kl},{\varphi}_j^{a_{\scalebox{0.45}{$R$}}}]=0.\nonumber
\end{eqnarray}
Using (\ref{eq:defejkl}), we obtain
\begin{eqnarray}
\forall j,k,l\:;\;\;\;[\varphi_{j,kl},{\varphi}_j^{a_{\scalebox{0.45}{$R$}}}]=0,\nonumber
\end{eqnarray}
where $\varphi_{j,kl}:=\langle k|^C\varphi_j^{a_{\scalebox{0.45}{$R$}}C}|l\rangle^C$. Therefore, due to the irreducibility of the KI decomposition, we have that
\begin{eqnarray}
\varphi_j^{a_{\scalebox{0.45}{$R$}}}=\pi_j^{a_{\scalebox{0.45}{$R$}}}:=I_j^{a_{\scalebox{0.45}{$R$}}}\!/({\rm dim}{\mathcal H}_j^{a_{\scalebox{0.45}{$R$}}}),\nonumber
\end{eqnarray}
 and consequently, that   
\begin{eqnarray}
e_{j,kl}=({\rm dim}{\mathcal H}_j^{a_{\scalebox{0.45}{$R$}}})^{\frac{1}{2}}\varphi_{j,kl},\nonumber
\end{eqnarray}
which implies the irreducibility of $\{e_{j,kl}\}_{k,l}$.

From (\ref{eq:krause}), for any $\hat{\tau}=\Gamma\tau\Gamma^\dagger\:(\tau\in{\mathcal S}({\mathcal H}^A))$, we have
\begin{eqnarray}
\langle{j}|^{a_{\scalebox{0.45}{$0$}}}\hat{\mathcal E}({\hat\tau})|j'\rangle^{a_{\scalebox{0.45}{$0$}}}=({\rm id}^{a_{\scalebox{0.45}{$L$}}}\otimes{\mathcal E}_{jj'}^{a_{\scalebox{0.45}{$R$}}})(\langle{j}|^{a_{\scalebox{0.45}{$0$}}}{\hat\tau}|j'\rangle^{a_{\scalebox{0.45}{$0$}}}),\nonumber
\end{eqnarray}
where ${\mathcal E}_{jj'}$ is a linear map on ${\mathcal L}({\mathcal H}^{a_{\scalebox{0.45}{$R$}}})$ defined by
\begin{eqnarray}
{\mathcal E}_{jj'}(\cdot)=\sum_{kl}e_{j,kl}(\cdot)e_{j',kl}^\dagger.\nonumber
\end{eqnarray}
Thus we have
\begin{eqnarray}
\langle{j}|^{a_{\scalebox{0.45}{$0$}}}\hat{\mathcal E}_\infty({\hat\tau})|j'\rangle^{a_{\scalebox{0.45}{$0$}}}=({\rm id}^{a_{\scalebox{0.45}{$L$}}}\otimes{\mathcal E}_{\infty,jj'}^{a_{\scalebox{0.45}{$R$}}})(\langle{j}|^{a_{\scalebox{0.45}{$0$}}}{\hat\tau}|j'\rangle^{a_{\scalebox{0.45}{$0$}}}),\nonumber
\end{eqnarray}
where ${\mathcal E}_{\infty,jj'}$ is a linear map on ${\mathcal L}({\mathcal H}^{a_{\scalebox{0.45}{$R$}}})$ defined by
\begin{eqnarray}
{\mathcal E}_{\infty,jj'}:=\lim_{n\rightarrow\infty}\frac{1}{N}\sum_{n=1}^N{\mathcal E}_{jj'}^n.\nonumber
\end{eqnarray}
Therefore, from
\begin{eqnarray}
|\Psi_{K\!I}\rangle=\sum_{j\in J}\sqrt{p_j}\ket{j}^{a_{\scalebox{0.45}{$0$}}}\ket{j}^{b_{\scalebox{0.45}{$0$}}}\ket{\omega_j}^{a_{\scalebox{0.45}{$L$}}b_{\scalebox{0.45}{$L$}}}\ket{\varphi_j}^{a_{\scalebox{0.45}{$R$}}b_{\scalebox{0.45}{$R$}}C},\nonumber
\end{eqnarray}
we obtain
\begin{eqnarray}
&&\!\!\!\!\!\!\hat{\mathcal E}_\infty(|\Psi_{K\!I}\rangle\!\langle\Psi_{K\!I}|)\nonumber\\
&&\!\!\!\!\!\!\!\!\!\!=\sum_{jj'}|j\rangle\!\langle{j}|^{a_{\scalebox{0.45}{$0$}}}\hat{\mathcal E}_\infty(|\Psi_{K\!I}\rangle\!\langle\Psi_{K\!I}|)|j'\rangle\!\langle j'|^{a_{\scalebox{0.45}{$0$}}}\nonumber\\
&&\!\!\!\!\!\!\!\!\!\!=\sum_{jj'}\sqrt{p_jp_{j'}}|j\rangle\!\langle{j'}|^{a_{\scalebox{0.45}{$0$}}}\otimes|j,{\omega_j}\rangle\!\langle j',\omega_{j'}|^{b_{\scalebox{0.45}{$0$}}a_{\scalebox{0.45}{$L$}}b_{\scalebox{0.45}{$L$}}}\nonumber\\
&&\;\;\;\;\;\;\;\;\;\;\;\;\otimes\:{\mathcal E}_{\infty,jj'}^{a_{\scalebox{0.45}{$R$}}}(|\varphi_j\rangle\!\langle\varphi_{j'}|^{a_{\scalebox{0.45}{$R$}}b_{\scalebox{0.45}{$R$}}C}).\;\;\;\;\;\;\;\;\;\;\;\;\;\;\;\;\label{eq:einftykipsi}
\end{eqnarray}

Consider that we have ${\mathcal E}\circ{\mathcal E}_\infty={\mathcal E}_\infty$, and thus have ${\mathcal E}_\infty^{A}(\Psi_\infty^{ABC})=\Psi_\infty^{ABC}$. It follows that
\begin{eqnarray}
\forall k,l\:;\;\;\;[E_{kl}^A\otimes I^{BC},\Psi_\infty^{ABC}]=0,\nonumber
\end{eqnarray}
and consequently, that
\begin{eqnarray}
\forall k,l\:;\;\;\;[{\hat E}_{kl}^A\otimes I^{BC},\hat{\mathcal E}_\infty(|\Psi_{K\!I}\rangle\!\langle\Psi_{K\!I}|)]=0.\nonumber
\end{eqnarray}
Due to (\ref{eq:hatekl}) and (\ref{eq:einftykipsi}), this is equivalent to
\begin{eqnarray}
e_{j,kl}{\mathcal E}_{\infty,jj'}^{a_{\scalebox{0.45}{$R$}}}(|\varphi_j\rangle\!\langle\varphi_{j'}|^{a_{\scalebox{0.45}{$R$}}b_{\scalebox{0.45}{$R$}}C})={\mathcal E}_{\infty,jj'}^{a_{\scalebox{0.45}{$R$}}}(|\varphi_j\rangle\!\langle\varphi_{j'}|^{a_{\scalebox{0.45}{$R$}}b_{\scalebox{0.45}{$R$}}C})e_{j',kl}^\dagger\!\!\!\!\!\!\!\!\!\!\nonumber\\
\!\!\!(\forall j,j',k,l).\nonumber
\end{eqnarray}
Hence we have
\begin{eqnarray}
{\mathcal E}_{\infty,jj'}^{a_{\scalebox{0.45}{$R$}}}(|\varphi_j\rangle\!\langle\varphi_{j'}|^{a_{\scalebox{0.45}{$R$}}b_{\scalebox{0.45}{$R$}}C})=\begin{cases}\pi_{j}^{a_{\scalebox{0.45}{$R$}}}\otimes\varphi_{j}^{b_{\scalebox{0.45}{$R$}}C}&(j=j')\\
0&(j\neq j')
\end{cases}\nonumber
\end{eqnarray}
due to the irreducibility of $\{e_{j,kl}\}_{k,l}$. From (\ref{eq:einftykipsi}), we obtain
\begin{eqnarray}
&&\hat{\mathcal E}_\infty(|\Psi_{K\!I}\rangle\!\langle\Psi_{K\!I}|)\nonumber\\
&&=\sum_{j\in J}p_j\proj{j,j,\omega_j}^{a_{\scalebox{0.45}{$0$}}b_{\scalebox{0.45}{$0$}}a_{\scalebox{0.45}{$L$}}b_{\scalebox{0.45}{$L$}}}\otimes\pi_{j}^{a_{\scalebox{0.45}{$R$}}}\otimes\varphi_{j}^{b_{\scalebox{0.45}{$R$}}C}.\;\;\;\;\;\;\;\;\nonumber
\end{eqnarray}
Since $\hat{\mathcal E}_\infty(|\Psi_{K\!I}\rangle\!\langle\Psi_{K\!I}|)$ and $\Psi_\infty^{ABC}$ are equivalent up to local isometries on $A$ and $B$, we finally obtain
\begin{eqnarray}
S(\Psi_\infty^{ABC})\!\!\!\!&=&\!\!\!\!H(\{p_j\})+\sum_jp_j\left(S(\pi_j^{a_{\scalebox{0.45}{$R$}}})+S(\varphi_{j}^{b_{\scalebox{0.45}{$R$}}C})\right)\nonumber\\
&=&\!\!\!\!H(\{p_j\})+2\sum_jp_jS(\varphi_j^{a_{\scalebox{0.45}{$R$}}})\nonumber\\
&=&M_{A|B}(\Psi^{ABC}),\nonumber
\end{eqnarray}
where we used the fact that $\varphi_j^{a_{\scalebox{0.45}{$R$}}}=\pi_j^{a_{\scalebox{0.45}{$R$}}}$ and that $\varphi_{j}^{a_{\scalebox{0.45}{$R$}}b_{\scalebox{0.45}{$R$}}C}$ is a pure state.
\hfill$\blacksquare$

\begin{rmk}

From (\ref{def:einfty2}) and (\ref{eq:miss}), it is straightforward to verify that the statement of Theorem \ref{thm:comp} does not depend on a particular choice of a purification of $\Psi^A$. That is, for any purification $|\Omega\rangle^{AA'}$ of $\Psi^A$, we have $M_{A|B}(\Psi^{ABC})=S(\Omega_\infty^{AA'})$, where $\Omega_\infty^{AA'}:={\mathcal E}_\infty^A(|\Omega\rangle\!\langle\Omega|^{AA'})$. A purification of $\Psi^A$ is simply obtained by
\begin{eqnarray}
|\Omega\rangle^{AA'}=(\Psi^A)^{\frac{1}{2}}\sum_{k=1}^{d_A}|k\rangle^A|k\rangle^{A'},\nonumber
\end{eqnarray}
and its matrix representation is given by
\begin{eqnarray}
\proj{\Omega}^{AA'}=\sum_{klmn}[\Lambda_1]_{kl,mn}|k\rangle\!\langle l|^A\otimes|m\rangle\!\langle n|^{A'}.\nonumber
\end{eqnarray}
The matrix elements of the generalized inverse matrix $\Lambda_1^{-1}$ are given by 
\begin{eqnarray}
\![\Lambda_1^{-1}]_{kl,mn}\!\!=\!\!\langle k|(\Psi^A)^{-\frac{1}{2}}|m\rangle\langle n|(\Psi^A)^{-\frac{1}{2}}|l\rangle.\nonumber
\end{eqnarray}
In addition, the dimension of the eigensubspace of $\Lambda_\infty$ corresponding to the eigenvalue 1 is at least 1, since we have ${\mathcal E}_\infty(I)=I$ due to the self-adjointness of $\mathcal E$. These facts justify the algorithm described in Section \ref{sec:results}.
\end{rmk}

\hfill

\section{Proof of Inequality (\ref{eq:markovqcmi})}\label{app:prfmarkovqcmi}

The first inequality in (\ref{eq:markovqcmi}) is proved as follows. For an arbitrary $n$ and $\epsilon>0$, let ${\mathcal V}_n:\tau\mapsto2^{-nR}\sum_{k=1}^{2^{nR}}V_k\tau V_k^{\dagger}$ be a random unitary operation on $A^n$, and let $\Upsilon^{A^nB^nC^n}$ be a Markov state conditioned by $B^n$ such that
\begin{eqnarray}
\left\|{\mathcal V}_n(\rho^{\otimes n})-\Upsilon^{A^nB^nC^n}\right\|_1\leq\epsilon.\label{eq:defmarkovianizing2}
\end{eqnarray}
Let $|\psi\rangle^{ABCD}$ be a purification of $\rho^{ABC}$, and $E$ be a quantum system with dimension $2^{nR}$. Defining an isometry $W:A^n\rightarrow EA^n$ by $W=\sum_{k=1}^{2^{nR}}|k\rangle^{E}\otimes V_k^{A^n}$, a Stinespring dilation of ${\mathcal V}_n$ is given by ${\mathcal V}_n(\tau)={\rm Tr}_E[W\tau W^\dagger]$. Then a purification of $\rho_n'^{ABC}:={\mathcal V}_n(\rho^{\otimes n})$ is given by $|\psi_n'\rangle^{EA^nB^nC^nR^n}:=W(|\psi\rangle^{ABCR})^{\otimes n}$. For this state, we have
\begin{eqnarray}
nR\geq S(E)_{\psi'_n}&=&S(A^nB^nC^nR^n)_{\psi_n'}\nonumber\\
&\geq&S(A^nB^nC^n)_{\psi_n'}-S(R^n)_{\psi_n'}\nonumber\\
&=&S(A^nB^nC^n)_{\rho_n'}-S(R^n)_{\psi^{\otimes n}}\nonumber\\
&=&S(A^nB^nC^n)_{\rho_n'}-nS(ABC)_{\rho},\;\;\;\;\label{eq:qcmient1}
\end{eqnarray}
where the second line follows from (\ref{eq:subadditivity2}). From (\ref{eq:defmarkovianizing2}), we also have
\begin{eqnarray}
&&S(A^nB^nC^n)_{\rho_n'}\nonumber\\
&\geq&S(A^nB^nC^n)_{\Upsilon}-n\eta(\epsilon)\log{(d_Ad_Bd_C)}\nonumber\\
&=&S(A^nB^n)_{\Upsilon}+S(B^nC^n)_{\Upsilon}-S(B^n)_{\Upsilon}\nonumber\\
&&\;\;\;-n\eta(\epsilon)\log{(d_Ad_Bd_C)}\nonumber\\
&\geq&S(A^nB^n)_{\rho_n'}+S(B^nC^n)_{\rho_n'}-S(B^n)_{\rho_n'}\nonumber\\
&&\;\;\;-4n\eta(\epsilon)\log{(d_Ad_Bd_C)}\nonumber\\
&\geq&S(A^nB^n)_{\rho^{\otimes n}}+S(B^nC^n)_{\rho^{\otimes n}}-S(B^n)_{\rho^{\otimes n}}\nonumber\\
&&\;\;\;-4n\eta(\epsilon)\log{(d_Ad_Bd_C)}\nonumber\\
&=&n\left(S(AB)_{\rho}+S(BC)_{\rho}-S(B)_{\rho}\right)\nonumber\\
&&\;\;\;-4n\eta(\epsilon)\log{(d_Ad_Bd_C)}.\label{eq:qcmient2}
\end{eqnarray}
Here, the second line follows by Inequality (\ref{eq:fannes}); the third line because of $\Upsilon$ being a Markov state conditioned by $B^n$; the fourth line by Inequality (\ref{eq:fannes}); and the fifth line by the von Neumann entropy being nondecreasing under random unitary operations, in addition to $\rho_n'^{B^nC^n}=(\rho^{BC})^{\otimes n}$. From (\ref{eq:qcmient1}) and (\ref{eq:qcmient2}), we obtain
\begin{eqnarray}
R\geq I(A:C|B)_\rho-4\eta(\epsilon)\log{(d_Ad_Bd_C)},\nonumber
\end{eqnarray}
which concludes the proof.\hfill$\blacksquare$

\end{document}